\documentclass[a4paper,fleqn,usenatbib]{mnras}

\usepackage{newtxtext,newtxmath}
\usepackage[T1]{fontenc}
\usepackage{ae,aecompl}
\usepackage{graphicx}	
\usepackage{amsmath}	
\usepackage{amssymb}	
\usepackage{multirow}      
\usepackage{pdflscape}
\usepackage{times}
\usepackage[dvipsnames]{xcolor}
\usepackage{xspace}

\newcommand{\mlc}{\multicolumn}
\newcommand{\mlr}{\multirow}

\newcommand{\hip}{\ensuremath{\mathbf{x}}\xspace}
\newcommand{\hhi}{\ensuremath{\mathbf{\hat{x}}}\xspace}
\newcommand{\dat}{\ensuremath{\{\mathcal{D}_i\}}\xspace}
\newcommand{\pos}[2]{\ensuremath{\pi_N\left(#1\,\middle|\, #2\right)}\xspace}
\newcommand{\pri}[1]{\ensuremath{\pi_0\left(#1\right)}\xspace}
\newcommand{\lik}[3][]{\ensuremath{\mathcal{L}_{#1}\left(#2\,\middle|\, #3\right)}\xspace}

\newcommand{\unit}[2][]{\ensuremath{#1\,\textrm{#2}}\xspace}
\newcommand{\SL}{\textsc{starlight}\xspace}

\newcommand{\tg}{TGASPEX\xspace}
\newcommand{\dynbas}{DynBaS\xspace}
\newcommand{\tform}{\ensuremath{t_\text{form}}\xspace}

\newcommand{\mburst}{\ensuremath{M_\text{burst}}\xspace}

\newcommand{\logm}{\ensuremath{\log{M_*/\text{M}_\odot}}\xspace}
\newcommand{\mwla}{\ensuremath{\left<\log{t_*/\text{yr}}\right>_M}\xspace}
\newcommand{\lwla}{\ensuremath{\left<\log{t_*/\text{yr}}\right>_L}\xspace}
\newcommand{\mwlz}{\ensuremath{\left<\log{Z_*/\text{Z}_\odot}\right>_M}\xspace}
\newcommand{\lwlz}{\ensuremath{\left<\log{Z_*/\text{Z}_\odot}\right>_L}\xspace}
\newcommand{\extv}{\ensuremath{A_V}\xspace}

\newcommand{\dlogm}{\ensuremath{\Delta\log{M_*}}\xspace}
\newcommand{\dmwla}{\ensuremath{\Delta\left<\log{t_*}\right>_M}\xspace}
\newcommand{\dlwla}{\ensuremath{\Delta\left<\log{t_*}\right>_L}\xspace}
\newcommand{\dmwlz}{\ensuremath{\Delta\left<\log{Z_*}\right>_M}\xspace}
\newcommand{\dlwlz}{\ensuremath{\Delta\left<\log{Z_*}\right>_L}\xspace}
\newcommand{\dextv}{\ensuremath{\Delta\extv}\xspace}

\newcommand{\clogm}{\ensuremath{\delta\log{M_*}}\xspace}
\newcommand{\cmwla}{\ensuremath{\delta\left<\log{t_*}\right>_M}\xspace}
\newcommand{\clwla}{\ensuremath{\delta\left<\log{t_*}\right>_L}\xspace}
\newcommand{\cmwlz}{\ensuremath{\delta\left<\log{Z_*}\right>_M}\xspace}
\newcommand{\clwlz}{\ensuremath{\delta\left<\log{Z_*}\right>_L}\xspace}
\newcommand{\cextv}{\ensuremath{\delta\extv}\xspace}

\newcommand{\glwla}{\ensuremath{\delta_\text{G05}\left<\log{t_*}\right>_L}\xspace}

\newcommand{\glwlz}{\ensuremath{\delta_\text{G05}\left<\log{Z_*}\right>_L}\xspace}

\newcommand{\Oi}{\ensuremath{\left[\textrm{O~\textsc{i}}\right]\lambda 6300}\xspace}
\newcommand{\Oii}{\ensuremath{\left[\textrm{O~\textsc{ii}}\right]\lambda\lambda 3726,3729}\xspace}
\newcommand{\Oiii}{\ensuremath{\left[\textrm{O~\textsc{iii}}\right]\lambda\lambda 4959,5007}\xspace}
\newcommand{\Hei}{\ensuremath{\textrm{He~\textsc{i}}\lambda 5876}\xspace}

\newcommand{\Nii}{\ensuremath{\left[\textrm{N~\textsc{ii}}\right]\lambda\lambda 6548,6583}\xspace}
\newcommand{\Sii}{\ensuremath{\left[\textrm{S~\textsc{ii}}\right]\lambda\lambda 6717,6731}\xspace}

\newcommand{\Dn}{\ensuremath{\textrm{D}4000}\xspace}
\newcommand{\Ha}{\ensuremath{\textrm{H}\alpha}\xspace}
\newcommand{\Hb}{\ensuremath{\textrm{H}\beta}\xspace}
\newcommand{\Hg}{\ensuremath{\textrm{H}\gamma}\xspace}
\newcommand{\Hdg}{\ensuremath{\textrm{H}\delta_A\!+\!\textrm{H}\gamma_A}\xspace}
\newcommand{\MgFe}{\ensuremath{\left[\textrm{Mg}_2\textrm{Fe}\right]}\xspace}
\newcommand{\MgbFe}{\ensuremath{\left[\textrm{MgFe}\right]'}\xspace}

\defcitealias{Bruzual2003}{BC03}
\defcitealias{Chen2012}{C12}
\defcitealias{Gallazzi2005}{G05}
\defcitealias{Magris2015}{M15}

\title[Galaxy properties from J-PAS narrow-band photometry]{Galaxy properties from J-PAS narrow-band photometry}

\author[A. Mej\'ia-Narv\'aez, G. Bruzual, G. Magris et al.]{%
A. Mej\'ia-Narv\'aez$^{1,2}$\thanks{E-mail: mejia@cida.gob.ve}, %
G. Bruzual$^{3}$, %
G. Magris C.$^{1}$, %
J. S. Alcaniz$^{4}$, %
N. Ben\'itez$^{5}$, \and%
S. Carneiro$^{6}$, %
A. J. Cenarro$^{7}$, %
D. Crist\'obal-Hornillos$^{7}$, %
R. Dupke$^{4,8}$, %
A. Ederoclite$^{7}$, \and%
A. Mar\'in-Franch$^{7}$, %
C. Mendes de Oliveira$^{9}$, %
M. Moles$^{7}$, %
L. Sodre Jr.$^{9}$, %
K. Taylor$^{4,9}$, \and%
J. Varela$^{7}$, and
H. V\'azquez Rami\'o$^{7}$ \and%
\\
$^{1}${Centro de Investigaciones de Astronom\'ia, AP 264, M\'erida 5101-A, Venezuela.}\\
$^{2}${Posgrado en F\'isica Fundamental, Universidad de Los Andes, M\'erida, Venezuela.}\\
$^{3}${Instituto de Radioastronom\'ia y Astrof\'isica, IRyA, UNAM, Campus Morelia, A.P. 3-72, C.P. 58089, Morelia, Michoac\'an, M\'exico.}\\
$^{4}${Observat\'orio Nacional-MCT, Rua Jos\'e Cristino, 77. CEP 20921-400 Rio de Janeiro-RJ, Brazil.}\\
$^{5}${Instituto de Astrof\'isica de Andaluc\'ia (IAA-CSIC), Glorieta de la Astronom\'ia s/n, E-18008 Granada, Spain.}\\
$^{6}${Instituto de F\'isica, Universidade Federal da Bahia, 40210-340 Salvador, Bahia, Brazil.}\\
$^{7}${Centro de Estudios de F\'isica del Cosmos de Arag\'on, Plaza San Juan 1, E-44001 Teruel, Spain.}\\
$^{8}${Department of Astronomy, University of Michigan, Ann Arbor, MI 48109, USA.}\\
$^{9}${Instituto de Astronomia, Geof\'isica e Ci\^encias Atmosf\'ericas, Universidade de S\~ao Paulo, Cidade Universit\'aria, 05508-090 S\~ao Paulo, Brazil.}}

\begin{document}

\maketitle

\begin{abstract}

We study the consistency of the physical properties of galaxies retrieved from SED-fitting as a
function of spectral resolution and signal-to-noise ratio (SNR). Using a selection of physically
motivated star formation histories, we set up a control sample of mock galaxy spectra representing
observations of the local Universe in high-resolution spectroscopy, and in 56 narrow-band and 5
broad-band photometry. We fit the SEDs at these spectral resolutions and compute their corresponding
the stellar mass, the mass- and luminosity-weighted age and metallicity, and the dust extinction. We
study the biases, correlations, and degeneracies affecting the retrieved parameters and explore the
r\^ole of the spectral resolution and the SNR in regulating these degeneracies. We find that
narrow-band photometry and spectroscopy yield similar trends in the physical properties derived, the
former being considerably more precise. Using a galaxy sample from the SDSS, we compare more
realistically the results obtained from high-resolution and narrow-band SEDs (synthesized from the
same SDSS spectra) following the same spectral fitting procedures. We use results from the
literature as a benchmark to our spectroscopic estimates and show that the prior PDFs, commonly
adopted in parametric methods, may introduce biases not accounted for in a Bayesian framework. We
conclude that narrow-band photometry yields the same trend in the age-metallicity relation in the
literature, provided it is affected by the same biases as spectroscopy; albeit the precision
achieved with the latter is generally twice as large as with the narrow-band, at SNR values typical
of the different kinds of data.

\end{abstract}

\begin{keywords}
galaxies: formation -- galaxies: evolution -- galaxies: stellar content
\end{keywords}

\section{Introduction}
Galaxies are one of the most fundamental building blocks of the visible Universe. Understanding the
processes of their formation and evolution is fundamental to constrain theories aimed to explain the
development of the large-scale structure observed today. Towards this goal we rely on the
\emph{light} arriving from distant galaxies, gathered either through photometric or spectroscopic
detectors. Spectroscopic observations provide detailed information on a galaxy spectral energy
distribution (SED), but due to their high telescope cost, the bulk of the observations collected
until now are photometric.\footnote{Energy distributions measured through a series of conveniently
located photometric bands can be regarded as \emph{ultra-low resolution} SEDs.} In the last decade a
number of photometric sky surveys have been completed. The Sloan Digital Sky Survey
\citep[SDSS;][]{York2000} in the optical wavelength range, the panchromatic Great Observatories
Origins Deep Survey \citep[GOODS;][]{Giavalisco2004}, and the Cosmic Assembly Near-infrared Deep
Extragalactic Legacy Survey \citep[CANDELS;][]{Koekemoer2011} provide imaging data from the UV to
the NIR. The availability of these surveys triggered the development of techniques to extract
reliable information on the physical properties of galaxies from their SEDs.

It has long been known that the SED encodes information on a galaxy stellar content
\citep{Morgan1956, Wood1966, Faber1972, Tinsley1972}. Since the pioneering work of
\citet{Morgan1956} on inverse spectral synthesis (IS hereafter), and \citet{Tinsley1972} on stellar
population synthesis (SPS), many authors have explored this subject and developed powerful
techniques to decode the star formation history (SFH) of unresolved galaxies from their SEDs. SPS
\citep{Bruzual2003, Maraston2005, Conroy2010b} provides the simple stellar population (SSP) models
feeding both models of galaxy formation and evolution \citep[e.\,g.,][]{DeLucia2007, Chen2012,
Vogelsberger2014}, and IS codes \citep[e.\,g.,][]{CidFernandes2005}. SED fitting has thus become a
standard procedure to extract physical information \emph{directly} from galaxy observations, while
the adequacy of the SSP models to reproduce such observations is tested on each fit.

Despite this success, prevailing uncertainties on key aspects of the theories of star formation,
e.\,g., the universality of the initial mass function, \citep[][]{Bastian2010, Conroy2013b}, and
stellar evolution, e.\,g., the thermally pulsing phase of AGB stars \citep{MacArthur2010, Kriek2010,
Zibetti2013}, and on the properties of the interstellar medium and their dependence on galaxy type
through cosmic time, e.\,g., the properties of dust present in the ISM \citep{Kobayashi2013,
Kriek2013}, propagate through the integrated spectral analysis, as also do possible instrumentation
and calibration errors in spectroscopic or photometric galaxy surveys \citep[e.\,g.][]{Conroy2010a}.
The net result of these uncertainties is to blur the conclusions drawn from SED-fitting studies
alone.

Apart from the availability of large data samples mentioned above, the use of photometric
observations is attractive for a number of reasons. In spectroscopic surveys the flux calibration,
the sky emission, along with the limitations introduced by the aperture and multiplexing, are common
issues which introduce unwanted sources of systematics in the derived stellar content of galaxies
\citep[see][for reviews]{Walcher2011, Conroy2013a}. Photometric surveys, on the other hand, are in
principle free from these issues. Furthermore, broad-band optical galaxy colours show little
sensitivity to the IMF slope \citep{Hansson2012} and to complex abundance patterns, such as the
enhancement of $\alpha$ elements in early-type galaxies \citep{Greggio1997, Maraston2003}.
SED-fitting of photometric data can then proceed safely assuming a universal IMF and the widely
implemented solar abundance pattern. Nonetheless, this advantage is defeated by the lack of ability
of most photometric observations to provide any clue on the metal content of galaxies
\citep[however, see][]{Bell2000, MacArthur2010}, given the high spectral resolution required to
reach an accurate estimate of this parameter \citep[e.\,g.][]{Pforr2012}. Likewise, estimating
photometric redshifts usually demands the implementation of sophisticated methods in order to
overcome multiple degeneracies and achieve results comparable to spectroscopic redshifts
\citep[e.\,g.][]{Benitez2000, Oyaizu2008}.

The ideal galaxy survey designed to provide reliable estimations of the stellar content of
unresolved galaxies would thus combine the strengths of spectroscopy and photometry, namely:
spectral resolution and wavelength sampling good enough to allow for accurate determination of
physical properties (including metallicity and redshift), and large and deep sky coverage to allow
for large volume-limited samples. Photometric surveys with these characteristics and goals already
exist, e.\,g. COMBO-17 \citep{Wolf2003} and ALHAMBRA \citep{Moles2008}, allowing for the study of
the global physical properties of distant galaxies \citep[see e.\,g.][]{Diaz2015}, but still
restricted to certain aspects of galaxy formation and evolution because of limited sky coverage
and/or coarse wavelength sampling, due to the use of a small number of intermediate width
($\text{FWHM}>\unit[200]{\AA}$) passbands.

The Javalambre-PAU Astrophysical Survey \citep[J-PAS][]{Benitez2014, Dupke2015} will gather data for
$\sim1/5$ of the sky in $54$ narrow-band ($\text{FWHM}=\unit[145]{\AA}$) and $2$ broad-band filters,
covering the optical range from $\sim\unit[3500]{}$ up to $\unit[10000]{\AA}$
\citep{Marin-Franch2015}. The main goal of the J-PAS collaboration is to measure the baryonic
acoustic oscillations (BAOs), for which the instruments are designed to provide photo-$z$ estimates
for a large number of galaxies with accuracy $\approx 0.003(1+z)$, comparable to spectroscopic
redshifts. Therefore, this survey will be a unique laboratory for galaxy formation and evolution
studies. Given the wavelength coverage of J-PAS, it is anticipated that the galaxy mass, age, and
metallicity estimates will supersede the limitations imposed by traditional photometry, but the
figures of merit still remain to be derived. On the eve of the start of J-PAS, in this paper we
study to what extent the properties of galaxies can be derived from its narrow-band photometric data
using a non-parametric SED fitting code such as \dynbas, described by \citet[][hereafter
\citetalias{Magris2015}]{Magris2015}. We select a test sample of $\sim10^4$ nearby galaxies
($z<0.1$) from the SDSS-DR7 \citep{York2000, Abazajian2009} and synthesize the photometry through
the $56$ narrow-band response functions. This sample spans a variety of SFHs, allowing us to
confront the well-studied age-metallicity relation (AMR) derived from spectroscopic data
\citep[see][and references therein]{Gallazzi2005, Sanchez2012, Gonzalez2014}, with our photometric
derivation of the AMR.

In \S2 and \S3 we describe the galaxy samples and the SED fitting method, respectively. In \S4 we
compare the merits of SED-fitting galaxy spectra at the resolution of spectroscopic and narrow-band
photometric data sets, using mock data. In \S5 we analyse the insights on stellar metallicity
obtained from spectroscopic SED fits, and, using this result as a benchmark, we study the
corresponding results for the narrow-band fits, both based on observed data. In \S6 we present our
conclusions.

\section{Galaxy samples}\label{samples}
\begin{figure}
\includegraphics[scale=1.0]{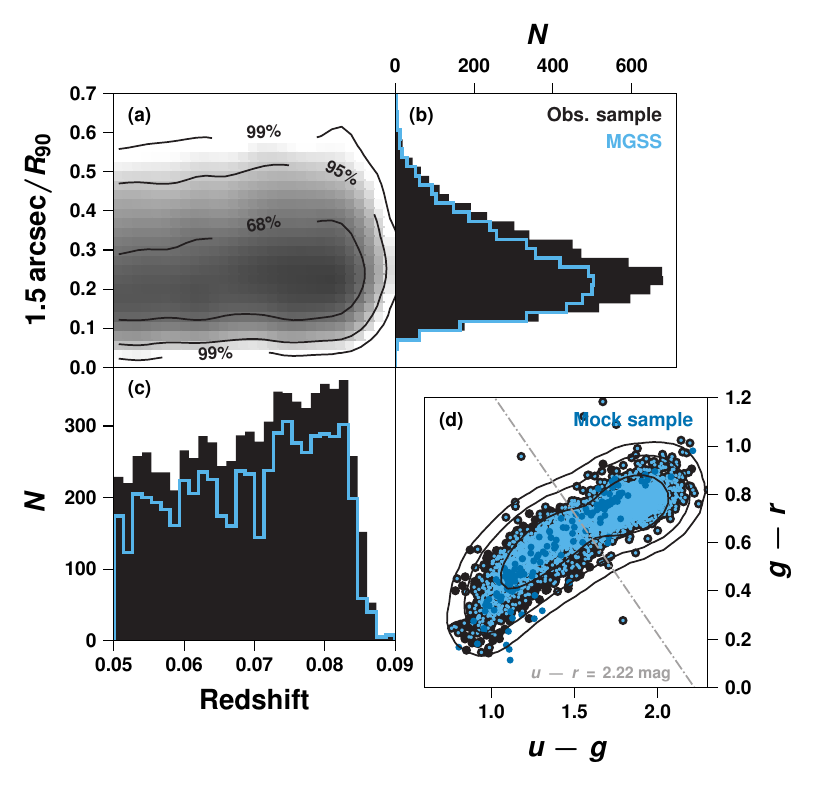}
\caption{\emph{(a)} The fraction of light entering the $\unit[1.5]{arcsec}$ fibre radius with
respect to the $\unit[90]{per cent}$ petrosian radius ($\unit[1.5]{arcsec}/R_{90}$) as a function of
redshift is shown as evidence of the absence of aperture bias in the observed galaxy sample. The
corresponding $y$-axis \emph{(b)} and $x$-axis \emph{(c)} marginal distributions are shown (black),
along with the MPA-Garching subset (MGSS, light blue histograms). \emph{(d)} $u-g$ vs. $g-r$
colour-colour distribution of the observed galaxy sample described in \S\ref{samples} (black dots
and $1\sigma$, $2\sigma$, and $3\sigma$ confidence regions as contours), the MGSS (light blue dots),
and the mock galaxy sample (dark blue dots). The colour separator $u-r=\unit[2.22]{mag}$
\citep{Strateva2001} is shown as a dot-dashed line.}\label{fig:samples}
\end{figure}
\begin{figure}
\includegraphics[scale=1]{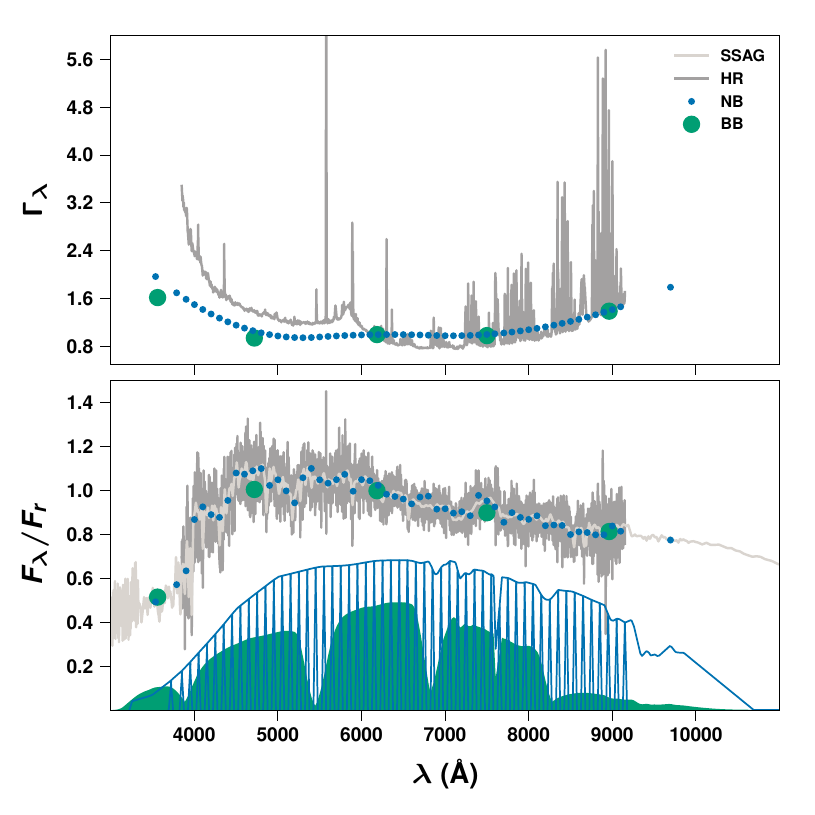}
\caption{\emph{Top:} Standard deviation spectrum template, $\Gamma_\lambda$, for the three spectral
resolutions used in this paper. The dark grey line is the mean value of the noise reported for all
the galaxies up to $z=0.1$ in the SDSS DR7, normalised at the effective wavelength of the
$r'$-passband. This curve, smoothed and interpolated at the central wavelength points of the J-PAS
passbands response function (blue), is used to calculate the standard deviation for the simulated
narrow-band (NB) data. For the broad-band (BB) photometric resolution we use the average of the SDSS
imaging reported noise, converted to flux (green). Alghough a different behaviour in the NB
instrumental noise may be expected, the assumed shape reflects the simple fact that the quantum
efficiency of most CCD-based optical instruments decrease towards the wavelength extreme values
\citep{Howell2006}. \emph{Bottom:} An example SED from the mock sample without noise (light grey)
and after adding instrumental noise (dark grey). The NB and the BB versions of the same mock galaxy
along with the corresponding passbands throughput (blue and green, respectively) are also
shown.}\label{fig:mock-seds}
\end{figure}
\subsection{The observed galaxy sample}\label{sec:real-sample}
We draw a sample of $\sim7\,$k galaxy spectra from the SDSS-DR7 \citep{Abazajian2009} by requiring
that the SEDs have: \textit{(i)} $z\leq0.09$ to remain in the optical range;  \textit{(ii)}
signal-to-noise ratio $>10$ to minimise degeneracies in the physical parameters recovered from the
spectral fits; and \textit{(iii)} a fraction of good pixels $\gtrsim95\,$per cent. A narrow
wavelength range ($\approx 20$ \AA) centred at the $\Oii$, $\Hg$, $\Hb$, $\Oiii$, $\Hei$, $\Oi$,
$\Nii$, $\Ha$ and $\Sii$ emission lines is masked out regardless of their presence in the target
SED. Finally, the sample is divided into star-forming galaxies (SFGs) and passive galaxies (PaGs)
according to the colour separator $u-r=\unit[2.22]{mag}$ \citep{Strateva2001}. The black dots in
Fig.~\ref{fig:samples}\textit{(d)} shows the colour-colour distribution of the resulting sample, the
contours correspond to the $1\sigma$, $2\sigma$, and $3\sigma$ confidence regions. A subset (MGSS
hereafter) of galaxies from our sample was studied by \citet[\citetalias{Gallazzi2005}
hereafter]{Gallazzi2005} and is highlighted in light blue in Fig.~\ref{fig:samples}. The
MGSS\footnote{ available at \url{http://www.mpa.mpa-garching.mpg.de/SDSS/}.} is suitable for setting
up a benchmark to compare with our spectroscopic estimates of galaxy properties (see
\S\ref{sec:spec-amr}).

To test the consistency between the parameters derived from spectroscopic data and narrow-band
photometric data, we synthesize narrow-band observations from SDSS galaxy spectra as seen at $z=0$
using the J-PAS throughput \citep{Marin-Franch2015} shown in Fig.\ref{fig:mock-seds}. The mean flux
expected through the $k$th passband is given by
\begin{equation}\label{eq:photo-flux}
F_k = \left.\int_\lambda f_\lambda T_k(\lambda)\text{d}\lambda\middle/\int_\lambda T_k(\lambda)\text{d}\lambda\right.,
\end{equation}
where $f_\lambda$ is the rest-frame SDSS galaxy SED, and $T_k(\lambda)$ is the response function of
the $k$th passband. A measurement of the error in $F_k$ is provided by conventional error
propagation \citep{Bevington2003} using the relation
\begin{equation}\label{eq:photo-error}
\sigma_k^2 = \int_\lambda T_k(\lambda)^2\sigma_\lambda^2\text{d}\lambda,
\end{equation}
where $\sigma_\lambda$ is the standard deviation in $f_\lambda$.

In the ideal case, the SDSS spectra cover the rest-frame wavelength range
$3800\,$---$\,\unit[9200]{\AA}$, whereas the J-PAS filters span
$\sim3500\,$---$\,\unit[10000]{\AA}$. Hence, around five passbands are naturally masked out.
Moreover, those passbands where more than $\unit[10]{per cent}$ of the pixels are reported as `bad'
by the SDSS pipeline are removed from the final narrow-band SEDs. To minimise the impact of missing
passbands during SED-fitting, we compute the narrow-band fluxes using the original SDSS spectra,
without emission line masking. Then we remove the nebular emission effect \emph{a posteriori}, using
a rather simple algorithm to mask out the passband only when an emission line is detected. Overall,
the mean number of passbands with potential information on the stellar age and metallicity masked
out (including the emission affected passbands) is $\sim15$.

\subsection{The mock galaxy sample}\label{sec:mock-sample}
To assess the physical properties retrieved from high-resolution spectroscopic, narrow-band and
broad-band \citep[$u'g'r'i'z'$,][]{Doi2010} spectral fits (hereafter HR, NB and BB, respectively),
we select a sample of $134$ mock galaxies from the Synthetic Spectral Atlas of Galaxies (SSAG
herefater) built by \cite{SSAG2014}, used by \citetalias{Magris2015}, and described in
Appendix~\ref{sec:mockpar}. We require that the selected galaxies (dark blue dots in
Fig.~\ref{fig:samples}\textit{(d)}) reproduce the observed properties of galaxies, such as the
bimodality in the colour $u-r$ and in the $\unit[4000]{\AA}$-break index distributions
\citep{Strateva2001, Kauffmann2003, Baldry2004}. Finally, to emphasise the difference between the
mock SEDs at the HR and NB spectral resolutions and their observed counterparts, due to instrumental
artefacts and emission line masking, we label the latter as HR$^*$ and NB$^*$, respectively.

In order to simulate \textit{observed} SEDs at the spectral resolution of the NB and BB photometry,
and HR spectroscopy, we proceed as follows: \textit{(i)} from the mock galaxy SED we compute the
photometric flux through the passband using Eq.~\eqref{eq:photo-flux}.\footnote{In this case we use
a version of the SED in which the stellar kinematic effects are not included.} \textit{(ii)} To
mimic more realistically galaxy observations, we add Gaussian random noise to the computed fluxes,
assuming observationally motivated uncorrelated error sources. For this purpose, we use as a
template for generating the standard deviation on $F_\lambda$, the averaged error spectra normalized
to one at the $r'$-passband, $\Gamma_\lambda$, for galaxies in the SDSS up to $z=0.1$, smoothed and
degraded in resolution if needed to match each SED type as shown in Fig.~\ref{fig:mock-seds}. Then
the actual standard deviation is simply $\Gamma_\lambda F_\lambda/\text{SNR}r$, given a value for
the $\text{SNR}r$ at the effective wavelength of the $r'$-passband, as in
\citetalias{Magris2015}.\footnote{Different regions of the optical spectra trace different stellar
populations. Hence, properly weighting each region during spectral fitting by using an
observationally based $\sigma_\lambda$ spectrum is crucial to furnish realistic estimates of the
uncertainties on the physical properties of the target galaxy.} We note that the flux adds linearly
while the noise adds quadratically when synthesizing the flux using the Eq.~\eqref{eq:photo-flux}.
Therefore, to ensure a fair comparison between the mock sample results and those drawn from the
observed sample as a function of the spectral resolution, we assume different values of the
signal-to-noise ratio for the different flavours of SEDs: $\text{SNR}r=20,\,45,$ and $140$ for HR,
NB, BB, respectively. To account for statistical variations, the procedure to add noise is repeated
$20$ times for each SED. The final mock galaxy sample then comprises $2680$ SEDs, representative of
the local Universe. Once the process of adding observational noise is completed, each mock galaxy
SED is passed as input to our spectral fitting code at their HR, NB and BB spectral resolutions,
i.\,e. without any spectral masking. The differences between the HR, NB and the HR$^*$, NB$^*$ data
sets, respectively, are noteworthy in the context of a potential comparison between the mock and the
observed sample results; we will discuss further in \S\ref{sec:real-deal}, where is more
appropriate.

Despite the motivation stated above, the chosen $\text{SNR}r$ values deserve further justification
in the context of current galaxy surveys. The assumed value of $\text{SNR}r=20$ for the mock
spectroscopy can be justified straightforwardly if compared to the observed sample described in
\S\ref{sec:real-sample} which is characterized by a median $\text{SNR}=18.8$ across the wavelength
range. The value of $\text{SNR}r=140$ chosen for the BB photometry, on the other hand, corresponds
to a standard deviation of $\unit[0.007]{mag}$,\footnote{Here we use the fact that
$\sigma_\text{magnitude}\propto1/\text{SNR}$ \citep{Howell2006}.} which is comparable to the median
error $\unit[0.006]{mag}$ in $r'$-passband for the SDSS observed sample. Both of these values are
roughly typical of galaxies with $r\sim\unit[18]{mag}$ within an aperture of $\unit[3]{arcsec}$. To
justify the $\text{SNR}r=45$ chosen for the NB mock SEDs, we select three galaxies from the SDSS
spectroscopic sample with magnitude $r\sim19$, $18$, and $\unit[17]{mag}$ and $\text{SNR}\sim10$,
$16$, and $32$ across all the wavelength range, respectively. For these galaxies we compute the
expected $\text{SNR}$ at the NB resolution using the formula
\begin{equation}\label{eq:snr}
\text{SNR} = \frac{\bar{F}\bar{T}t}{\sqrt{\bar{F}\bar{T}t + N_\text{pixel}N_\text{read}\sigma_\text{read}^2}},
\end{equation}
and the instrumental setting for the J-PAS \citep{Benitez2014}, where $\bar{F}$ is the mean photon
count per second for each source, $\bar{T}=0.5$ is the mean camera response,
$t=t_\text{exp}N_\text{filter}N_\text{read}$ is the total exposure time,
$t_\text{exp}=\unit[60]{sec}$ is the exposure time per filter per number of readouts,
$N_\text{read}=4$ is the number of readouts of the filter array, $N_\text{filter}=56$ is the number
of filters, $N_\text{pixel}=\uppi N_\text{filter}\left(D/2/p_s\right)^2$ is the total number of
pixels within a $D=\unit[3]{arcsec}$ aperture, assuming a pixel scale
$p_s=\unit[0.227]{arcsec\,pixel}^{-1}$. We consider no sky nor dark current contributions to the
total noise (i.\,e. bright sources only). For the three sources with $r\sim19$, $18$, and
$\unit[17]{mag}$, we find an exposure time per filter per readout, $t_\text{exp}\sim180$, $75$,
$\unit[40]{sec}$. Thus the assumed $\text{SNR}r=45$ is observationally plausible only for bright
sources in the local Universe according to the planned ($t_\text{exp}=\unit[60]{sec}$) J-PAS
configuration. By running the same calculation using the SDSS imaging set up, namely: $\bar{T}=0.5$,
$t_\text{exp}=\unit[60]{sec}$, $N_\text{filter}=5$, $D=\unit[3]{arcsec}$,
$p_s=\unit[0.396]{arcsec\,pixel}^{-1}$; we find the brightest ($r=\unit[17]{mag}$) source requires a
shorter exposure in order to reach a $\text{SNR}r=140$, while the $r=18$ and $\unit[19]{mag}$
sources would require $>1$ and $>3$ scans, respectively. Interestingly, the SDSS spectroscopic set
up: $\bar{T}=0.3$, $N_\text{read}=5$, $3$ and $3$ (for the $r\sim19$, $18$, and $\unit[17]{mag}$
sources, respectively), $t_\text{exp}=\unit[900]{sec}$ and $N_\text{pixel}=2RD/p_s$, with mean
spectral resolution $R=1800$, $D=\unit[3]{arcsec}$ and $p_s=\unit[0.396]{arcsec\,pixel}^{-1}$,
requires a total exposure time for the brightest source about $2$ times shorter than J-PAS imaging
to reach a $\text{SNR}=20$. This may suggest that planning a project like J-PAS is a pointless task.
However there are at least two key aspects that we have ignored during this exercise. First,
photometric surveys do not suffer from aperture effects. Hence, depending on the observing
conditions, J-PAS should require a shorter exposure $t_\text{exp}$ to reach a $\text{SNR}=20$,
characteristic of a spectroscopic survey like SDSS. Second and most important, multi-object imaging
is limited mainly by the field of view of the camera, whereas multiplexed spectroscopy usually
suffers from engineering limitations. E.\,g., the J-PAS camera has a field of view of
$\unit[4.7]{square degrees}$, capable of imaging $\sim10^4$ galaxies in one exposure, whereas the
SDSS-DR7 spectroscopic camera, having a similar field of view, is limited to $640$ objects per
exposure. See \S1 in \citet{Benitez2014} for several other factors favoring photometric surveys over
spectroscopic ones.

Overall, the assumed values of the SNR can be reached only for bright sources in the local Universe
and our mock sample is thus suitable as a basis of comparison with the observed sample. Given the
known impact of data quality on the galaxy physical properties inferred through SED-fitting, in
\S\ref{sec:noise-effect} we discuss the effects of the $\text{SNR}r$ on the reliability of the
stellar content derived for the three spectral resolutions used in this study.

\section{Galaxy Properties from SED-fitting}
\subsection{On SED-fitting}\label{sec:sed-fitting}
Spectral fitting algorithms can be classified according to the assumptions made to model the galaxy
SFH as parametric and non-parametric. The parametric methods assume a fixed functional form for the
SFH and prescribe physically motivated prior distributions for the intervening parameters
\citep[e.\,g.,][]{Kauffmann2003, Kriek2009, Hansson2012}, whereas the non-parametric methods
\citep[\citetalias{Magris2015}]{CidFernandes2013, Diaz2015} make no restrictions on the functional
shape of the SFH, although some prior assumptions are usually made regarding the time sampling of
the SFH in order to minimize potential degeneracies. The parametric approach is generally tackled
within the Bayesian framework, where our knowledge about a set of parameters $\hip$ \emph{posterior}
to data consideration, $\pos{\hip}{\dat}$, arises naturally from updating our degree of belief
(including prejudices) \emph{prior} to any data consideration, $\pri{\hip}$, through the
\emph{likelihood} of some assumed model to describe the data, $\lik{\dat}{\hip}$. According to
Bayes' theorem this is
\begin{equation}\label{eq:bayes}
\pos{\hip}{\dat} \propto \lik{\dat}{\hip}\times\pri{\hip},
\end{equation}
where $\dat$ is the $N$ dimensional data set (e.\,g., the integrated SED or a number of spectral
indices). There are several reasons why this approach is particularly advantageous to the spectral
analysis of galaxies. Here we mention two of the most important ones in this context: first, our
current knowledge about the data sets and the physics intervening in galaxy formation and evolution
allows us to assign in a straightforward manner both the likelihood and the prior distribution; and
second, there is no limit (in principle) to the dimensionality of the vector $\hip$, regardless of
the size of the data set. This last advantage has been particularly exploited by several authors,
who have focused their studies on data sets consisting of a small number ($N\sim10$) of specific
spectral features or photometric data \citep[e.\,g.,][]{Kauffmann2003, Brinchmann2004, Tremonti2004,
Gallazzi2005, Maraston2010}.

In the non-parametric problem, on the other hand, the observed integrated SED $F_\lambda^\text{obs}$
of the target galaxy is modelled by maximizing the likelihood. In general it is plausible to assume
that the uncertainties on the data are Gaussian and uncorrelated, and that they are described by the
proposed model $F_\lambda^\text{mod}(\hip)$, so that the likelihood is given by,
\begin{equation}\label{eq:likelihood}
\lik{F_\lambda^\text{obs}}{\hip} = \prod_{\lambda}^{N_\lambda}\frac{1}{\sqrt{2\uppi{\sigma_\lambda}^2}}\exp{\left\{-\frac{\left[F_\lambda^\text{obs} - F_\lambda^\text{mod}(\hip)\right]^2}{2{\sigma_\lambda}^2}\right\}},
\end{equation}
where $\sigma_\lambda$ is the standard deviation on $F_\lambda^\text{obs}$. The problem of finding
$\hhi$ such that the function above reaches its maximum, is equivalent to minimizing the merit
function,
\begin{equation}\label{eq:chi-square}
-2\log{\mathcal{L}}+\text{const} \equiv \chi^2\left(\hip\right) = \sum_\lambda^{N_\lambda}\frac{\left[F_\lambda^\text{obs} - F_\lambda^\text{mod}(\hip)\right]^2}{{\sigma_\lambda}^2}.
\end{equation}
Furthermore, if the proposed model takes the form
\begin{equation}\label{eq:gen-model}
F_\lambda^\text{mod}(\hip) = \sum_{k=1}^{N_\text{SSPs}}w^k\left(\hip\right)F_\lambda^k\left(\hip\right),
\end{equation}
and the weights $w^k\left(\hip\right)$ are linear on $\hip$, then $\chi^2$ is a quadratic function
on the parameter vector $\hip$, and the problem reduces to solving a linear system of $N_\lambda$
equations with $N_\text{SSPs}$ unknowns \citep{CidFernandes2005, Ocvirk2006a, Tojeiro2007}.

The main advantage of the non-parametric approach as posed above comes from the simplicity of
linearity. This approach has been widely used when the data sets are large enough ($N_\lambda\geq
N_\text{SSPs}$), so that the system is (over)determined (e.\,g. GASPEX, TGASPEX in
\citetalias{Magris2015}). However even in the overdetermined regime, the linear problem is
ill-conditioned and the solution $\hhi$ heavily depends on the observed data noise and on model
ingredient uncertainties \citep[see discussion in][]{Ocvirk2006a}. Furthermore, it does not provide
a straightforward calculation of the uncertainties affecting the best estimate $\hhi$
\citep[however, see][for inspiration]{Diaz2015}, nor does it allow for the inclusion of parameters
which, even if physically irrelevant, are needed to fully describe the data set.\footnote{In the
Bayesian framework these are the so-called nuisance parameters.} On the other hand, current Bayesian
studies in the SED-fitting literature lack a thorough assessment of two relevant aspects that may
bias galaxy SED interpretations. First, despite the fact that a parametrization of the SFH is not
required in the Bayesian framework \citep[see e.\,g.,][]{daCunha2008, Pacifici2012, Chevallard2016},
most authors have opted to assume that star formation is continuous in time after its initial onset,
and that the SFH can be described by an exponentially declining \citep[$\tau$-model,][]{Wuyts2009,
Lee2009}, exponentially increasing \citep[inverted $\tau$-model,][]{Maraston2010}, or a `delayed
exponential' \citep[e.\,g.,][]{Hansson2012} function of time. Second and most important, if a
non-objective prior $\pri{\hip}$ is adopted, it needs to be carefully chosen when the data set is
not very informative about the parameters through the likelihood (e.\,g. small and/or low-SNR data
set), a regime where it is known that the prior may play an important r\^ole in shaping the
posterior probability distribution \citep[see][also our discussion in
\S\ref{sec:fitted-ds}]{Benitez2000, Gallazzi2008, Chevallard2016}. Both methodologies, Bayesian and
maximum likelihood, must agree in their solutions in the limiting case when the adopted likelihood
is able to fully span the space of the data set, i.\,e., when the contribution of the prior
distribution becomes irrelevant. The reader is referred to \citet{Walcher2011} for a more thorough
discussion on SED fitting approaches.

When introducing a new algorithm, most authors apply their SED fitting procedure to synthetic galaxy
spectra of known SFH to compare the true and derived values of parameters like the galaxy stellar
mass, mean stellar age, stellar metallicity, and SFR. We will call this procedure
theoretical-theoretical assessment (TTA). \citetalias{Magris2015} showed from their TTA that
different non-parametric codes using the same SSP ingredients yield, in general, different estimates
of the stellar populations present in a galaxy (regardless of the goodness-of-fit), introducing
uncertainties in the interpretation of the target galaxy SED. \citet{Wuyts2009, Lee2009, Lee2010,
Pforr2012, Pforr2013, Mitchell2013} have studied the propagation of these uncertainties into the
stellar properties derived from \emph{photometric} energy distributions using parametric SED fitting
methods. \cite{Pforr2012, Mitchell2013, Hayward2015} have shown that the $\tau$-models are prone to
introducing biases in the retrieved stellar properties when the assumed functional shape of the SFH
is in high disagreement with the target SFH (e.\,g., bursty galaxies). \citet{Pforr2012} show that
even using inverted $\tau$-models, the stellar mass, age, and reddening by dust in nearby ($z=0.5$)
galaxies with complex SFHs, cannot be retrieved in a robust manner. Moreover, the correct
application of parametric methods to derive physical properties out of \emph{incomplete} optical
spectra (whether photometric or isolated spectroscopic features), usually requires high-quality data
and/or independent determinations of some of the physical properties, which are not always available
\citep[see e.\,g., \citetalias{Gallazzi2005};][]{Kriek2010, Castellano2014}. In this paper we use
the non-parametric DynBaS fitting code (\S\ref{sec:dynbas}) as an alternative to overcome the
limitations of assuming a parametrization for modelling the SED of galaxies from photometric data
sets \citep[see also][]{Diaz2015}.
\subsection{The \dynbas non-parametric spectral fitting code}\label{sec:dynbas}
Dynamical Basis Selection (\dynbas, \citetalias{Magris2015}) is a non-parametric SED fitting code
designed to recover the stellar population content of galaxies. The target SED is reconstructed
using the model in Eq.~\ref{eq:gen-model} with,
\begin{subequations}\label{eq:dynbas-model}
\begin{align}
w^k\left(\hip\right) &= a_{ij}^k, \label{eq:coeff}\\
F_\lambda^k\left(\hip\right) &= F_\lambda^k\left(t_i,Z_j\right), \label{eq:ssps}
\end{align}
\end{subequations}
where $i=1,\ldots,N_\text{ages}$, and $j=1,\ldots,N_\text{metallicities}$. The \dynbas code keeps at
a minimum the number of spectral components, by combining up to three SSPs ($N_\text{SSPs}=1,2,$ or
$3$), \emph{dynamically selected} for each target galaxy, to yield the absolute minimum of the merit
function, $\chi^2$, in the region of parameter space sampled in the fit. The summation in
Eq.~\eqref{eq:chi-square} runs over the (not masked) $N_\lambda$ wavelength points in the target SED
$F^\text{obs}_\lambda$. Whether the optimal solution $F_{\lambda}^\text{mod}$ is assembled combining
$1$, $2$ or $3$ SSPs, depends on the peculiarities of the target SED.

It is worth noting that the \dynbas code is not intended to retrieve the full time resolution SFH of
galaxies, but rather to deliver robust estimates of their global physical properties, averaged over
their stellar content. The current implementation of \dynbas is suitable for the analysis of stellar
populations coarsely sampled spectroscopically and/or with low $\text{SNR}$, formed through complex
(cf. \S\ref{sec:spec-amr}) or simple SFHs \citep[e.\,g.,][]{Cabrera-Ziri2016}. Using $\text{SNR}=20$
mock spectroscopy, \citetalias{Magris2015} showed that \dynbas recovers such global properties with
a bias and precision equivalent to those obtained with methods aimed at SFH recovery (e.\,g. \SL,
\tg); a result suggesting that in fact galaxies resulting from complex SFHs can be described by the
combination of young, intermediate, and old stellar populations \citep{CidFernandes2005}.

Indeed several authors have implemented non-parametric models like Eq.~\eqref{eq:gen-model} with
$N_\text{SSPs}\sim5$. For instance, \citet{Ocvirk2006a} implemented the STECMAP code which sets
$N_\text{SSPs}$ during the fitting by regulating the prior PDF contribution to the posterior PDF
depending on the $\text{SNR}$ of the data set. \citet{Tojeiro2007} implemented VESPA, a SED-fitting
code that adjusts the number of parameters retrieved from the fit depending on the $\text{SNR}$ of
the data. \citeauthor{Tojeiro2007} found that a spectroscopic sample of SDSS galaxies with
$\text{SNR}>10$ is well represented by the combination of $N_\text{SSPs}\sim5$. Both codes show a
dependency of the number of parameters retrieved on the wavelength range, where the inclusion of
additional spectral information allows for a higher parameter space resolution. The general trend
behind the results drawn from STECMAP, VESPA, and \dynbas is that the amount of physical information
retrieved in a robust manner is a function of the size of the data set and its quality, described by
some property $Q$. In fact the number of free parameters on the retrieved SFH should obey a function
of the form $N_\text{SSPs}=N_\text{SSPs}(N_\lambda,Q)$. The authors above have successfully
implemented models with $N_\text{SSPs}(Q)$, with $Q\equiv \text{SNR}$. The complete function
$N_\text{SSPs}(N_\lambda,Q)$ remains understudied in the spectral fitting literature regarding
non-parametric and parametric methods. In an upcoming paper we will address the dependency of
$N_\text{SSPs}$ on both the data set quality and the wavelength sampling in the context of current
multi-band photometric surveys of galaxies.
\subsection{Deriving physical parameters}\label{sec:derphys}
\begin{figure*}
\includegraphics[scale=1.0]{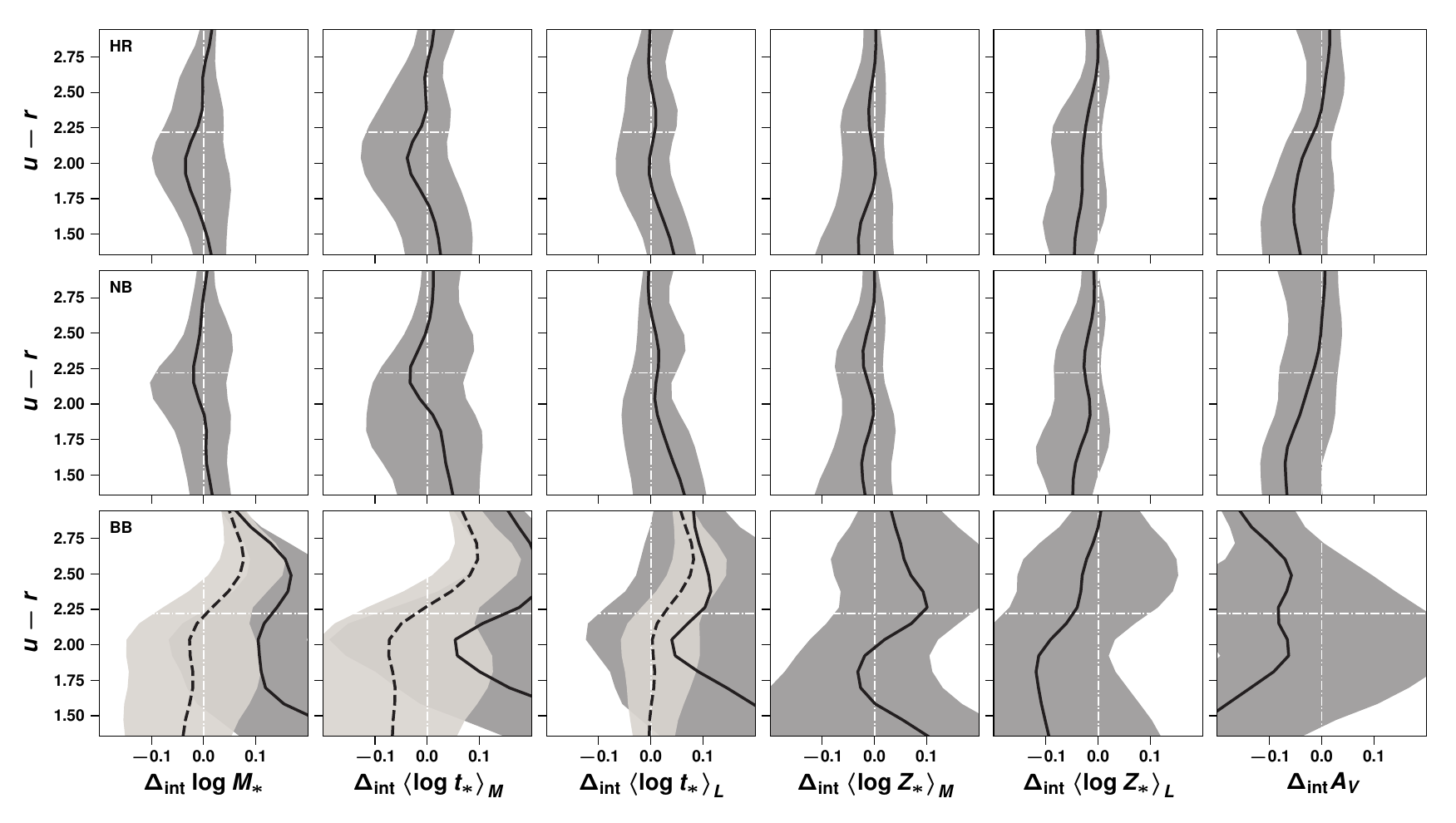}
\caption{The bias (black line) and precision (grey region) computed for the noiseless data as a
function of the $u-r$ galaxy colour at the three spectral resolutions, as indicated in the first
column. The behaviour of these residuals at the HR and NB resolutions is notably similar across
all the colour range, though it should be noted that NB tends to yield more biased and imprecise
results, most notably for SFGs. The BB results are largely biased and imprecise, a weakness that
can be overcome by adding relevant information to the fit, as shown by the dashed line, for which
the dust extinction and metallicity were fixed to their true values. See text for
details.}\label{fig:noiseless-residuals}
\end{figure*}
\begin{figure*}
\includegraphics[scale=1.0]{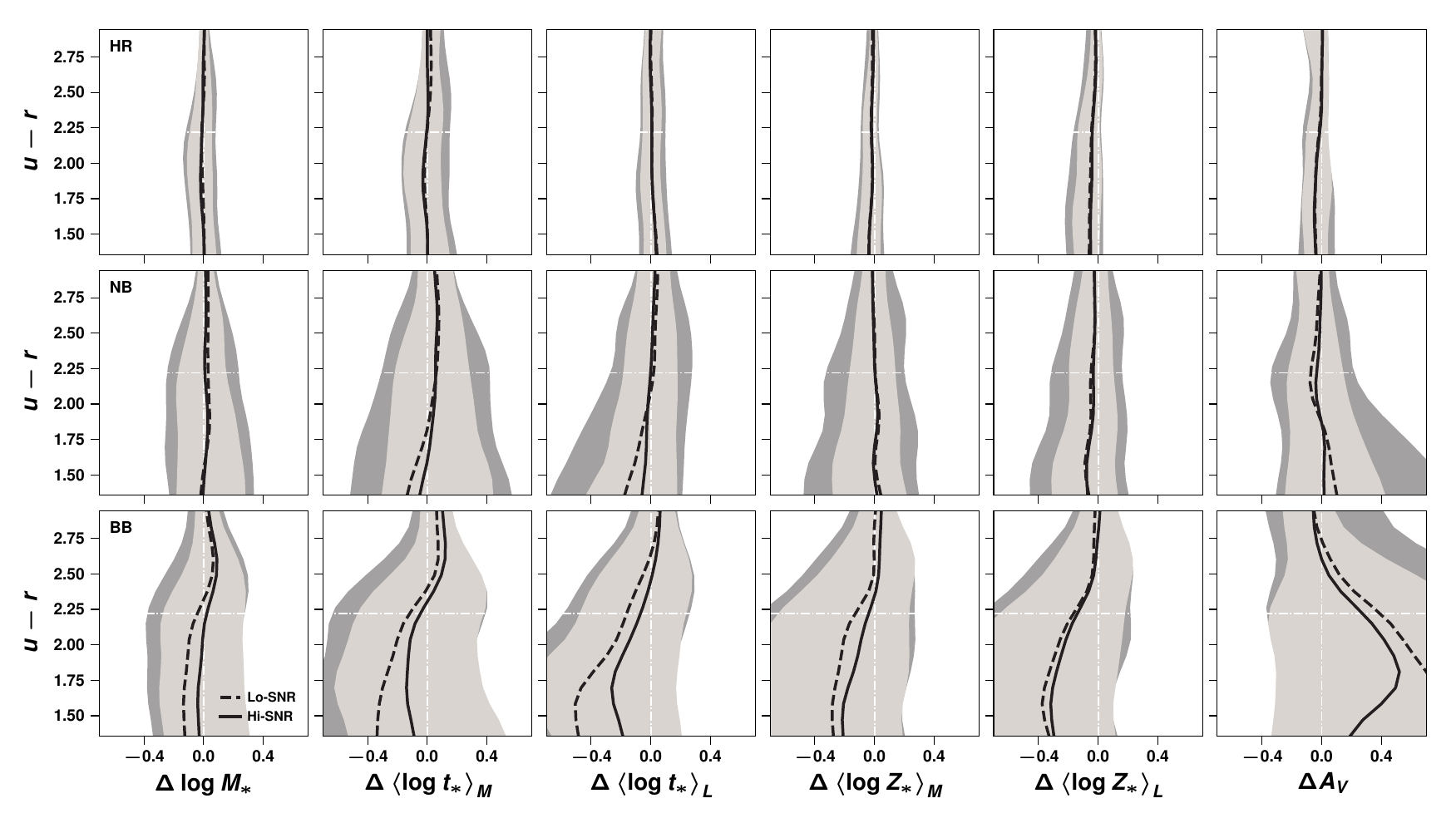}
\caption{The residuals bias (black lines) and precision (shaded regions) as a function of the $u-r$
colour for fixed values of the $\text{SNR}r=20$ (Lo-SNR) and $45$ (Hi-SNR) at the three spectral
resolutions, as indicated in the first column. As expected, the precision in the recovered
parameters tends to decrease (increases in numerical value) as the $\text{SNR}r$ decreases. The bias
shows little correlation with the $\text{SNR}r$ at high spectral resolution. However, as the
spectral resolution decreases, the (absolute) bias at lower values of the $\text{SNR}r$ increases.
Both effects become more important towards blue SFGs. See text for
details.}\label{fig:residuals-vs-snr}
\end{figure*}
To build $F^\text{mod}_\lambda$ in Eqs.~\eqref{eq:gen-model} and \eqref{eq:dynbas-model} we adopt
the same BC03xm SSP models as in the mock sample (Appendix~\ref{sec:mockpar}). If
$F^\text{obs}_\lambda$ represents spectroscopic data, $F^\text{mod}_\lambda$ is broadened using a
Gaussian kernel, $G(0,\sigma_v)$, to account for the effect of stellar kinematics. The best estimate
of the line-of-sight velocity dispersion (LOSVD), $\sigma_v$, is assumed to be in the range
$0\,$---$\,\unit[400]{km~s}^{-1}$ and computed iteratively by fitting the narrow ($\unit[100]{\AA}$)
wavelength range around the Ca~\textsc{ii} H\&K lines.\footnote{This effect is neglected
altogether when fitting photometric data.} The extinction model used in the SSAG
(Appendix~\ref{sec:mockpar}) is a two-phase model \citep{Charlot2000}: birth cloud and diffuse dust,
characterised by $\tau_V$ and $\mu\tau_V$, respectively. It has been shown by \citet{Tojeiro2007}
that due to degeneracies it is not possible to accurately recover these two parameters
simultaneously from SED-fitting solutions. Therefore, we model the effect of starlight absorption by
the diffuse dust as a single parameter curve $S_\lambda(\mu\tau_V)$, where $\mu\tau_V$ is related to
the total extinction in the $V$-passband as $A_V=1.086\mu\tau_V$ (single screen model; see
\citetalias{Magris2015}). The best fitting $A_V$ is assumed to be in the range
$0.0\,$---$\,\unit[1.5]{mag}$ and computed iteratively in a predefined discrete grid designed to
ensure the global minimum of $\chi^2$ in Eq.~\eqref{eq:chi-square}. For $S_\lambda$ in this work we
adopt the \citet{Charlot2000} extinction curve for fitting the mock sample SEDs, which is consistent
with the SFH recipe, hence the uncertainties introduced by the starlight absorption model assumed
make no contribution in our analysis. For the observed sample, on the other hand, we adopt instead
the more conventional \citet{Cardelli1989} extinction curve with a Milky Way parametrization
($R_V=3.1$). The different treatment of dust in the SSAG and in our fitting model may introduce
systematics in our results. The same is true for the time scale of star formation, treated
differently in the SSAG than in our fits (see Appendix \ref{sec:mockpar}). We explore the incidence
these differences may have in our results with nearly noiseless mock galaxies in
\S\ref{sec:noise-effect}.

Out of the box, the SSP model flux is expressed in units of L$_\odot\,$\AA${}^{-1}\,$M$_\odot^{-1}$.
If $F^\text{obs}_\lambda$ is in units of L$_\odot\,$\AA${}^{-1}$, the coefficients $a_{ij}^k$ in
Eq.~\eqref{eq:coeff} are in units of M$_\odot$. Then $\sum_ka_{ij}^k$ is the estimate of the
luminous mass present in the target galaxy, hereafter referred to as the stellar mass. Determining
or even defining $\tform$, the formation time of a galaxy, or its metallicity $Z$, is a difficult
task when the integrated spectrum is constantly rejuvenated by young populations in SFGs.
$\left<x\right>_M$ and $\left<x\right>_L$, the mass- and luminosity-weighted mean values of property
$x$ of the stellar populations present in the model galaxy, are used as \emph{proxies} for $x$, and
defined as
\begin{subequations}\label{eq:mean-props}
\begin{align}
\left<x\right>_M &=\frac{1}{M_*}\sum_{k=1}^{N_\text{SSPs}} a^k x_k,\\
\left<x\right>_L &=\frac{1}{L_{*,r}}\sum_{k=1}^{N_\text{SSPs}}a^k F^k_r x_k,
\end{align}
\end{subequations}
where for simplicity we have dropped the indices $ij$ in $a_{ij}^k$, and
\begin{subequations}\label{eq:mean-weights}
\begin{align}
M_{*}   &=\sum_{k=1}^{N_\text{SSPs}}a^k,\\
L_{*,r} &=\sum_{k=1}^{N_\text{SSPs}}a^k F^k_r,
\end{align}
\end{subequations}
are the model galaxy stellar mass and luminosity in the $r'$-passband, which we use as a reference
luminosity, respectively. $\left<x\right>_M$ is biased towards the value of $x$ of the most massive
star-formation event that took place in the target galaxy, while $\left<x\right>_L$ is biased
towards the value of $x$ of the most luminous population dominating the $r'$-band.

As an entry point to our TTA, after fitting the spectrum of a mock galaxy we define the residual
$\Delta x$ of property $x$ as its \emph{true value} $x_\text{SSAG}$ subtracted from the value $x$
retrieved from the fit, i.\,e.,
\begin{equation}\label{eq:residual}
\Delta x\equiv~x-x_\text{SSAG}.
\end{equation}
The values of $\logm$, $\mwla$, $\lwla$, $Z$, and $\extv$ in Table~\ref{tab:mock-parameters} are the
\emph{true values} of these properties used in the fits below.

The $\text{median}(\Delta x)$ of the distribution of residuals is used to estimate the \emph{bias
(accuracy)}, and the semi-difference of the $84$th and $16$th percentiles to estimate the
\emph{precision} of our results. We adopt the bias and precision as a measure of the systematic and
the random uncertainty in the determination of the physical parameters, respectively. We warn the
reader that according to this definition, increasing values of precision characterise less precise
results. Moreover these estimates of the bias and precision are lower limits, since results from
SED-fitting real observations are prone to suffer from additional uncertainties that for the sake of
simplicity are not included in the TTA, e.\,g., differences between the adopted model stellar
ingredients and the actual ingredients present in observed galaxies; the assumed shape and
universality of the IMF; instrumental and calibration errors \citep[for a thorough revision see][and
references therein]{Conroy2009, Conroy2010a}.

We define the \emph{biasless estimator} $\tilde{x}$ and the \emph{discrepancy} $\delta x$ of
parameter $x$ as
\begin{subequations}\label{eq:discrepancy}
\begin{align}
\tilde{x} &\equiv x-\text{median}(\Delta x), \\
\delta x  &\equiv\tilde{x}^\text{NB}-\tilde{x}^\text{HR},
\end{align}
\end{subequations}
respectively, to measure possible differences between the NB photometric and HR spectroscopic
estimates of $x$. In contrast to the residual $\Delta x$, the discrepancy $\delta x$ measures the
combined bias and (im)precision arising from the NB and the HR parameter estimations. We note in the
case of observed data we cannot compute $\Delta x$, therefore we use the same value as for the TTA
to provide with a value of $\tilde{x}$. By removing the predicted bias, we reduce the possible
sources of discrepancy to either unpredicted bias (neglected for simplicity in the TTA) and/or
random sources of error.
\subsection{The impact of the instrumental noise}\label{sec:noise-effect}
Maximum likelihood methods are known to produce biased results if the adopted $\lik{\dat}{\hip}$
does not account for the full data set space \citep{Smith2015}. In \S\ref{sec:mock-sample} we stated
that our motivation behind the different assumptions on the $\text{SNR}r$ values at the HR, NB and
BB spectral resolutions is to ensure a comparison as independent as possible on the level of noise
in the data. We also showed that the adopted values, namely: $\text{SNR}r=20$ for HR spectroscopy,
$45$ for NB photometry, and $140$ for BB photometry, are plausible in the context of current surveys
only for bright sources in the local Universe. Given the ill-defined condition of the problem at
hand, the overly simplified model assumed as compared to the SSAG recipe, and the dependency of the
derived physical properties on data quality \citep[e.\,g.][]{Ocvirk2006a}, it is worth exploring the
impact of instrumental noise in these properties, i.\,e., the trends in the residuals arising from
fitting the data sets as a function of instrumental noise level. For this purpose we use the same
SFHs from the mock sample, adopting three fixed values of the $\text{SNR}$, namely: $\sim1000$
(essentially noiseless data), $20$ (Lo-SNR), and $45$ (Hi-SNR), regardless of the spectral
resolution. As in the fiducial mock sample, we compute $20$ realisations of the noise for the Lo-SNR
and Hi-SNR samples. No noise realisation is performed for the noiseless sample. The resulting sample
SEDs at the three spectral resolutions are fed to \dynbas, and the recovered physical properties are
compared to the true values according to the Eq.~\eqref{eq:residual} in order to compute the
intrinsic residuals, $\Delta_\text{int}x$, and the conventional residuals $\Delta x$ for the
noiseless and the Lo-, and Hi-SNR samples, respectively.\footnote{We remark $\Delta_\text{int}x$ and
$\Delta x$ are different in the sense that the former is a measure of the inaccuracy due to the
assumed model and methodology alone, whereas the latter measures the combination of both these
effects and the contributions of the simulated instrumental noise.}

Fig.~\ref{fig:noiseless-residuals} shows the behaviour of the intrinsic residuals as a function of
the $u-r$ colour for the physical properties studied in this paper. The trends seen in the HR and NB
samples show similarities across all the colour range, although the bias (solid line) and the
(im)precision (dark grey region) derived from the NB fits are marginally larger, most notably in the
case of SFGs. On the other hand, the BB photometry exhibits larger bias and highly imprecise results
for all physical properties. \citet{Pforr2012, Mitchell2013} have found that BB photometry does not
provide clues on the stellar metallicity in a robust manner. They also showed that by excluding
stellar properties from the fits (i.\,e. reducing the $\hip$ size), the results can be notably
improved. In this same spirit, we repeated the BB fits fixing both, the dust extinction and the
stellar metallicity, to their corresponding true values for each galaxy in the sample. The resulting
bias (dashed line) and precision (light grey region) were computed in this case using the model with
$N_\text{SSPs}=2$, which was found to be remarkably better than the $N_\text{SSPs}=3$, as expected
given the size of the data set. Interestingly, $\lwla$, i.\,e. the luminosity-weighted-age of the
stellar population dominating the galaxy light in the $r'$-passband, was retrieved with nearly no
bias and a precision comparable to NB and HR in the case of SFGs. However, the mass related
properties ($\logm$ and $\mwla$) are both underestimated for the same type of galaxies, most likely
due to the outshining effect \citep{Maraston2010, Sorba2015}. For the PaGs, the three retrieved
properties are overestimated, a sign of the mass-age degeneracy (see discussion in
\S\ref{sec:multdeg} for a full analysis). Likewise, Fig.~\ref{fig:residuals-vs-snr} shows the bias
for the Lo-SNR (dashed line) and the Hi-SNR (solid line) samples, which in general do not depend
strongly on data quality. The precision (dark shaded region for Lo-SNR and light shaded region for
Hi-SNR) shows a mild increment (decrement in numerical value) with increasing $\text{SNR}r$, an
expected result. It should be noted that these effects show an increasing impact toward lower
spectral resolution and bluer SFGs, where the outshining effect is expected to contribute to both
the bias and the (im)precision of the results, the latter contribution being due to the sole
presence of the instrumental noise.

It is interesting to note that whatever their origin (the assumed physics and/or the methodology
itself), the biases reflect the presence of the several degeneracies between mass, age, metallicity
and dust extinction (cf. Fig.~\ref{fig:degeneracies}), and are therefore susceptible of being
mitigated once the relevant information in the form of data and/or \emph{a priori} assumptions are
taken into account during the spectral fitting, as demonstrated above.

\section{Fitting the Mock Sample: Critical View}\label{sec:main-tta}
\begin{figure*}
\includegraphics[scale=1.0]{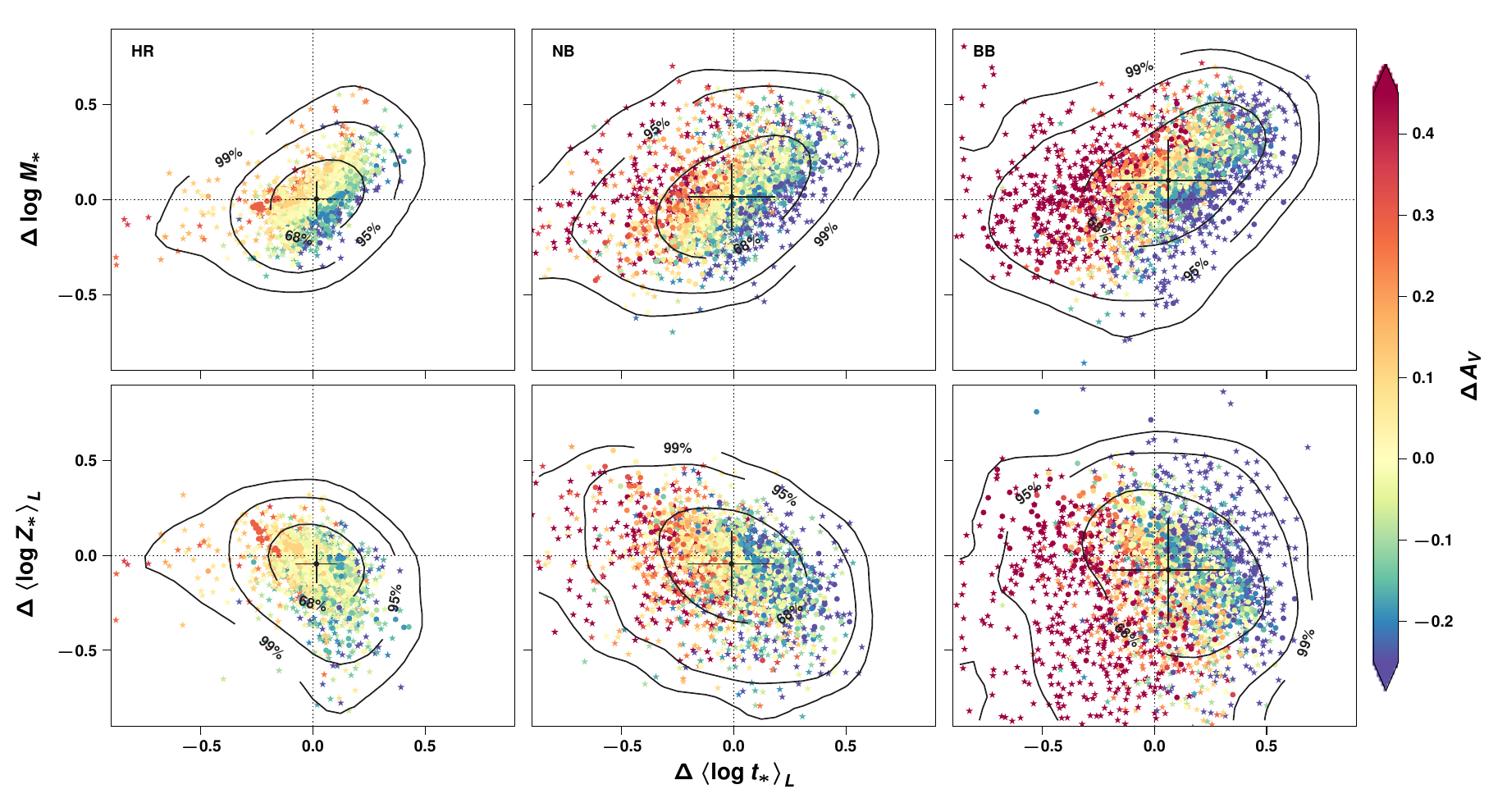}
\includegraphics[scale=1.0]{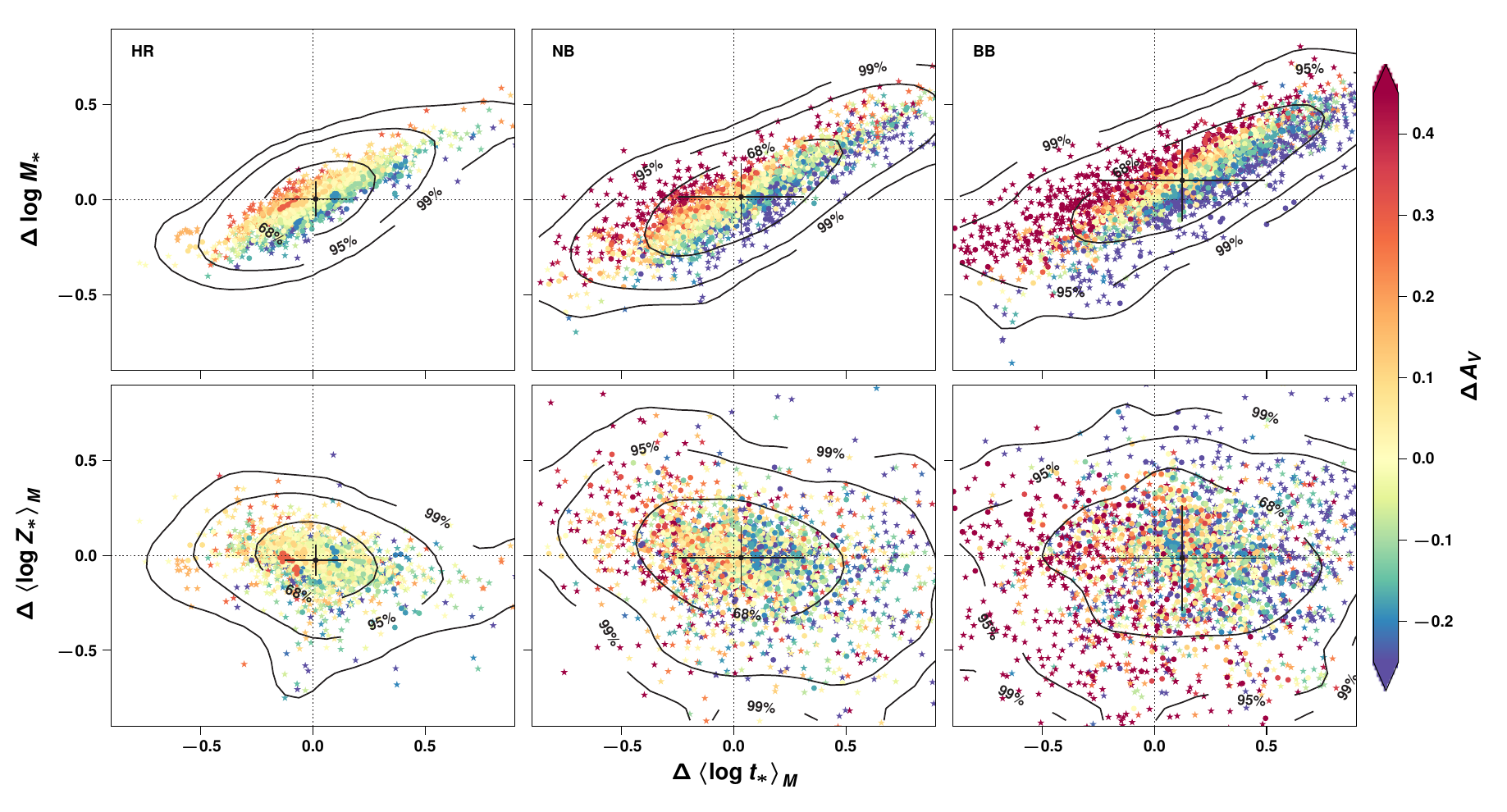}
\caption{\textit{Top:} The residuals for the several physical properties we discuss in this paper in
the framework of the TTA results are shown. Each point represents a galaxy in the mock sample and
the shape of the points characterise the galaxy type (SFGs: stars; PaGs: circles). The error cross
on each frame is centred at the bias and is twice as large as the precision. The observed trends are
interpreted to be due to the several degeneracies, while the different scattering around those
trends, as pointed out by the $68$, $95$ and $\unit[99]{per cent}$ confidence regions (contours) and
the precision, depends on the spectral resolution. \textit{Bottom:} Same as top frame, but for the
mass-weighted mean age and metallicity. The strong correlation between stellar mass and
mass-weighted age points out the presence of the mass-age degeneracy, regardless of the spectral
resolution, whilst the \emph{strength} of the age-metallicity and age-dust extinction degeneracies
are clearly correlated with the spectral resolution. SFGs are usually more scattered in all planes
than PaGs, i.\,e., it is more likely to find a SFG beyond the $\unit[95]{per cent}$ confidence
region than a PaG. See text for details.}\label{fig:residual-corrs}
\end{figure*}
In this section we explore the reliability of SED fitting to infer the physical properties of the
target galaxy. We do so by examining the distributions of residuals $\Delta x$
(Eq.~\ref{eq:residual}), where $x$ denotes any of the following variables, $\logm$ (log stellar
mass), $\mwla$ (mass-weighted mean log age), $\lwla$ (luminosity-weighted mean log age), $\mwlz$
(mass-weighted mean log metallicity), $\lwlz$ (luminosity-weighted mean log metallicity), and
$\extv$ (extinction in the $V$-band), obtained after fitting with the \dynbas code described in
\S\ref{sec:dynbas} the SEDs in our fiducial mock galaxy sample built in \S\ref{sec:mock-sample}.

The content of the next two subsections is rather technical. The uninterested reader can skip
directly to \S\ref{sec:partsumm} where we summarise the main results of \S\ref{sec:residual-pdfs}
and \S\ref{sec:multdeg}.
\subsection{Model Assessment}\label{sec:residual-pdfs}
In the \emph{top frames} of Fig.~\ref{fig:residual-corrs} we show the distributions of residuals for
$\logm$ and  $\lwlz$ \emph{vs.} $\lwla$ and $\extv$ (colour scale). The \emph{bottom frames} show
the corresponding distributions for $\logm$ and  $\mwlz$ \emph{vs.} $\mwla$ and $\extv$. The
corresponding $68$, $90$, and $\unit[99]{per cent}$ confidence regions of the PDF are shown as
contours. The first, second, and third columns refer to HR, NB, and BB, spectral resolutions,
respectively. The error cross on each frame is centred at the bias and its length is twice the
precision, as defined above. The BB photometry is presented for comparison with similar previous
studies \citep[see e.\,g.][]{Pacifici2012, Hansson2012}.

In the ideal case, these residual distributions will show highly concentrated PDFs centred at
$\Delta x=0$. Instead, our residual distributions are characterised by the
$\text{bias}\pm\text{precision}$ values listed in Table~\ref{tab:residuals}. From this table it is
clear that the precision increases (the numerical value decreases) with increasing spectral
resolution. In most cases the bias decreases with increasing spectral resolution as well, but
unexpectedly there are cases where BB and/or NB photometry seem to outmatch the performance of HR
spectroscopy in terms of the bias parameter, e.\,g., $\dmwlz$ for All Gals. In the forthcoming
sections we will see that such cases are related to the preponderance of instrumental noise over
physical effects (e.\,g. degeneracies).
\begin{table*}
\caption{Bias and precision after fitting the mock sample in the different spectral
resolutions.}\label{tab:residuals}
\begin{tabular}{lrrrrrrrrr}
\hline
                          &           &\multicolumn{2}{c}{HR}& &\multicolumn{2}{c}{NB}& &\multicolumn{2}{c}{BB} \\
                                       \cline{3-4}                 \cline{6-7}                 \cline{9-10}      \\
 Parameter                & Gal. Type &      bias & precision & &      bias & precision & &      bias & precision\\
\hline
\multirow{2}{*}{\dlogm}   &      SFGs &     0.001 &     0.101 & &     0.011 &     0.215 & &     0.091 &     0.264\\
                          &      PaGs &     0.005 &     0.071 & &     0.019 &     0.125 & &     0.110 &     0.140\\
                          & All Gals. &     0.004 &     0.092 & &     0.015 &     0.178 & &     0.102 &     0.216\\
\hline
\multirow{2}{*}{\dmwla}   &      SFGs &     0.003 &     0.159 & &     0.002 &     0.339 & &     0.101 &     0.454\\
                          &      PaGs &     0.028 &     0.110 & &     0.065 &     0.193 & &     0.143 &     0.192\\
                          & All Gals. &     0.012 &     0.139 & &     0.033 &     0.281 & &     0.124 &     0.370\\
\hline
\multirow{2}{*}{\dlwla}   &      SFGs &     0.018 &     0.098 & &    -0.023 &     0.214 & &     0.047 &     0.289\\
                          &      PaGs &     0.009 &     0.086 & &     0.014 &     0.163 & &     0.081 &     0.187\\
                          & All Gals. &     0.015 &     0.093 & &    -0.010 &     0.195 & &     0.062 &     0.255\\
\hline
\multirow{2}{*}{\dmwlz}   &      SFGs &    -0.030 &     0.093 & &    -0.010 &     0.212 & &    -0.043 &     0.320\\
                          &      PaGs &    -0.016 &     0.060 & &    -0.010 &     0.126 & &     0.047 &     0.184\\
                          & All Gals. &    -0.024 &     0.080 & &    -0.010 &     0.171 & &    -0.013 &     0.280\\
\hline
\multirow{2}{*}{\dlwlz}   &      SFGs &    -0.057 &     0.119 & &    -0.069 &     0.202 & &    -0.142 &     0.318\\
                          &      PaGs &    -0.025 &     0.074 & &    -0.020 &     0.131 & &    -0.008 &     0.184\\
                          & All Gals. &    -0.044 &     0.104 & &    -0.044 &     0.175 & &    -0.076 &     0.273\\
\hline
\multirow{2}{*}{\dextv}   &      SFGs &    -0.032 &     0.113 & &    -0.014 &     0.256 & &    -0.032 &     0.504\\
                          &      PaGs &    -0.008 &     0.068 & &    -0.020 &     0.139 & &    -0.018 &     0.204\\
                          & All Gals. &    -0.024 &     0.098 & &    -0.016 &     0.210 & &    -0.028 &     0.387\\
\hline
\end{tabular}
\end{table*}
There are regions in Fig.~\ref{fig:residual-corrs} where $\logm$, $\lwla$, $\mwla$, $\lwlz$,
$\mwlz$, and $\extv$, are well determined, showing small residuals \emph{simultaneously}. We note
that regions of good $\extv$ determination (green symbols, with $\unit[-0.08<\dextv<0.08]{mag}$) are
usually extended in all frames of Fig.~\ref{fig:residual-corrs}, suggesting that $\extv$ is well
determined even if other properties are poorly constrained. Furthermore, there are three distinct
regions in the distribution of $\dextv$,  which are more eloquent at the BB resolution: \textit{(i)}
for most galaxies with $\dextv>\unit[0.30]{mag}$, $\lwla$ and $\mwla$ are underestimated, $\logm$ is
overestimated, and $\lwlz$ and $\mwlz$ are either under or overestimated; \textit{(ii)} most
galaxies with $\unit[-0.08<\dextv<0.16]{mag}$ show a linear trend with positive slope in the
$[\dlwla,\dlogm]$ and $[\dmwla,\dlogm]$ planes, and with negative slope in the $[\dlwla,\dlwlz]$
plane; and \textit{(iii)} for most galaxies with $\dextv>\unit[0.16]{mag}$, $\lwla$ and $\mwla$ are
overestimated, whilst $\lwlz$, $\mwlz$, and  $\logm$, are either under or overestimated. These
trends are essentially the same in the $[\dlwla,\dlogm]$ and $[\dmwla,\dlogm]$ planes at the HR
spectroscopy and NB photometry resolutions, regardless of the better precision in $\extv$ for the
former. However, we note that the fraction of galaxies with overestimated $\extv$ and $\lwlz$ shows
a tendency to increase with increasing spectral resolution. In the $[\dmwla,\dmwlz]$ plane, the
smooth dependence with spectral resolution apparent in the $[\dlwla,\dlwlz]$ plane breaks into two
distinct trends: \textit{(i)} at the BB resolution there is no evident correlation between $\dmwla$,
$\dmwlz$, and $\dextv$; this also holds in the $[\dlwla,\dlwlz]$ plane, at least outside the
$\unit[68]{per cent}$ confidence region; and \textit{(ii)}, for HR spectroscopy and NB photometry,
the shape of the confidence regions reveals trends among the parameters. The stronger trends lay in
the $[\dmwla,\dlogm,\dextv]$ volume, and remain strong for all spectral resolutions inside all
confidence regions.
\subsection{Multiple degeneracies}\label{sec:multdeg}
It is clear that the residuals discussed in \S\ref{sec:residual-pdfs} are correlated with each other
in several planes with a remarkable dependence on the wavelength sampling, as signaled by the shape
of the confidence regions and the size of the error crosses (see Fig.~\ref{fig:residual-corrs} and
Table~\ref{tab:degeneracies}). We interpret these correlations as due to degeneracies in the galaxy
properties for different combinations of the galaxy mass, age, metallicity and/or dust extinction.
As expected, the stronger correlations appear in the age-metallicity (with the clear exception of
BB), age-dust extinction, and age-mass planes. The strong dependence of the precision on the
wavelength resolution points towards an interesting result: the dominant degeneracies depend not
only on the galaxy colours, but also on the overall stellar population information provided by the
data itself, in this case a function of the wavelength resolution.

Fig.~\ref{fig:degeneracies} presents a quantitative comparison of the different degeneracies
apparent in Fig.~\ref{fig:residual-corrs}. The vertices of the polygons in
Fig.~\ref{fig:degeneracies} \emph{(left column)} are plotted at a radius equal to the absolute value
of the correlation coefficient $\rho$ of the residuals in Table~\ref{tab:degeneracies}, measured in
the planes $[\dlwla,\dlwlz]$ (right vertex), $[\dlwla,\dextv]$ (left vertex), $[\dlogm,\dlwla]$ (top
vertex), and $[\dlogm,\dmwla]$ (bottom vertex). The larger the radius, the stronger the degeneracy
in the corresponding plane. The lines joining the vertices have no meaning and are shown only to
guide the eye. To emphasise the dependence of the degeneracies on the SFH, we show different
polygons for SFGs (blue), PaGs (orange), and the whole sample (shaded region). In what follows we
will consider a correlation/degeneracy to be relevant if $\lvert\rho\rvert>0.5$. We will further
distinguish between a weak correlation/degeneracy ($0.5<\lvert\rho\rvert\leq0.7$), and a strong
correlation/degeneracy ($0.7<\lvert\rho\rvert\leq1.0$). In this sense the mass-age degeneracy is the
strongest, followed by the age-dust extinction degeneracy (with a strong dependence on galaxy type
and spectral resolution), and then by the age-metallicity degeneracy (also with a strong dependence
on galaxy type).
\subsubsection{The mass-age degeneracy}
The mass:luminosity-weighted-age degeneracy affects mostly PaGs. Its strength varies between
$\rho=0.61$ and $0.72$ from BB to NB photometry, showing a negligible weakening towards HR
spectroscopy ($\rho=0.70$). The mass:mass-weighted-age parameters show the strongest degeneracy
($\rho>0.8$) independently of spectral resolution and galaxy type. This degeneracy is consequence of
the logarithmic scale of time evolution of the galaxy integrated optical SED. Above
$\approx\unit[1]{Gyr}$, the mass-to-light ratio evolves at essentially the same pace across the
optical range, i.\,e., the shape of the SED is practically time independent within the
uncertainties, allowing equally good fits for almost any combination of old stellar populations
reaching up to the age of the Universe, with little dependence on the stellar metallicity.
\subsubsection{The age-dust extinction degeneracy}
For SFGs the age-dust extinction degeneracy decreases as proper age tracers become more prominent
with increasing resolution, varying from strong ($\rho=-0.80$) for BB photometry, to weak
($\rho=-0.64$) for NB photometry, to irrelevant ($\rho=-0.47$) for HR spectroscopy. Conversely, PaGs
dominated by passively evolving old stellar populations, usually have small amounts of interstellar
dust. Features like the $\unit[4000]{\AA}$-break, prominent in the spectra of PaGs, allow to date
these galaxies and break the age-dust extinction degeneracy even at the BB resolution, provided that
the reddening by dust is not parallel to the evolution to redder colours with age. Therefore PaGS
show a weak age-dust extinction degeneracy with $\rho\sim-0.60$ for BB and NB photometry.
\subsubsection{The age-metallicity degeneracy}
\begin{figure}
\center
\includegraphics[scale=1.0]{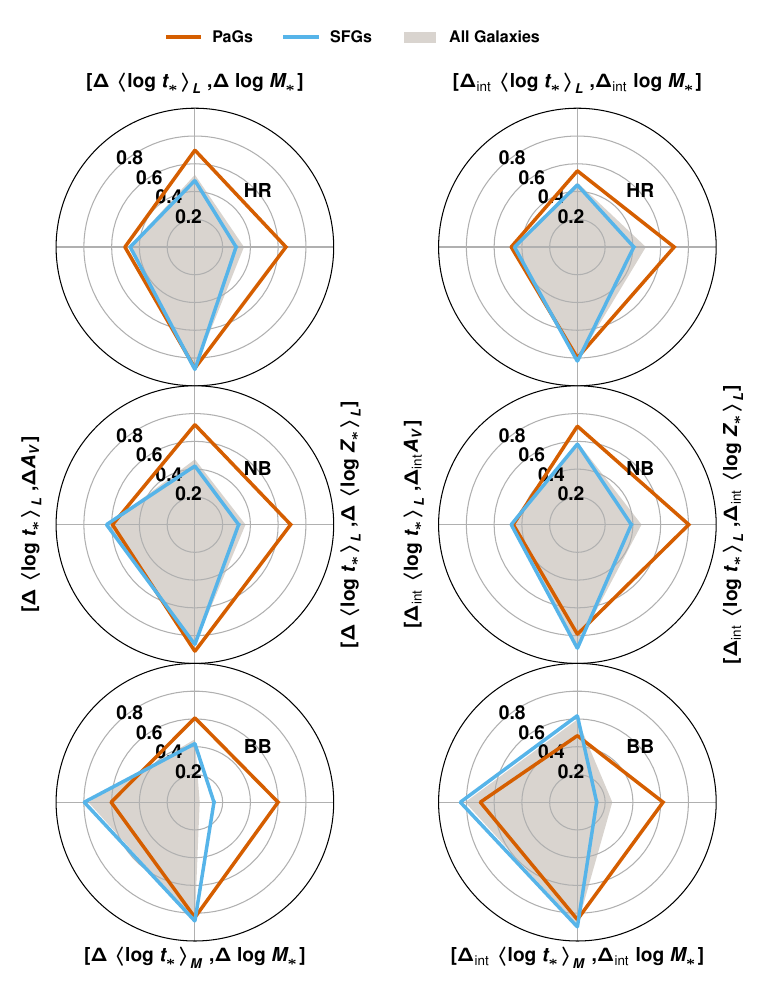}
\caption{\emph{Left column:} The strength of the age-metallicity, age-dust extinction and mass-age
degeneracies are shown as the radii of the polygons, respectively labelled as $[\dlwla,\dlwlz]$,
$[\dlwla,\dextv]$, and $[\dmwla,\dlogm]$ and $[\dlwla,\dlogm]$ at the HR (top), the NB (middle) and
BB (bottom) spectral resolutions. The degeneracies affecting PaGs (orange) are almost identical at
all SED resolutions. Conversely, the age-dust extinction degeneracy affecting SFGs (blue) shows an
important evolution with SED resolution, most notably in the transition between BB and NB, with the
latter being marginally equivalent to spectroscopic degeneracies. See text for details. \emph{Right
column:} same as left column, but for the intrinsic residuals. The strength of degeneracies
affecting HR spectroscopy and NB photometry show little dependence on the presence of instrumental
noise. See discussion in \S\ref{sec:noise-effect}.}
\label{fig:degeneracies}
\end{figure}

The luminosity-weighted age and luminosity-weighted metallicity tend to be degenerated for PaGs. Its
weak strength marginally increases with resolution from $\rho=-0.60$ for BB, to $\rho=-0.69$ for NB
photometry, dropping again to $\rho=-0.66$ for HR spectroscopy. The mass-weighted age:mass-weighted
metallicity degeneracy (not shown in Fig.~\ref{fig:degeneracies}) is marginally relevant
($\rho=-0.56$) for HR spectroscopy. At the BB resolution, the luminosity-weighted metallicity is
poorly constrained \citep[e.\,g.][]{Pacifici2012, Pforr2012, Hansson2012, Mitchell2013}. The
\emph{simultaneous} absence of age and metallicity tracers in SFGs, in which the young stellar
populations may outshine the underlying old ones, increases the imprecision in the determination of
these parameters, conspiring to hide the age-metallicity degeneracy at the BB resolution.
Conversely, for PaGs the spectral range $\lambda>\unit[7000]{\AA}$ provides information on the
stellar metallicity, and the age-metallicity degeneracy appears at comparable strength for the three
spectral resolutions.
\subsection{Summary of Section~\ref{sec:main-tta}}\label{sec:partsumm}
In \S\ref{sec:residual-pdfs} and \S\ref{sec:multdeg} we made progress in characterising,
understanding, and quantifying the ubiquitous degeneracies appearing in galaxy properties derived
from SED-fitting. Here we summarise the more important conclusions from these subsections.
\begin{description}
\item[\textit{(i)}] The values of the retrieved galaxy properties are usually biased when compared
to the known true values, specially at the BB resolution. At the HR spectroscopy and NB photometry
resolutions, the biases remarkably diminish, being negligible for PaGs. These systematics may have
several origins, namely: physical assumptions in the spectral modelling, the adopted methodology,
instrumental effects (e.\,g., spectral resolution, signal-to-noise ratio) and/or the ability of some
galaxies to hide their past SFH from the fitting procedure. We note however, that the
goodness-of-fit is always ensured through the existing degeneracies among the retrieved galaxy
parameters.
\item[\textit{(ii)}] In fact the biases in the residual distributions and the correlations among
them are mainly consequence of the ability of SFGs to \emph{rejuvenate} the stellar population while
hiding the underlying older ones (i.\,e. the outshining effect) and the several well-known
degeneracies between stellar mass, stellar age, stellar metallicity, and dust extinction, and as
such are susceptible to be mitigated by introducing additional information in the spectral fitting
procedure, for instance, in the form of a more comprehensive likelihood and/or with independent
determinations of some of the physical properties (as shown in \S\ref{sec:noise-effect}). The
statistical dispersion has its origin on both the outshining effect, and the observational
uncertainties added to our mock SEDs.
\item[\textit{(iii)}] The strength of the degeneracies depends heavily on the spectral resolution of
the fitted SED and the galaxy type. PaGs exhibit mass-age, age-metallicity, and age-dust extinction
degeneracies, with negligible dependence on the spectral resolution. SFGs are more prone to suffer
from the age-dust extinction degeneracy, whose strength decreases smoothly with increasing spectral
resolution. For SFGs the age-metallicity degeneracy is \emph{hidden} for the BB resolution, becoming
progressively stronger with increasing spectral resolution.
\item[\textit{(iv)}] Physical properties retrieved from photometric data show in general larger
biases and statistical dispersions than the ones retrieved from spectroscopy, albeit PaGs show
comparable biases and degeneracy strength for all spectral resolutions. Interestingly enough, the
biases and the degeneracy strengths of the properties derived from HR spectroscopy and NB photometry
are strikingly similar, a fact that deserves further exploration.
\end{description}
\begin{table}
\caption{Correlation coefficients of residuals versus spectral resolution.}\label{tab:degeneracies}
\centering
\begin{tabular}{llrrr}
\hline
                             Plane &  Gal. type &   HR &   NB & BB\\
\hline
\multirow{3}{*}{$[\dlwla,\dlogm]$} &       SFGs &    0.48 &    0.42 &    0.42\\
                                   &       PaGs &    0.70 &    0.72 &    0.61\\
                                   &  All Gals. &    0.52 &    0.47 &    0.45\\
\hline
\multirow{3}{*}{$[\dlwla,\dlwlz]$} &       SFGs &   -0.30 &   -0.32 &    0.14\\
                                   &       PaGs &   -0.66 &   -0.69 &   -0.60\\
                                   &  All Gals. &   -0.35 &   -0.36 &    0.03\\
\hline
\multirow{3}{*}{$[\dlwla,\dextv]$} &       SFGs &   -0.47 &   -0.64 &   -0.80\\
                                   &       PaGs &   -0.50 &   -0.59 &   -0.60\\
                                   &  All Gals. &   -0.47 &   -0.63 &   -0.77\\
\hline
\multirow{3}{*}{$[\dmwla,\dlogm]$} &       SFGs &    0.88 &    0.86 &    0.86\\
                                   &       PaGs &    0.87 &    0.91 &    0.83\\
                                   &  All Gals. &    0.88 &    0.87 &    0.85\\
\hline
\multirow{3}{*}{$[\dmwla,\dmwlz]$} &       SFGs &   -0.32 &   -0.26 &    0.12\\
                                   &       PaGs &   -0.56 &   -0.46 &   -0.31\\
                                   &  All Gals. &   -0.34 &   -0.29 &    0.06\\
\hline
\multirow{3}{*}{$[\dmwla,\dextv]$} &       SFGs &   -0.20 &   -0.39 &   -0.48\\
                                   &       PaGs &   -0.33 &   -0.43 &   -0.45\\
                                   &  All Gals. &   -0.22 &   -0.39 &   -0.48\\
\hline
\end{tabular}
\end{table}

\section{Recovered galaxy properties}\label{sec:real-deal}
\subsection{Consistency of galaxy property determinations}\label{sec:discrepancy}
Given the remarkable resemblance in the strength of the degeneracies affecting NB and HR
(\S\ref{sec:multdeg}), it is worth exploring how similar or discrepant are the stellar population
properties retrieved from both data sets. The objective of this comparison is to quantify the
differences between both determinations and establish under what circumstances the degree of
agreement is better or worse.

\begin{figure*}
\includegraphics[scale=1.0]{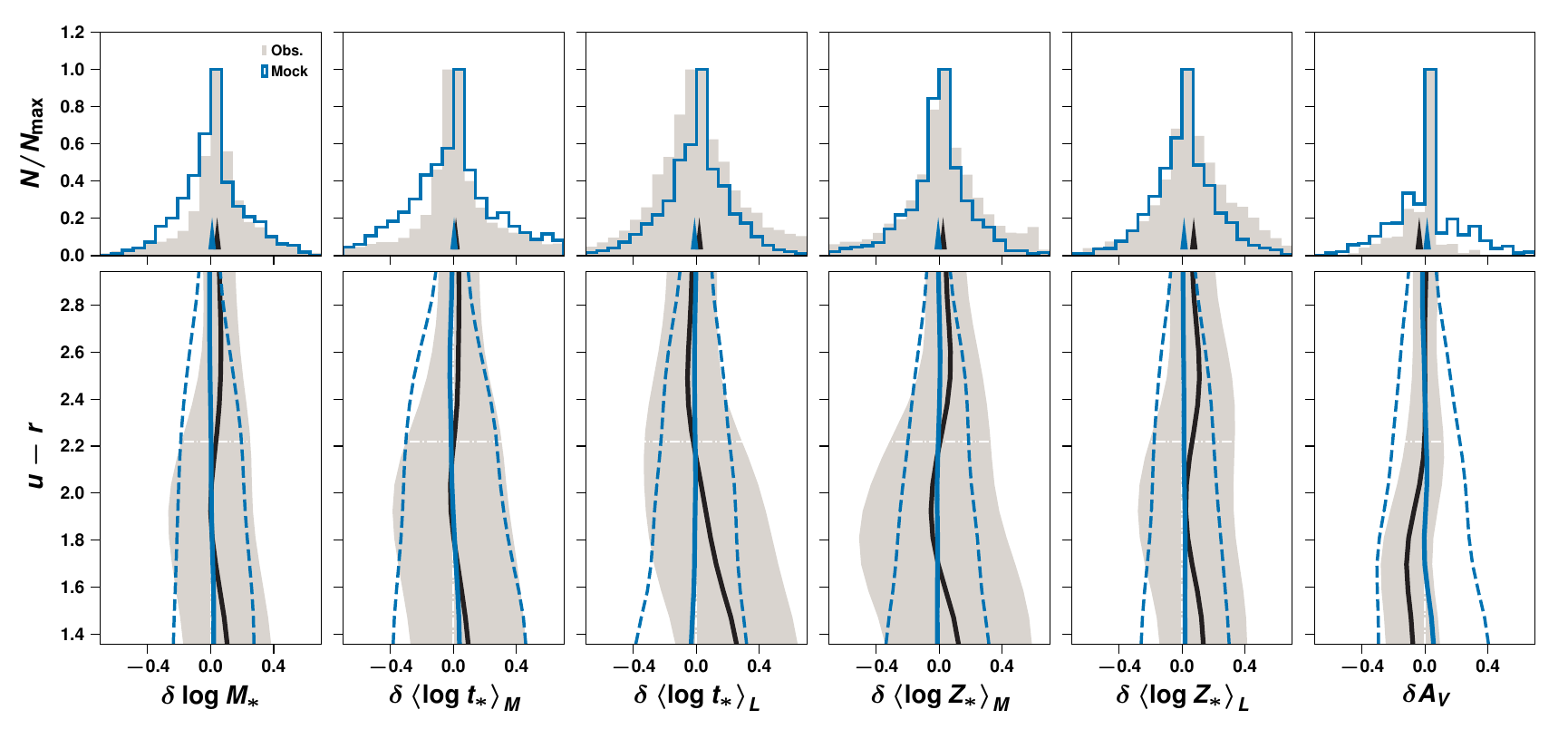}
\caption{\emph{Top:} Distribution of the discrepancy $\delta x$ (Eq.~\ref{eq:discrepancy}) for the
indicated galaxy parameters in the mock and the observed galaxy samples. \emph{Bottom:} Mean and
RMSD discrepancy for the indicated parameters as a function of the $u-r$ colour. Several effects
contribute to the discrepancy in the observed sample affecting mainly SFGs. See text for details.}
\label{fig:tta-vs-obs}
\end{figure*}

The top row of Fig.~\ref{fig:tta-vs-obs} shows the distributions of the discrepancy $\delta x$
defined in Eq.~(\ref{eq:discrepancy}) for the indicated physical property. The histograms outlined
in blue correspond to the mock sample, the shaded histograms to the observed sample. The blue
(black) arrow indicates the mean value of $\delta x$ for the mock (observed) sample. In the bottom
frames of Fig.~\ref{fig:tta-vs-obs} the continuous lines show the mean $\delta x$ as a function of
the galaxy $u-r$ colour, in blue (black) for the mock (observed) sample. The dashed blue line
(shaded region) shows the root-mean-square deviation (RMSD) discrepancy for the mock (observed)
samples. In Table~\ref{tab:discrepancies} we list the mean and RMSD $\delta x$ computed for the
SFGs, the PaGs, and the whole mock and observed samples.
Observed galaxy data are usually subject to different sources of uncertainty and/or preprocessing
due to incomplete model ingredients (e.\,g. gas emission), which may translate into systematic
errors. In particular, the procedure of emission line masking described in \S\ref{sec:real-sample}
has removed the Balmer lines carrying information on the stellar content of SFGs in both, HR$^*$ and
NB$^*$ data sets. If these deficiencies in the observed SEDs (as opposed to the mocks) equally
propagate from the NB$^*$ and the HR$^*$ data into the physical properties and the spectral
resolution play minor a r\^ole, the histograms of the discrepancies for the observed and the mock
samples should be similar and highly peaked at $\delta x=0$, with small RMSD values. In
Fig.~\ref{fig:tta-vs-obs} (top row) this behaviour is observed for $\logm$ and $\mwla$, regardless
of galaxy type (according to the $u-r$ colour, bottom row), and in $\lwla$ and $\extv$ for PaGs. The
mean discrepancy for $\lwla$, $\mwlz$, $\lwlz$, and $\extv$ for SFGs, exhibits larger departures
from a perfect match as can be seen in Fig.~\ref{fig:tta-vs-obs}.

The low values of $\delta x$ for the parameters mentioned above suggest that these parameters
\emph{can be retrieved from NB and HR data with a similar level of accuracy using a non-parametric
SED-fitting method, but with a larger precision in the case of HR data}. In the rest of this section
we use the galaxy age-metallicity relation (AMR) to explore the veracity of the last statement.
\subsection{The G05 spectroscopic AMR as a benchmark}\label{sec:spec-amr}
The most accepted theory of galaxy formation in the framework of the $\Lambda$CDM cosmology assumes
that the onset of star formation in galaxies takes place inside dark matter clumps, which
subsequently merge by gravitational attraction with neighbouring clumps in a process dubbed
\emph{hierarchical galaxy formation} \citep{White1991, Baugh1996, Kauffmann1996}. In the local
Universe ($z<0.5$) we only see the aftermath of this process. Population synthesis and spectral
modelling techniques allow us to infer the stellar content of galaxies observed by current surveys
\citep[e.\,g.][]{MacArthur2009, Maraston2010, Lee2011, Kriek2013}. From such studies several
correlations among galaxy properties have emerged \citep[see][and references therein]{Renzini2006}.
The mass-metallicity \citep[MMR,][]{Tremonti2004} and the age-metallicity
\citep[AMR,][]{Worthey1994} relations are examples of correlations that any model of galaxy
formation and evolution should predict. See \citet{Cappellari2016} for a recent review.

The first hint of a relationship between age and metallicity in galaxies was found in early-type
systems using the Lick spectral indices \citep{Gonzalez1993, Worthey1994, Bernardi1998}. Later on,
\citetalias{Gallazzi2005} presented an assessment of the AMR for a carefully selected sample of
galaxies in the local Universe probed by the SDSS, spanning a wide range of galaxy colours which
includes SFGs. \citetalias{Gallazzi2005} conclude that the most massive galaxies have on average the
older and more metal rich stellar populations, in concordance with previous studies. Likewise, the
metallicity of SFGs depends strongly on the stellar mass of the system: on average, smaller galaxies
are dominated by younger and lower metallicity populations. In this section we use the
\citetalias{Gallazzi2005} results as a benchmark to explore to what extent we can recover their AMR
using our SED fitting and parameter recovery methodology. Since our sample was selected to ensure a
fair comparison between HR$^*$ and NB$^*$ over the full range of galaxy colours in the local
Universe, a different selection criterium that the one defining the \citetalias{Gallazzi2005}
sample, we may expect some differences in the behaviour seen in our determination of the AMR and
that of \citetalias{Gallazzi2005} and others \citep[see e.\,g.,][]{Panter2008, Gonzalez2014}. To
minimise these differences, we select from our observed sample a subset of $5925$ galaxies studied
in \citetalias{Gallazzi2005}, referred to as the MPA-Garching subset (MGSS, in light blue in
Fig.~\ref{fig:samples}).

\begin{figure*}
\includegraphics[scale=1.0]{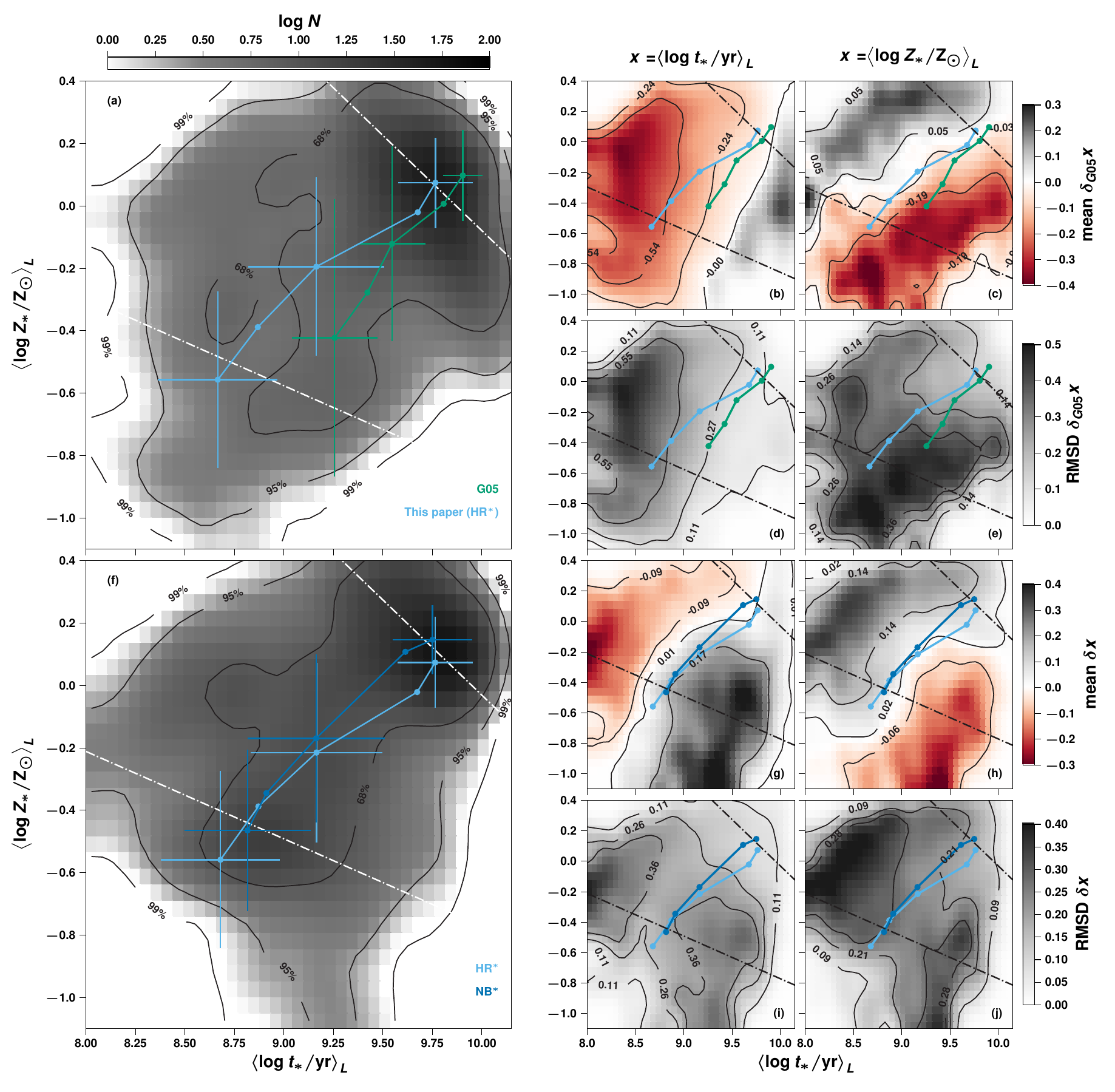}
\caption{The AMR for our observed galaxy sample determined from HR$^*$ and NB$^*$. \textit{(a,f)}
The grey-shaded areas show the AMR for the MGSS retrieved using HR$^*$, and for the complete sample
of observed galaxies retrieved using NB$^*$, respectively. The contours enclose the $1$, $2$ and
$3\sigma$ confidence regions of the PDF. \textit{(b,d,g,i)} Maps of the mean and RMSD discrepancy
for $\lwla$ for the sample in frames \textit{(a)} and \textit{(f)}, respectively. \textit{(c,e,h,j)}
Same as \textit{(b,d,g,i)} but for $\lwlz$. In frame \textit{(a)} the green and light blue dots and
crosses indicate the mean and RMSD values for the galaxies in the MGSS inside five $u-r$ colour bins
in the range $1.3\leq u-r\leq3.0$ for \citetalias{Gallazzi2005} and this paper, respectively,
derived from HR$^*$. In frame \textit{(f)} the dark blue and light blue dots and crosses indicate
the mean and RMSD values for the galaxies in the complete observed sample, derived from NB$^*$ and
HR$^*$, respectively. The colour bins are the same as in frame \textit{(a)}. In all frames, the
dot-dashed lines show the slope of the age-metallicity degeneracy computed from the residual
correlations shown in Fig.~\ref{fig:residual-corrs} using the SFGs and the PaGs subsamples. See text
for details.}\label{fig:amr}
\end{figure*}

\begin{figure*}
\includegraphics[scale=1.0]{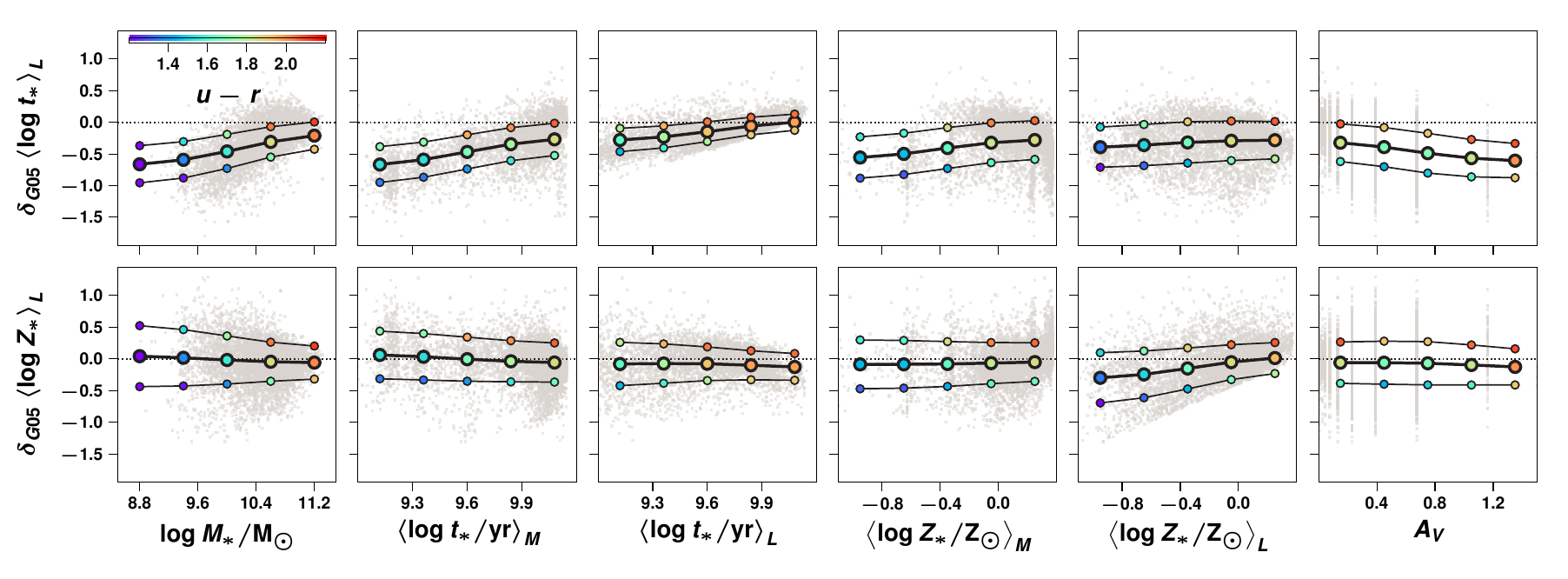}
\caption{Behaviour of $\glwla$ and $\glwlz$ as a function of the physical properties derived from
the HR$^*$ fitting. The grey dots show the distribution of all galaxies in the MGSS. The big dots
represent the mean $\delta_\text{G05}x$ in five bins spanning each physical property range, and
colour-coded according to the mean $u-r$ colour to reflect the predominant galaxy type in each bin.
The small dots represent the standard deviation in $\delta_\text{G05}x$, colour-coded according to
the standard deviation in $u-r$ to indicate the galaxy type variation within the same bins.
\emph{Top row:} $\glwla$ shows significant trends with all properties except with $\lwlz$.
\emph{Bottom row:} $\glwlz$ shows trends only with the $\lwlz$ retrieved by \dynbas, as shown in
Fig.~\ref{fig:amr}\emph{(a)}. The behaviour of these trends seems to indicate that several
degeneracies are at play. See text for details.}\label{fig:dis-vs-par}
\end{figure*}

Since stellar chemical evolution takes place in a generational fashion where massive stars evolve
rapidly returning chemically enriched material to the ISM from which the next generations of stars
form, whilst less massive stars evolve slowly enough to survive several generations of massive
stars, the stellar metallicity is expected to be related to the current SFR (traced by short-lived
stars) and to the stellar mass (traced by long-lived stars). Indeed, the theoretical framework
describing the large-scale evolution of the metal content locked in stars and present in the
interstellar and intergalatic media primarily relate these three physical properties
\citep{Tinsley1980, Madau2014}. Hence most `fossil' studies seeking to shed light on such interplay,
jointly relate the AMR to the stellar mass or equivalently the MMR to the current SFR
\citep[\citetalias{Gallazzi2005};][]{Sanchez2013, Salim2014}. Nonetheless, in this section we are
mainly concerned in providing an objective as possible comparison of the AMRs derived in
\citetalias{Gallazzi2005} and in this paper. Therefore we choose to use the $u-r$ colour as a proxy
for the mass \citep[see also $u-r\,$---$\,\logm$ trend in Fig.~\ref{fig:dis-vs-par}]{Bell2001,
Taylor2010}, which, in contrast to the stellar mass, is independent of the spectral analysis
method.\footnote{Although $k$-corrections were computed using the spectroscopic redshift, those have
small amplitudes for the observed sample.}
\begin{table}
\caption{Discrepancy $\delta x$ between the NB and spectroscopic stellar property
determinations.}\label{tab:discrepancies}
\begin{tabular}{lrrrrrr}
\hline
                          &           &\multicolumn{2}{c}{Mock}& &\multicolumn{2}{c}{Observed} \\
                                       \cline{3-4}               \cline{6-7}            \\
 Parameter                & Gal. type &      mean &      RMSD & &      mean &      RMSD \\
\hline
\multirow{3}{*}{\clogm}   &      SFGs &     0.016 &     0.224 & &     0.019 &     0.271\\
                          &      PaGs &    -0.004 &     0.150 & &     0.065 &     0.130\\
                          & All Gals. &     0.009 &     0.202 & &     0.041 &     0.215\\
\hline
\multirow{3}{*}{\cmwla}   &      SFGs &     0.019 &     0.357 & &     0.001 &     0.369\\
                          &      PaGs &    -0.024 &     0.230 & &     0.031 &     0.179\\
                          & All Gals. &     0.004 &     0.319 & &     0.016 &     0.292\\
\hline
\multirow{3}{*}{\clwla}   &      SFGs &    -0.010 &     0.262 & &     0.087 &     0.385\\
                          &      PaGs &    -0.012 &     0.186 & &    -0.052 &     0.197\\
                          & All Gals. &    -0.011 &     0.239 & &     0.019 &     0.315\\
\hline
\multirow{3}{*}{\cmwlz}   &      SFGs &    -0.013 &     0.251 & &    -0.020 &     0.444\\
                          &      PaGs &     0.006 &     0.158 & &     0.071 &     0.190\\
                          & All Gals. &    -0.006 &     0.224 & &     0.025 &     0.347\\
\hline
\multirow{3}{*}{\clwlz}   &      SFGs &     0.013 &     0.228 & &     0.043 &     0.300\\
                          &      PaGs &     0.012 &     0.159 & &     0.107 &     0.193\\
                          & All Gals. &     0.013 &     0.207 & &     0.074 &     0.255\\
\hline
\multirow{3}{*}{\cextv}   &      SFGs &     0.022 &     0.289 & &    -0.068 &     0.169\\
                          &      PaGs &    -0.001 &     0.162 & &    -0.002 &     0.078\\
                          & All Gals. &     0.014 &     0.253 & &    -0.036 &     0.136\\
\hline
\end{tabular}
\end{table}
Fig.~\ref{fig:amr}\emph{(a)} shows our derivation of the AMR (grey-shaded area). The $1$, $2$ and
$3\sigma$ confidence regions from the corresponding PDF are shown as contours. The green and dark
blue dots and crosses indicate the mean and RMSD values for the galaxies inside five $u-r$ colour
bins in the range $1.3\leq u-r\leq3.0$ for \citetalias{Gallazzi2005} and this paper, respectively.
All these quantities have been computed for the MGSS. The dot-dashed lines in this figure show the
slope of the age-metallicity degeneracy computed from the residual correlations shown in
Fig.~\ref{fig:residual-corrs}, using the SFGs and PaGs subsamples. We adopt the mode as the best
fitting parameter estimator of the \citetalias{Gallazzi2005} results for consistency with this
paper.

Our AMR spans wider ranges in $\lwla$ and $\lwlz$ than \citetalias{Gallazzi2005}'s at the
$u-r$-colour blue end of the trends. Furthermore, our estimates of the $\lwlz$ show smaller RMSD
values for these galaxy types. These results may be suggesting that our methodology is capable of
resolving the AMR for SFGs to a higher degree than \citetalias{Gallazzi2005}'s. However, the unknown
true trend and intrinsic scatter in the AMR poses an important difficulty in assessing the veracity
of the former statement (see \S\ref{sec:fitted-ds}, where we turn back to this issue). Our results
imply marginally younger and more metal poor PaGs as compared to the \citetalias{Gallazzi2005}
results. Both trends rapidly diverge in $\lwla$ and $\lwlz$ at the transition from PaGs to SFGs. To
quantify these differences we define the discrepancy between the \citetalias{Gallazzi2005} and our
results as $\delta_\text{G05}x\equiv x^{\text{HR}^*}-x^\text{G05}$.\footnote{Since the biases from
\citetalias{Gallazzi2005} are unknown to us, we cannot define the discrepancy as in
Eq.~\eqref{eq:discrepancy}. Therefore $\delta_\text{G05}x$ includes the systematics from both
methodologies.} Figs.~\ref{fig:amr}\emph{(b)\,--\,(e)} show maps of the mean and RMSD
$\delta_\text{G05}x$ discrepancy for $x=\lwla$ and $\lwlz$. We note in
Figs.~\ref{fig:amr}\emph{(b,c)} that the gradients (the normal vectors to the contours) in the mean
$\lwla$ and $\lwlz$ have opposite signs, an evidence of the age-metallicity degeneracy.
Figs.~\ref{fig:amr}\emph{(d,e)} show that the higher RMSD values for $\lwla$ and $\lwlz$ occur in
the region occupied by SFGs. From the reddest (PaGs) to the bluest (SFGs) $u-r$ colour bins in
Fig.~\ref{fig:amr}\emph{(b,c)}, $\glwla$ varies from $\unit[-0.15]{}$ to $\unit[<-0.54]{}$, and
$\glwlz$ from $\unit[-0.02]{}$ to $\unit[-0.17]{}$. Our prediction of the systematic effect
introduced by the age- and metallicity-related correlations (e.\,g. the age-metallicity and the
age-dust extinction degeneracies) is not enough to explain this amount of discrepancy (cf. mean
discrepancy for observed sample in Table~\ref{tab:discrepancies}), although we cannot disregard
completely their r\^ole as a possible source thereof.
\begin{figure*}
\includegraphics[scale=1.0]{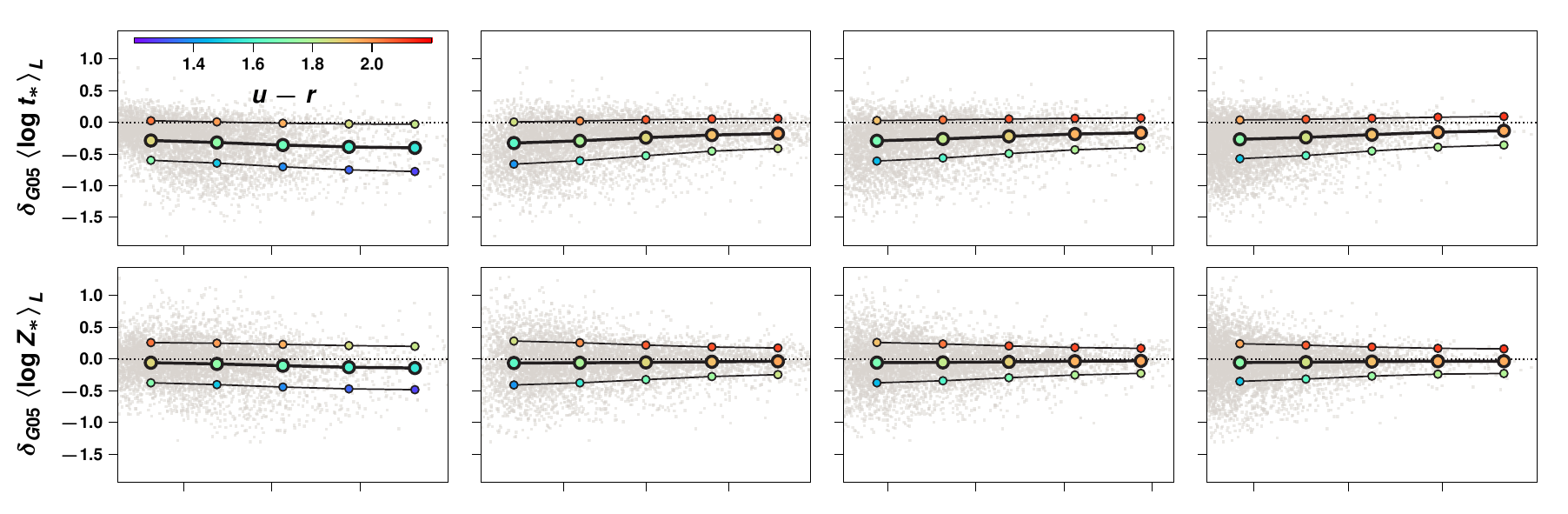}
\includegraphics[scale=1.0]{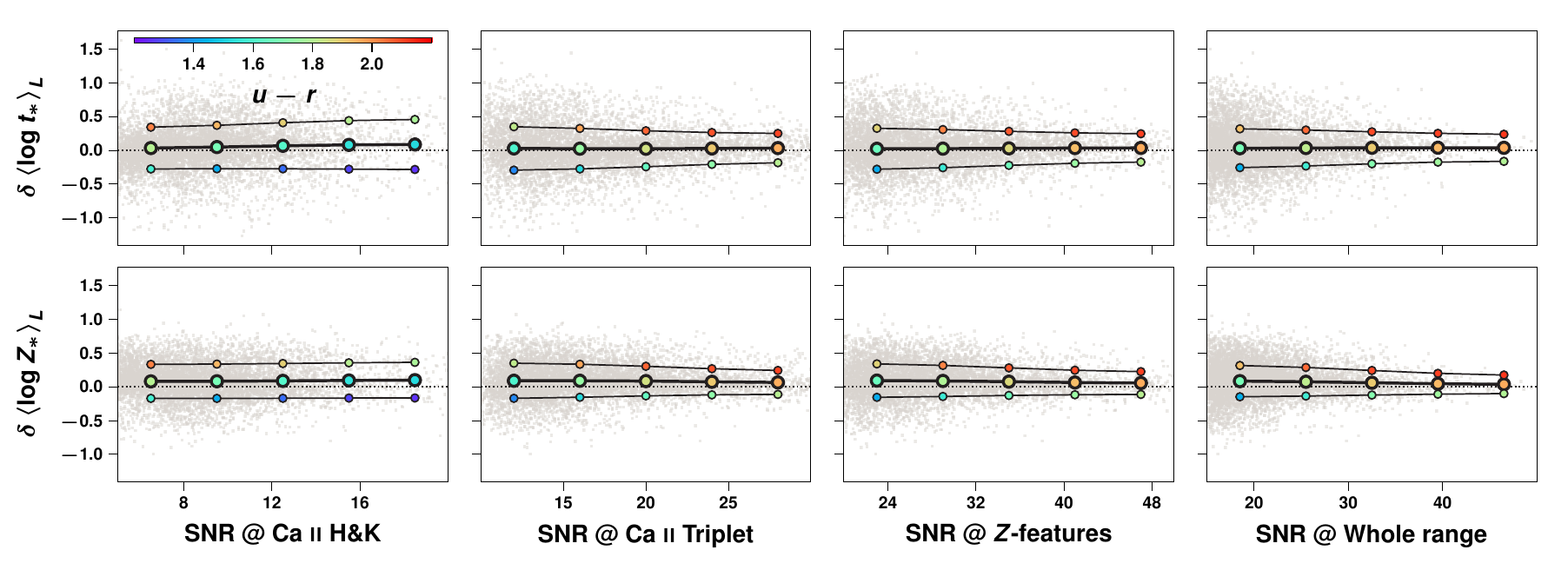}
\caption{Behaviour of the $\delta_\text{G05}x$ discrepancy as a function of median SNR in four
spectral regions, from left to right: Ca~\textsc{ii} H\&K lines, Ca~\textsc{ii} triplet,
$\lambda>\unit[7000]{\AA}$ ($Z$-feature), and the whole spectral range sampled by the SDSS. The grey
dots show the distribution of all galaxies in the MGSS. The big dots represent the mean
$\delta_\text{G05}x$ in five bins spanning each SNR range, colour-coded according to the mean $u-r$
colour to reflect the predominant galaxy type in each bin. The small dots represent the standard
deviation in $\delta_\text{G05}x$ colour-coded according to the standard deviation in $u-r$ to
indicate the galaxy type variation within the same bins. \emph{Top two rows:} $\clwla$ and $\clwlz$
for the galaxies in the MGSS, cf. Fig.~\ref{fig:amr}\emph{(a)}. There are no trends with the SNR in
any spectral range strong enough to explain the full amplitude of $\delta_\text{G05}x$ observed in
SFGs. It should be noted, though, that $\glwla$ is usually larger (in absolute value) towards
low-SNR blue galaxies. A correlation between the mean galaxy type (big colour-coded dots) and the
SNR is clearly visible in the Ca~\textsc{ii} triplet, the $\lambda>\unit[7000]{\AA}$, and the whole
spectral range, i.\,e., the smaller the SNR, the bluer the $u-r$ colour, although the galaxy type
variation is still large in some cases as indicated by the small colour-coded dots. Interestingly,
the trend is inverted in the Ca~\textsc{ii} H\&K lines, i.\,e., the smaller the SNR, the redder
the $u-r$ colour. \emph{Bottom two rows:} $\clwla$ and $\clwlz$ for the galaxies in the complete
observed sample, cf. Fig.~\ref{fig:amr}\emph{(f)}. The lack of significant trends of $\clwla$ and
$\clwlz$ with SNR in all the spectral ranges, reveals that the SNR is not a strong source of
systematics between the HR$^*$ and the NB$^*$ results, in agreement with our predictions in
\S\ref{sec:noise-effect}.}\label{fig:dis-vs-snr}
\end{figure*}
In Fig.~\ref{fig:dis-vs-par} we show the $\glwla$ (top panels) and $\glwlz$ (bottom panels)
discrepancies as a function of the physical properties retrieved by \dynbas by fitting the HR$^*$
from the MGSS. The big dots represent the mean discrepancy in five bins spanning each physical
property range, and are colour-coded according to the mean $u-r$ colour to reflect the predominant
galaxy type in each bin. The small dots represent the standard deviation in $\delta_\text{G05}x$,
colour-coded according to the standard deviation in $u-r$ to indicate the galaxy type variation
within the bin. The mean tendencies seem to be mainly a consequence of the mass-age and the
age-extinction degeneracies, which are not present in \citetalias{Gallazzi2005} determinations.
Nonetheless, these degeneracies can explain at most about half the amplitude of $\glwla$. It is
interesting to note that $\glwlz$ shows a trend with $\lwlz$ but not with $\lwla$, as would be
expected if the age-metallicity degeneracy is affecting the results derived from both data sets in a
relative fashion. The strong correlations with galaxy $u-r$ colour, most notably in $\logm$ and
$\lwla$, may be indicating that other sources of the $\delta_\text{G05}x$ discrepancy are needed to
explain its amplitude.
\subsection{Possible sources of the $\delta_\text{G05}x$ discrepancy}\label{sec:sources}
Since the expected discrepancy (from the TTA) for SFGs does not account for the observed systematics
in Fig.~\ref{fig:amr}, this may indicate that one or more additional sources of this discrepancy are
operating behind the scene. Here we discuss three possible causes of the discrepancies, namely:
\emph{(i)} the SNR of the observed SEDs; \emph{(ii)} differences in the SPS models used in the SED
fits; and \emph{(iii)}, differences in the nature of the data sets/methodologies adopted to estimate
the physical parameters.
\subsubsection{The impact of the SNR}
The fact that the statistical dispersion in $\delta_\text{G05}x$ is higher in the locus of SFGs
could be pointing to a relation of $\delta_\text{G05}x$ with the SNR of the SDSS spectra for these
galaxies, since on average the SNR is lower in SFG spectra, and there exists a dependency of the
bias and precision on the instrumental noise level, as shown in \S\ref{sec:noise-effect}. In the
series of panels in the top two rows of Fig.~\ref{fig:dis-vs-snr} we show the behaviour of the
$\glwla$ and $\glwlz$ discrepancies with the median SNR in several spectral regions: the
Ca~\textsc{ii} H\&K lines, the Ca~\textsc{ii} triplet, the $\lambda>\unit[7000]{\AA}$
($Z$-features) region, and the full spectral range covered by the SDSS spectra, computed for the
MGSS. The big dots represent the mean discrepancy in five bins spanning each SNR range, and are
colour-coded according to the mean $u-r$ colour to reflect the predominant galaxy type in each bin.
The small dots represent the standard deviation in $\delta_\text{G05}x$, colour-coded according to
the standard deviation in $u-r$ to indicate the galaxy type variation within the bin. In the
Ca~\textsc{ii} triplet, the $Z$-feature, and the whole spectral range diagrams, blue SFGs tend to
have low SNR and larger departures from $\delta_\text{G05}x=0.0$. In fact, $\glwla$ is slightly
anticorrelated with SNR, i.\,e. bluer SFGs have the lowest SNR and the largest $\glwla$, and
\emph{vice versa}. Curiously, in the Ca~\textsc{ii} H\&K lines region, $\delta_\text{G05}x$ is
directly correlated with SNR, i.\,e., the smaller the SNR, the smaller $\delta_\text{G05}x$, and
\emph{vice versa}. In the Ca~\textsc{ii} H\&K line region in the SDSS spectra, the H line is
missing while the K line is always present. Since the relative strength of the H\&K lines is
sensitive to stellar age, this could be a source of systematics introduced in our stellar property
estimates from the synthetic narrow-band photometry. We will turn back to this finding in
\S\ref{sec:jpas-amr}.
\subsubsection{The impact of the stellar library used in the SPS models}
G05 used the standard \citetalias{Bruzual2003} models whereas we use the BC03xm version of these
models described in Appendix \ref{sec:mockpar}. Both sets of models are based on the Padova 1994
stellar evolutionary tracks \citep{Alongi1993, Bressan1993, Fagotto1994a, Fagotto1994b,
Girardi1996}. In the optical range of interest to us these models differ only in the stellar library
used to build the galaxy SEDs. The STELIB library \citep{LeBorgne2003} is used in the
\citetalias{Bruzual2003} models and the MILES library
\citep{SanchezBlazquez2006,Falcon2011,Prugniel2011} in the BC03xm models. The main reason supporting
our choice is the very small number of stars with $Z$ away from $\unit[1.0]{Z$_\odot$}$ in the
STELIB library. The number of these stars in the MILES library is much larger (roughly by a factor
of 10), and the spectra have higher resolution and higher SNR than in STELIB. Moreover,
\citetalias{Bruzual2003} report that the evolutionary tracks and colours are calibrated for
$[\text{Fe/H}]=+0.25$, while the spectral features (including those used by
\citetalias{Gallazzi2005}) are calibrated using solar abundance models (see Appendix in
\citetalias{Bruzual2003} for details). As noted by \citetalias{Gallazzi2005}, this limitation is
particularly important for PaGs, since the \citetalias{Bruzual2003} models predict redder colours
than observed for $Z\sim\unit[2.5]{Z$_\odot$}$. \citetalias{Gallazzi2005} addressed this issue by
allowing for negative internal dust extinction to account for this \emph{colour excess} in the
models. In their analysis \citetalias{Gallazzi2005} found a strong age-metallicity degeneracy for
PaGs. Although we use a different stellar spectral library in our analysis, our results agree with
\citetalias{Gallazzi2005}'s in the presence of this degeneracy (cf. Fig.~\ref{fig:degeneracies}),
indicating a library-independent behaviour of this degeneracy (at least for the two stellar
libraries being tested).
\subsubsection{The impact of the adopted methodology and data set}\label{sec:fitted-ds}
\begin{figure}
\includegraphics[scale=1.0]{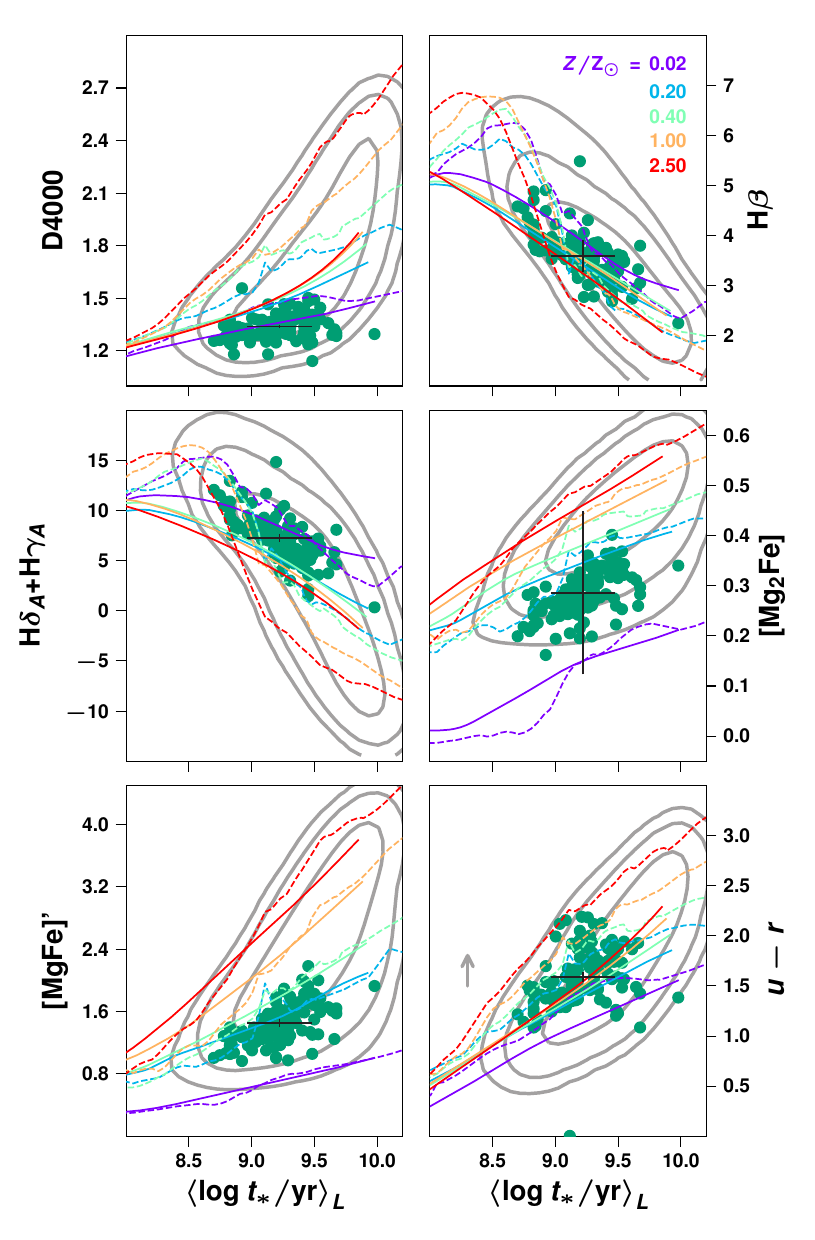}
\caption{The joint prior PDF in luminosity-weighted mean age computed from the SSAG and projected
onto the data set space $\dat=\left\{\Dn, \Hb, \Hdg, \MgFe, \MgbFe\right\}$ is shown as contours
corresponding to the $1$, $2$ and $3\sigma$ confidence regions. The projection onto the dereddened
stellar continuum as measured by the $u-r$ colour and the reddening vector (arrow) averaged over the
SSAG are also shown. The colour-coded lines indicate the indices strengths as a function of the age
for the standard \citetalias{Bruzual2003} SSP (dashed) and as a function of the luminosity-weighted
mean age for $\tau=\unit[5]{Gyr}$ (continuous) models for the different metallicities used in
\citetalias{Gallazzi2005}. A subset of galaxies from the MGSS for which the $\delta_\text{G05}x$ in
age and metallicity is large is shown as green dots, and placed in age according to
\citetalias{Gallazzi2005} estimates. The cross in each plane is centred at the the mean $\lwla$ and
the mean index strength and has a length of $2\sigma$, as measured from the joint posterior PDF in
$\lwla$ reported by \citetalias{Gallazzi2005} (horizontal direction) and the SDSS spectroscopic data
(vertical direction). In the planes corresponding to the $\Dn$, $\Hb$, and $\Hdg$ spectral indices
the subset of galaxies matches regions of high probability in the prior PDF. Interestingly, in the
planes corresponding to $\MgFe$ and $\MgbFe$, the subset of galaxies seems off the high probability
locus, favoring old and metal poor stellar populations. See text for
details.}\label{fig:g05-indices}
\end{figure}
In this section we evaluate the nature of the discrepancies introduced by the use of different data
sets and spectral modelling methodologies in \citetalias{Gallazzi2005} and this paper.
\citetalias{Gallazzi2005} inferred stellar properties by simultaneously modelling the five spectral
indices $\dat=\left\{\Dn, \Hb, \Hdg, \MgFe, \MgbFe \right\}$, suitable for studies of the recent SFH
in galaxies. The main motivation behind the \citetalias{Gallazzi2005} choice of the composite \MgFe
and \MgbFe spectral indices was that, while providing information of the stellar metallicity, these
indices show little dependence on complex chemical abundance patterns such as the
$\alpha$-enhancement, known to be present mainly on early-type galaxies \citep{Conroy2014}. Thus,
\citetalias{Gallazzi2005} were able to reduce the number of free parameters, excluding
$\alpha$-enhanced models. In this paper we also adopt SSP models with solar abundance patterns,
therefore the $\alpha$-enhancement is probably a source of systematics given our spectroscopic data
set. It should be noted however that, as mention before, $\alpha$-enhanced abundances are expected
in galaxies dominated by red old stellar populations, that as shown by the trends in
Fig.~\ref{fig:amr}\emph{(a)} have the smaller $\delta_\text{G05}x$ discrepancies. We can therefore
rule out the over abundance of $\alpha$ elements as a major source of systematics in
$\delta_\text{G05}x$. Other possible source of the $\delta_\text{G05}x$ discrepancy may be the
nebular infilling of the Balmer lines, which is expected to affect star-forming systems. Likewise,
the masking of these lines in our spectroscopic data set is possibly adding to the discrepancies as
well. However, as shown below, these effects are probably playing a minor r\^ole.

\citetalias{Gallazzi2005} use a Bayesian approach to determine the stellar content of galaxies (see
\S\ref{sec:sed-fitting}), i.\,e. they update the prior knowlegde on the physical properties of
galaxies through their data set (as indicated in Eq.~\ref{eq:bayes}), which under the plausible
assumption of Gaussian uncorrelated uncertainties yields the likelihood
\begin{equation}\label{eq:g05-likelihood}
\lik{\dat}{\hip} \propto \exp{\left\{-\sum_{i=1}^5\frac{\left[\mathcal{D}_i-\mu_i\left(\hip\right)\right]^2}{2{\sigma_i}^2}\right\}},
\end{equation}
where $\mu_i\left(\hip\right)$ represents the proposed model described by the parameters $\hip$.
From the equation above we can see that the relative importance of the likelihood in shaping the
posterior probability distribution depends on both the size of the data set, in this case $N=5$, and
its quality, $\text{SNR}\sim\mathcal{D}_i/\sigma_i$. For a fixed SNR, the likelihood scales with
$N$, whereas for fixed $N$ the likelihood scales with the SNR. Whenever the data set size is small
and/or low-quality, the adopted prior $\pri{\hip}$ dominates the posterior probability distribution.
It is therefore worth exploring how the combination of data set plus prior probability distribution
in \citetalias{Gallazzi2005} may contribute to the $\delta_\text{G05}x$ discrepancy.

We compute the prior PDF, $\pri{\hip}$, from a version of the SSAG using the same
\citetalias{Bruzual2003} models as in \citetalias{Gallazzi2005}, without including dust extinction
nor kinematics effects. In Fig.~\ref{fig:g05-indices} we show such prior PDF projected onto the
\citetalias{Gallazzi2005} data set space (grey contours). We show also its projection onto the
dereddened $u-r$ colour axis, which we choose as a proxy for the stellar continuum. The arrow
represents the averaged reddening vector from the original SSAG recipe, including dust extinction.
The age dependency predicted for the indices at the five values of $Z$ in the range
$0.02\,$---$\,\unit[2.5]{Z$_\odot$}$ for the standard \citetalias{Bruzual2003} SSPs (dashed lines)
and for the $\tau=\unit[5]{Gyr}$ models (continuous lines) is shown to highlight the expected
boundaries on each plane. At a first glance, those regions with a higher prior probability density
seem to match regions where the density of $\tau$-models is large, an expected result given the SFR
parametrization in the SSAG. It is noteworthy though, that there is a lower limit in age at
$\unit[\sim9]{dex}$, where the prior PDF drops rapidly, despite the fact that SSPs and $\tau$ models
clearly allow for younger stellar populations.

To see if the behaviour observed in Fig.~\ref{fig:g05-indices} corresponds with the predictions of
\citetalias{Gallazzi2005}, we select from the MGSS $\sim100$ SFGs with $\lwla<\unit[9]{dex}$ (from
the HR$^*$ fitting) and mean $\glwla,\glwlz\approx-0.6,-0.2$. To mitigate the importance of
degeneracies in the kinematic effects and the nebular infilling, we also require this subset of
galaxies to belong to the lowest percentile of the LOSVD distribution
($\sigma_v<\unit[50]{km\,s}^{-1}$), and we measure the spectral indices from the best fitting model
SED, both retrieved by \dynbas from the HR$^*$ spectroscopy. In Fig.~\ref{fig:g05-indices} these
galaxies are represented by the green dots, located in age according to \citetalias{Gallazzi2005}'s
results. The cross in each plane is centred at the subset mean age and mean index strength and have
a length of $2\sigma$ according to \citetalias{Gallazzi2005} estimates and SDSS measurements in the
horizontal and the vertical directions, respectively. It is worth noting that these galaxies are
located near the locus of the younger stellar content in the prior PDF ($\unit[\sim9]{dex}$)
projected onto the $\Dn$, $\Hb$, and $\Hdg$ indices, whilst in the $\MgFe$ and $\MgbFe$ projections,
the same galaxies require slightly younger populations and/or higher index strengths in those
spectral features to match the same locus. This result could be indicating a major influence of the
prior PDF in determining the stellar age. In the $u-r$ colour projection, adding to the prior PDF
the averaged colour excess indicated by the arrow seems to be enough to make the green dots fall
onto the high probability locus, indicating that this colour index may not provide further
information to the posterior PDF. We remark, however, that the prior set by the SSAG is probably
different from the one adopted by \citetalias{Gallazzi2005}, and such differences may well affect
the conclusions we can draw from the behaviour seen in Fig.~\ref{fig:g05-indices}.

\begin{figure}
\includegraphics[scale=1.0]{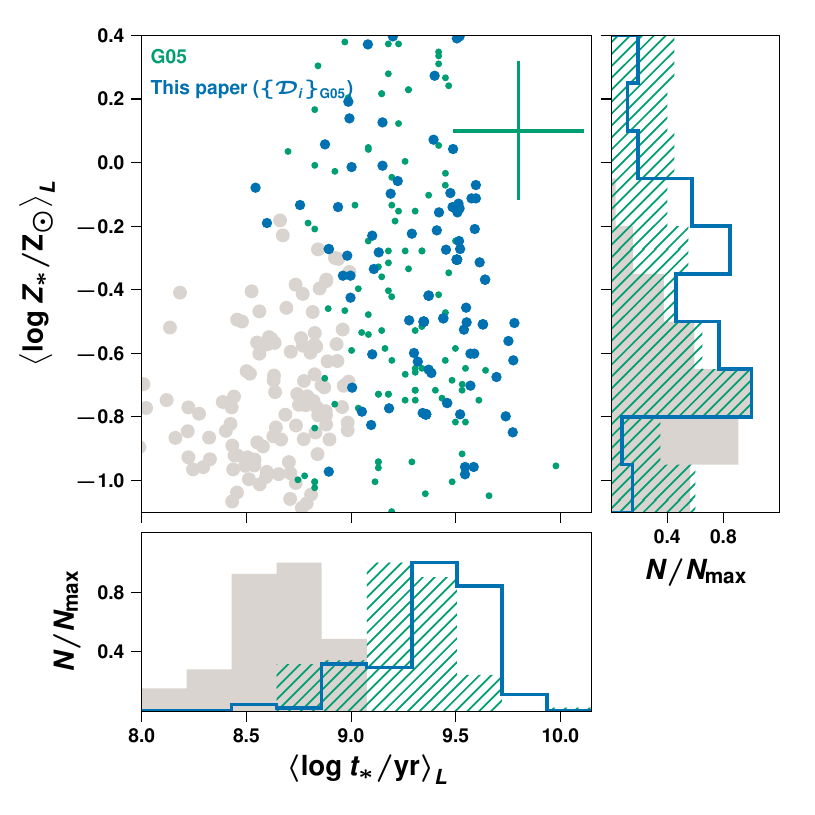}
\caption{The mode of the posterior PDF for each galaxy from the subset of $\sim100$ MGSS galaxies,
computed using the SSAG as prior, is shown in blue (labelled $\dat_\text{G05}$). The resuls from
\citetalias{Gallazzi2005} (green) are for comparison, demonstrating the same general behavior: high
dispersion in metallicity, with a tendency towards old and metal-poor stellar populations. The error
cross represents the typical $2\sigma$ range in age and metallicity in \citetalias{Gallazzi2005}
estimates. These results demonstrate that the derived galaxy properties are, to some extent,
independent of the chosen prior distribution, provided that the SSAG and \citetalias{Gallazzi2005}
adopt different priors. The results from this paper are also shown (grey) to illustrate the
discrepancy for this particular subset of galaxies. See text for details.}\label{fig:g05-posterior}
\end{figure}

To test if this is the case, we compute the posterior PDF by updating the knowlegde about $\hip$
enclosed by the SSAG through the likelihood in Eq.~\eqref{eq:g05-likelihood}, where $\mu_i(\hip)$
represents the SSAG's predictions of the same spectral indices as in \citetalias{Gallazzi2005}. The
distribution of the joint posterior PDF modes in $\lwla$ and $\lwlz$ for each galaxy is shown in
Fig.~\ref{fig:g05-posterior} (blue), along with the results from \citetalias{Gallazzi2005} (green),
and those we derived in this paper using \dynbas (grey). Despite the aforementioned difference in
the prior PDFs, we are able to retrieve trends in good agreement with \citetalias{Gallazzi2005}'s,
suggesting to some extent a degree of independency of the posterior PDF on the adopted prior,
regardless of the data set. However, the lack of stellar populations younger than $\unit[9]{dex}$ in
\citetalias{Gallazzi2005}, but present in the \dynbas results, could be pointing toward a limitation
in the parameter space sampling, according to the results presented in Fig.~\ref{fig:g05-indices}.
Incidentally, the results derived from the posterior PDF may be interpreted as either that the
indices $\Dn, \Hb,$ and $\Hdg$ place too strong constrains on stellar age given their typically high
SNR, therefore biasing the metallicity estimates through the age-metallicity degeneracy, or that the
$\MgFe$ and $\MgbFe$ indices are too weak to break this degeneracy in SFGs.
\subsection{The AMR determined from NB$^*$ photometry}\label{sec:jpas-amr}
In Fig.~\ref{fig:amr}\emph{(f)} we show the AMR inferred from our complete observed sample using the
NB$^*$ fits (grey-shaded area and contours). The light blue dots and crosses indicate the mean and
RMSD values for the galaxies in this sample in the same five $u-r$ colour bins of
Fig.~\ref{fig:amr}\emph{(a)}, computed using HR$^*$. The dark blue dots and crosses show the AMR for
the same galaxies derived from NB$^*$. Figs.~\ref{fig:amr}\emph{(g)\,--\,(j)} show the maps of the
mean and RMSD $\delta x$ discrepancy in $\lwla$ and $\lwlz$. Even though the main trends between the
photometric and spectroscopic results are essentially the same, cf. Figs.~\ref{fig:amr}\emph{(a)}
and \ref{fig:amr}\emph{(f)}, a closer inspection of the PDF (grey-shaded regions) reveals some
differences worth discussing. The PaGs in the two redder bins seem to have migrated towards higher
metallicities and slightly younger ages in the NB$^*$ AMR with respect to their position in the
spectroscopic AMR. This is consistent with our findings in \S\ref{sec:main-tta}, where we showed
that for this type of galaxies the age-metallicity degeneracy is more pronounced in the NB$^*$ than
in the HR$^*$ results, although the systematics found in the TTA cannot fully account for the
observed discrepancy. In the SFG region, the HR$^*$ results include a population of young and
metal-poor galaxies, reaching $\lwla$ $\unit[\sim8.5]{}$ and $\lwlz$ $\unit[\sim -0.8]{}$ at the
extremes of the bluest bin, which is absent in the NB$^*$ prediction. Instead, there are two groups
of galaxies in the NB$^*$ AMR not present in the HR$^*$ AMR: an extremely young and relatively
metal-rich subpopulation, ($\lwla$ $\unit[\sim8]{}$, $\lwlz$ $\unit[\sim-0.3]{}$), and a slightly
older and extremely metal-poor subpopulation, ($\lwla$ $\unit[>9.2]{}$, $\lwlz$ $\unit[<-1.0]{}$).
In Figs.~\ref{fig:amr}\emph{(g)\,--\,(h)} we can see that the gradient vectors in the mean maps of
$\delta x$ follow the age-metallicity degeneracy direction. The larger discrepancies are found
around $(8.0,-0.2)$, $(9.3,-1.1)$, and $(9.6,-0.5)$. Such regions have also the larger statistical
dispersion (RMSD, Figs.~\ref{fig:amr}\emph{(i)\,--\,(j)}). None of these discrepancies were
predicted in the TTA. Since the models and the method are the same for both data sets, this may
indicate that the origin of this discrepancy is related to observational artefacts.

In the bottom two rows of Fig.~\ref{fig:dis-vs-snr} we show the behaviour of $\clwla$ and $\clwlz$
with SNR in several spectral ranges for the NB$^*$ results. In general, SFGs have the lowest SNR at
the Ca~\textsc{ii} triplet, the $Z$-features, and the whole spectral ranges, and show the highest
discrepancy in $\clwla$ and $\clwlz$. The insensitivity of $\clwla$ and $\clwlz$ to the median SNR
in all the spectral ranges shown, reveals that the data quality (as measured by the SNR) is a not
strong source of systematics between the HR$^*$ and the NB$^*$ results. Therefore, the combination
of the lower spectral resolution of NB photometric data and the invisibility of some
age/metallicity-sensitive spectral features are likely the major sources of systematics. Such is the
case of the masking of the Balmer lines due to emission and the missing pixels for imperfect
spectroscopic observations. Particularly incident is the fact that the Ca~\textsc{ii} H line, part of
the Ca~\textsc{ii} H\&K spectral index and a stellar age tracer, is missing from spectroscopic
observations. In the blue region of the SED ($\unit[<4000]{\AA}$) we found that this lack produces a
strong systematic offset in the selected best fitting model, generally bluer than the observed SED.
This also explains the opposite behaviour of the discrepancies in stellar age and stellar
metallicity with SNR in the Ca~\textsc{ii} H\&K spectral range: the higher the SNR around this
feature, even if incomplete, increases its relative importance in the SED fit (since
$\chi^2_\lambda\propto1/\sigma_\lambda^2$) and the chances of mismatching the corresponding spectral
fitting result. Moreover, the lack of sensitivity to stellar metallicity of the reddest region of
the SED ($\unit[>7000]{\AA}$), due to missing pixels, contributes to the observed systematic
discrepancy in $\lwlz$ and $\mwlz$. This last fact is specially important for PaGs, for which the
photometric stellar metallicity estimates rely strongly on the availability of the reddest region of
the optical SED.

In our estimations from both HR$^*$ and NB$^*$ data, galaxies dominated by $\unit[<1]{Gyr}$ stars
are relatively rare, adding up to $\unit[\sim25]{per cent}$ in the observed sample. In the SSAG,
presumably representing a similar prior knowledge as that used in \citetalias{Gallazzi2005}, around
$\unit[20]{per cent}$ of the galaxies are dominated by such young populations and are characterised
by a specific SFR, i.\,e. the fraction of newly formed stellar mass per unit time averaged over a
given time scale, $\text{sSFR}\equiv\text{SFR}/M_*\unit[\approx0.18]{Gyr}^{-1}$. Given their
remarkable absence from the \citetalias{Gallazzi2005} results, the occurrence of these galaxies is
worth exploring in the context of current galaxy formation simulations. Recently, \citet{Guo2016}
presented a comprehensive comparison of three widely known simulations in the literature (see
references therein). The predicted distribution of the sSFR for galaxies with $\logm<9.75$ (blue
SFGs in our MGSS) is quite similar at $z=0$ among all explored simulations, characterised by a
median around $\sim0.06\,$---$\,\unit[0.16]{Gyr}^{-1}$, a long high-probability and slowly declining
tail toward lower values and a rapid drop toward upper values. This suggest that galaxies with
important recent star formation events, dominated by young stellar populations, may be in fact rare
in the nearby Universe. Indeed, the bulk of mass the budget seems to be in the form of old
metal-rich stars, albeit in the lowest mass bin young stars probably dominate \citep{Baldry2004,
Conselice2006}. This is also consistent with the reviewed references by \citet{Madau2014}, from
which estimations using UV/IR calibrations span $\text{sSFR}\sim0.10\,$---$\,\unit[0.30]{Gyr}^{-1}$
in the local Universe. From the MGSS, these galaxies ($\logm<9.75$ from \citetalias{Gallazzi2005}
estimations) are characterised by the following distributions, according to our NB$^*$
(\citetalias{Gallazzi2005}) results:
\begin{subequations}
\begin{align*}
\logm&\approx+9.69_{-0.34}^{+0.27}\quad\left(+9.63_{-0.23}^{+0.09}\right), \\
\lwla&\approx+8.76_{-0.93}^{+0.28}\quad\left(+9.08_{-0.26}^{+0.26}\right), \\
\lwlz&\approx-0.42_{-0.33}^{+0.25}\quad\left(-0.58_{-0.51}^{+0.75}\right).
\end{align*}
\end{subequations}
From the simulations explored by \citeauthor{Guo2016}, galaxies in this stellar mass range at $z=0$,
span a stellar metallicity range as wide as $\log{Z_*/\text{Z}_\odot}\approx-0.73\,$---$\,0.07$,
in agreement with our results and those from \citetalias{Gallazzi2005}. Nonetheless, the predictions
from galaxy formation simulations, specially in the case of low stellar mass galaxies, are still
highly contradictory among different implementations, most likely due to uncertainties regarding the
assumed physics of the star formation, the chemical enrichment and the regulating processes thereof
\citep[see][]{Guo2016, Naab2017}. As a matter of fact, the intrinsic scatter and the relative
importance of the global and local phenomena in shaping the AMR/MMR are subjects of an active debate
\citep[e.\,g.][]{Lara-Lopez2010, Rosales-Ortega2012, Sanchez2013, Gonzalez2014}. In the near future
the results from the spectral fitting will continue to prove valuable to both, fossil studies and
numerical simulations, in our path towards unraveling the nature of these relations in star-forming
systems.

\section{Summary and Conclusions}\label{sec:concl}

We have explored the consistency among the physical properties of galaxies retrieved from
SED-fitting data sets of different spectral resolution and quality. We perform our tests on two
different samples of galaxy spectra. For the first sample we draw $7$k high quality spectra from the
SDSS-DR7 including both PaGs ($u-r > \unit[2.22]{mag})$ and SFGs ($u-r \le \unit[2.22]{mag})$. More
than $\unit[80]{per cent}$ of these galaxies are in common with the sample studied by
\citetalias{Gallazzi2005} and define the MGSS. For the second sample we build a set of $2680$ mock
galaxy spectra representing the local Universe as realistically as possible by combining a randomly
selected collection of galaxy SFH's from \citetalias{Chen2012} with the \citetalias{Bruzual2003}xm
SPS models. For the mock sample we simulate broad-band (BB) and narrow-band (NB) observations across
the optical wavelength range, using the $u'g'r'i'z'$ and the $56$ J-PAS passbands response
functions, respectively. In the case of the observed sample a subset NB$^*$ of only $\sim 40$ can be
synthesized from the SDSS spectra, due to instrumental artefacts. We use the \dynbas code to fit the
high-resolution (HR) spectra in both samples, as well as their BB, NB and NB$^*$ versions. From each
spectral fit, we compute the stellar mass, the mass- and luminosity-weighted age and metallicity,
and the dust extinction for the associated galaxy. For the galaxies in the mock sample we know the
true value of each of these properties.

The galaxy properties retrieved from fitting the mock sample at the BB resolution are usually
biased. At the NB and HR resolutions the biases are practically nil for PaGs and show a slight
increase for the bluer SFGs. Such biases are produced at several levels by the modelling
methodology, the instrumental effects, the assumed physics and the ability of some galaxies to hide
their past SFH, and they manifest through the several well-known degeneracies between stellar mass,
stellar age, stellar metallicity, and dust extinction. The statistical dispersion, on the other
hand, arises from our simulation of observational effects through the addition of random noise to
the fitted mock SEDs, and from the ability of SFGs to rejuvenate their stellar population. In fact,
our results indicate that the instrumental noise is relatively unimportant as a source of the bias
at the HR and NB spectral resolutions, but becomes more important towards blue SFGs and at lower
spectral resolutions. We find that the strength of the several degeneracies is a function of the
spectral resolution and galaxy type, albeit exhibits little dependence on the level of instrumental
noise. PaGs show comparable biases and strong presence of the mass-age, age-metallicity, and
age-dust extinction degeneracies for all spectral resolutions. SFGs, on the other hand, are more
prone to show the age-dust extinction degeneracy at the BB resolution rather than at the HR, whilst
the age-metallicity degeneracy is hidden at the BB resolution and appears progressively with
increasing spectral resolution. In general, the biases and degeneracy strengths resulting from the
HR and NB fits are comparable.

The direct comparison of the parameters derived from HR$^*$ spectroscopy and NB$^*$ photometry for
the galaxies in the observed sample shows trends with galaxy colour. For PaGs these trends are
consistent with the relative presence of the age-metallicity, the mass-age, and, to a lesser extent,
the age-dust extinction degeneracies. For SFGs the trends are consistent with the relative presence
of the mass-age and the age-dust extinction degeneracies, and the absence of the age-metallicity
degeneracy. The existence/absence of the degeneracies is consistent with the results from the mock
sample fits, but the systematic discrepancies expected from the mock sample results are not large
enough to account for the discrepancies in the results derived from the observed samples. This
suggests that additional sources of systematic discrepancies between the NB$^*$ photometry and the
HR$^*$ spectroscopy results, not accounted for in the mock galaxy modelling, are at play. We
evaluated two possible sources: the decreasing SNR in the lower surface brightness regions of SFGs,
and the lack of stellar age and metallicity indicators in the spectral regions of interest affecting
both PaGs and SFGs. The latter is propagated from the artefacts in HR$^*$ data to the NB$^*$
synthetic photometry, showing the importance of sampling in relevant wavelength ranges to provide
with reliable estimates of the stellar contents in galaxies. Indeed the discrepancies in the
parameters determined from NB$^*$ photometry and HR$^*$ can be understood as due to these additional
sources of systematics.

Using the \citetalias{Gallazzi2005} results for the MGSS as a benchmark to compare with our
spectroscopic results, we show \emph{the r\^ole of the prior PDF in shaping the posterior PDF when
modelling small and/or low-SNR data sets, and how it may introduce biases not accounted for in a
Bayesian framework}. Despite the aforementioned sources of uncertainty, the same AMR trends can be
drawn using our methodology and the \citetalias{Gallazzi2005}'s, both consistent with more recent
results in the literature \citep[e.\,g.,][]{Panter2008, Gonzalez2014}. In general, the distributions
of the stellar mass and the mass-weighted age determined from HR$^*$ spectroscopy and NB$^*$
photometry are consistent across the full colour range. The luminosity-weighted age and dust
extinction distributions are consistent only for PaGs. The statistical dispersion in the parameters
derived for PaGs from HR$^*$ and NB$^*$ data are consistent with, or at least comparable to, the
expected dispersion from the mock sample fits. We conclude that \emph{NB photometry can provide the
stellar mass, stellar mass- and luminosity-weighted age, and dust extinction to an accuracy similar
to spectroscopic data sets, however the precision provided by the HR data sets still outmatches that
from NB photometry}, an effect that may be crucial in studies of distant galaxies.

\section*{Acknowledgments}

We thank an anonymous referee for very useful comments that made this paper more clear, a bit
longer, but hopefully more useful to the interested reader. Special thanks to Cecilia Mateu for
insightful discussions and useful suggestions. AMN acknowledges support from the Sociedad Mexicana
de F\'isica through its Program M\'exico-Centro Am\'erica y el Caribe para el Avance de la Ciencia,
la Tecnolog\'ia y la Innovaci\'on, and thanks the Centro de Investigaciones de Astronom\'ia (CIDA)
for a graduate student grant. AMN also thanks the warm hospitality of the Instituto de
Radioastronom\'ia y Astrof\'isica of the National Autonomous University of M\'exico (IRyA, UNAM) and
the Centro de Estudios de F\'isica del Cosmos de Arag\'on (CEFCA) during part of this research. GB
acknowledges support for this work from UNAM through grant PAPIIT IG100115.

Funding for the SDSS and SDSS-II has been provided by the Alfred P. Sloan Foundation, the
Participating Institutions, the National Science Foundation, the U.S. Department of Energy, the
National Aeronautics and Space Administration, the Japanese Monbukagakusho, the Max Planck Society,
and the Higher Education Funding Council for England. The SDSS Web Site is
\url{http://www.sdss.org/}.

The SDSS is managed by the Astrophysical Research Consortium for the Participating Institutions. The
Participating Institutions are the American Museum of Natural History, Astrophysical Institute
Potsdam, University of Basel, University of Cambridge, Case Western Reserve University, University
of Chicago, Drexel University, Fermilab, the Institute for Advanced Study, the Japan Participation
Group, Johns Hopkins University, the Joint Institute for Nuclear Astrophysics, the Kavli Institute
for Particle Astrophysics and Cosmology, the Korean Scientist Group, the Chinese Academy of Sciences
(LAMOST), Los Alamos National Laboratory, the Max-Planck-Institute for Astronomy (MPIA), the
Max-Planck-Institute for Astrophysics (MPA), New Mexico State University, Ohio State University,
University of Pittsburgh, University of Portsmouth, Princeton University, the United States Naval
Observatory, and the University of Washington.

\bibliographystyle{mnras}
\bibliography{library}

\begin{thebibliography}{}
\makeatletter
\relax
\def\mn@urlcharsother{\let\do\@makeother \do\$\do\&\do\#\do\^\do\_\do\%\do\~}
\def\mn@doi{\begingroup\mn@urlcharsother \@ifnextchar [ {\mn@doi@}
  {\mn@doi@[]}}
\def\mn@doi@[#1]#2{\def\@tempa{#1}\ifx\@tempa\@empty \href
  {http://dx.doi.org/#2} {doi:#2}\else \href {http://dx.doi.org/#2} {#1}\fi
  \endgroup}
\def\mn@eprint#1#2{\mn@eprint@#1:#2::\@nil}
\def\mn@eprint@arXiv#1{\href {http://arxiv.org/abs/#1} {{\tt arXiv:#1}}}
\def\mn@eprint@dblp#1{\href {http://dblp.uni-trier.de/rec/bibtex/#1.xml}
  {dblp:#1}}
\def\mn@eprint@#1:#2:#3:#4\@nil{\def\@tempa {#1}\def\@tempb {#2}\def\@tempc
  {#3}\ifx \@tempc \@empty \let \@tempc \@tempb \let \@tempb \@tempa \fi \ifx
  \@tempb \@empty \def\@tempb {arXiv}\fi \@ifundefined
  {mn@eprint@\@tempb}{\@tempb:\@tempc}{\expandafter \expandafter \csname
  mn@eprint@\@tempb\endcsname \expandafter{\@tempc}}}

\bibitem[\protect\citeauthoryear{{Abazajian} et~al.,}{{Abazajian}
  et~al.}{2009}]{Abazajian2009}
{Abazajian} K.~N.,  et~al., 2009, \mn@doi [ApJS] {10.1088/0067-0049/182/2/543},
  182, 543

\bibitem[\protect\citeauthoryear{{Alongi}, {Bertelli}, {Bressan}, {Chiosi},
  {Fagotto}, {Greggio}  \& {Nasi}}{{Alongi} et~al.}{1993}]{Alongi1993}
{Alongi} M.,  {Bertelli} G.,  {Bressan} A.,  {Chiosi} C.,  {Fagotto} F.,
  {Greggio} L.,   {Nasi} E.,  1993, A\&AS, 97, 851

\bibitem[\protect\citeauthoryear{{Baldry}, {Glazebrook}, {Brinkmann},
  {Ivezi{\'c}}, {Lupton}, {Nichol}  \& {Szalay}}{{Baldry}
  et~al.}{2004}]{Baldry2004}
{Baldry} I.~K.,  {Glazebrook} K.,  {Brinkmann} J.,  {Ivezi{\'c}} {\v Z}.,
  {Lupton} R.~H.,  {Nichol} R.~C.,   {Szalay} A.~S.,  2004, \mn@doi [ApJ]
  {10.1086/380092}, 600, 681

\bibitem[\protect\citeauthoryear{{Bastian}, {Covey}  \& {Meyer}}{{Bastian}
  et~al.}{2010}]{Bastian2010}
{Bastian} N.,  {Covey} K.~R.,   {Meyer} M.~R.,  2010, \mn@doi [ARA\&A]
  {10.1146/annurev-astro-082708-101642}, 48, 339

\bibitem[\protect\citeauthoryear{{Baugh}, {Cole}  \& {Frenk}}{{Baugh}
  et~al.}{1996}]{Baugh1996}
{Baugh} C.~M.,  {Cole} S.,   {Frenk} C.~S.,  1996, \mn@doi [MNRAS]
  {10.1093/mnras/283.4.1361}, \href
  {http://adsabs.harvard.edu/abs/1996MNRAS.283.1361B} {283, 1361}

\bibitem[\protect\citeauthoryear{{Bell} \& {de Jong}}{{Bell} \& {de
  Jong}}{2000}]{Bell2000}
{Bell} E.~F.,  {de Jong} R.~S.,  2000, \mn@doi [MNRAS]
  {10.1046/j.1365-8711.2000.03138.x}, 312, 497

\bibitem[\protect\citeauthoryear{{Bell} \& {de Jong}}{{Bell} \& {de
  Jong}}{2001}]{Bell2001}
{Bell} E.~F.,  {de Jong} R.~S.,  2001, \mn@doi [ApJ] {10.1086/319728}, 550, 212

\bibitem[\protect\citeauthoryear{{Ben{\'{\i}}tez}}{{Ben{\'{\i}}tez}}{2000}]{Benitez2000}
{Ben{\'{\i}}tez} N.,  2000, \mn@doi [ApJ] {10.1086/308947}, 536, 571

\bibitem[\protect\citeauthoryear{{Benitez} et~al.,}{{Benitez}
  et~al.}{2014}]{Benitez2014}
{Benitez} N.,  et~al., 2014, preprint,
  {(\href{https://arxiv.org/abs/1403.5237}{arXiv:1403.5237})}

\bibitem[\protect\citeauthoryear{{Bernardi}, {Renzini}, {da Costa}, {Wegner},
  {Alonso}, {Pellegrini}, {Rit{\'e}}  \& {Willmer}}{{Bernardi}
  et~al.}{1998}]{Bernardi1998}
{Bernardi} M.,  {Renzini} A.,  {da Costa} L.~N.,  {Wegner} G.,  {Alonso} M.~V.,
   {Pellegrini} P.~S.,  {Rit{\'e}} C.,   {Willmer} C.~N.~A.,  1998, \mn@doi
  [ApJ] {10.1086/311742}, 508, L143

\bibitem[\protect\citeauthoryear{{Bevington} \& Robinson}{{Bevington} \&
  Robinson}{2003}]{Bevington2003}
{Bevington} P.~R.,  Robinson D.~K.,  2003, {Data Reduction and Error Analysis
  for the Physical Sciences}.
McGraw-Hill

\bibitem[\protect\citeauthoryear{{Bressan}, {Fagotto}, {Bertelli}  \&
  {Chiosi}}{{Bressan} et~al.}{1993}]{Bressan1993}
{Bressan} A.,  {Fagotto} F.,  {Bertelli} G.,   {Chiosi} C.,  1993, A\&AS, 100,
  647

\bibitem[\protect\citeauthoryear{{Brinchmann}, {Charlot}, {White}, {Tremonti},
  {Kauffmann}, {Heckman}  \& {Brinkmann}}{{Brinchmann}
  et~al.}{2004}]{Brinchmann2004}
{Brinchmann} J.,  {Charlot} S.,  {White} S.~D.~M.,  {Tremonti} C.,  {Kauffmann}
  G.,  {Heckman} T.,   {Brinkmann} J.,  2004, \mn@doi [MNRAS]
  {10.1111/j.1365-2966.2004.07881.x}, 351, 1151

\bibitem[\protect\citeauthoryear{{Bruzual} \& {Charlot}}{{Bruzual} \&
  {Charlot}}{2003}]{Bruzual2003}
{Bruzual} G.,  {Charlot} S.,  2003, \mn@doi [MNRAS]
  {10.1046/j.1365-8711.2003.06897.x}, 344, 1000

\bibitem[\protect\citeauthoryear{{Cabrera-Ziri} \&
  {Mej{\'{\i}}a-Narv{\'a}ez}}{{Cabrera-Ziri} \&
  {Mej{\'{\i}}a-Narv{\'a}ez}}{2014}]{SSAG2014}
{Cabrera-Ziri} I.,  {Mej{\'{\i}}a-Narv{\'a}ez} A.,  2014, Synthetic Spectral
  Atlas of
  Galaxies~({\href{http://www.astro.ljmu.ac.uk/~asticabr/SSAG.html}{SSAG}})

\bibitem[\protect\citeauthoryear{{Cabrera-Ziri} et~al.,}{{Cabrera-Ziri}
  et~al.}{2016}]{Cabrera-Ziri2016}
{Cabrera-Ziri} I.,  et~al., 2016, \mn@doi [MNRAS] {10.1093/mnras/stv2977}, 457,
  809

\bibitem[\protect\citeauthoryear{{Cappellari}}{{Cappellari}}{2016}]{Cappellari2016}
{Cappellari} M.,  2016, \mn@doi [ARA\&A] {10.1146/annurev-astro-082214-122432},
  54, 597

\bibitem[\protect\citeauthoryear{{Cardelli}, {Clayton}  \& {Mathis}}{{Cardelli}
  et~al.}{1989}]{Cardelli1989}
{Cardelli} J.~A.,  {Clayton} G.~C.,   {Mathis} J.~S.,  1989, \mn@doi [ApJ]
  {10.1086/167900}, 345, 245

\bibitem[\protect\citeauthoryear{{Castellano} et~al.,}{{Castellano}
  et~al.}{2014}]{Castellano2014}
{Castellano} M.,  et~al., 2014, \mn@doi [A\&A] {10.1051/0004-6361/201322704},
  566, A19

\bibitem[\protect\citeauthoryear{{Chabrier}}{{Chabrier}}{2003}]{Chabrier2003}
{Chabrier} G.,  2003, \mn@doi [PASP] {10.1086/376392}, 115, 763

\bibitem[\protect\citeauthoryear{{Charlot} \& {Fall}}{{Charlot} \&
  {Fall}}{2000}]{Charlot2000}
{Charlot} S.,  {Fall} S.~M.,  2000, \mn@doi [ApJ] {10.1086/309250}, 539, 718

\bibitem[\protect\citeauthoryear{{Chen} et~al.,}{{Chen}
  et~al.}{2012}]{Chen2012}
{Chen} Y.-M.,  et~al., 2012, \mn@doi [MNRAS]
  {10.1111/j.1365-2966.2011.20306.x}, 421, 314

\bibitem[\protect\citeauthoryear{{Chevallard} \& {Charlot}}{{Chevallard} \&
  {Charlot}}{2016}]{Chevallard2016}
{Chevallard} J.,  {Charlot} S.,  2016, \mn@doi [MNRAS] {10.1093/mnras/stw1756},
  462, 1415

\bibitem[\protect\citeauthoryear{{Cid Fernandes}, {Mateus}, {Sodr{\'e}},
  {Stasi{\'n}ska}  \& {Gomes}}{{Cid Fernandes} et~al.}{2005}]{CidFernandes2005}
{Cid Fernandes} R.,  {Mateus} A.,  {Sodr{\'e}} L.,  {Stasi{\'n}ska} G.,
  {Gomes} J.~M.,  2005, \mn@doi [MNRAS] {10.1111/j.1365-2966.2005.08752.x},
  358, 363

\bibitem[\protect\citeauthoryear{{Cid Fernandes} et~al.,}{{Cid Fernandes}
  et~al.}{2013}]{CidFernandes2013}
{Cid Fernandes} R.,  et~al., 2013, \mn@doi [A\&A]
  {10.1051/0004-6361/201220616}, 557, A86

\bibitem[\protect\citeauthoryear{{Coelho}}{{Coelho}}{2014}]{Coelho2014}
{Coelho} P.~R.~T.,  2014, \mn@doi [MNRAS] {10.1093/mnras/stu365}, 440, 1027

\bibitem[\protect\citeauthoryear{{Conroy}}{{Conroy}}{2013}]{Conroy2013a}
{Conroy} C.,  2013, \mn@doi [ARA\&A] {10.1146/annurev-astro-082812-141017}, 51,
  393

\bibitem[\protect\citeauthoryear{{Conroy} \& {Gunn}}{{Conroy} \&
  {Gunn}}{2010}]{Conroy2010b}
{Conroy} C.,  {Gunn} J.~E.,  2010, \mn@doi [ApJ] {10.1088/0004-637X/712/2/833},
  712, 833

\bibitem[\protect\citeauthoryear{{Conroy}, {Gunn}  \& {White}}{{Conroy}
  et~al.}{2009}]{Conroy2009}
{Conroy} C.,  {Gunn} J.~E.,   {White} M.,  2009, ApJ, 699, 486

\bibitem[\protect\citeauthoryear{{Conroy}, {White}  \& {Gunn}}{{Conroy}
  et~al.}{2010}]{Conroy2010a}
{Conroy} C.,  {White} M.,   {Gunn} J.~E.,  2010, \mn@doi [ApJ]
  {10.1088/0004-637X/708/1/58}, 708, 58

\bibitem[\protect\citeauthoryear{{Conroy}, {Dutton}, {Graves}, {Mendel}  \&
  {van Dokkum}}{{Conroy} et~al.}{2013}]{Conroy2013b}
{Conroy} C.,  {Dutton} A.~A.,  {Graves} G.~J.,  {Mendel} J.~T.,   {van Dokkum}
  P.~G.,  2013, \mn@doi [ApJ] {10.1088/2041-8205/776/2/L26}, 776, L26

\bibitem[\protect\citeauthoryear{{Conroy}, {Graves}  \& {van Dokkum}}{{Conroy}
  et~al.}{2014}]{Conroy2014}
{Conroy} C.,  {Graves} G.~J.,   {van Dokkum} P.~G.,  2014, \mn@doi [ApJ]
  {10.1088/0004-637X/780/1/33}, 780, 33

\bibitem[\protect\citeauthoryear{{Conselice}}{{Conselice}}{2006}]{Conselice2006}
{Conselice} C.~J.,  2006, \mn@doi [MNRAS] {10.1111/j.1365-2966.2006.11114.x},
  373, 1389

\bibitem[\protect\citeauthoryear{{Da Cunha}, {Charlot}  \& {Elbaz}}{{Da Cunha}
  et~al.}{2008}]{daCunha2008}
{Da Cunha} E.,  {Charlot} S.,   {Elbaz} D.,  2008, \mn@doi [MNRAS]
  {10.1111/j.1365-2966.2008.13535.x}, 388, 1595

\bibitem[\protect\citeauthoryear{{De Lucia} \& {Blaizot}}{{De Lucia} \&
  {Blaizot}}{2007}]{DeLucia2007}
{De Lucia} G.,  {Blaizot} J.,  2007, \mn@doi [MNRAS]
  {10.1111/j.1365-2966.2006.11287.x}, 375, 2

\bibitem[\protect\citeauthoryear{{D{\'{\i}}az-Garc{\'{\i}}a}
  et~al.,}{{D{\'{\i}}az-Garc{\'{\i}}a} et~al.}{2015}]{Diaz2015}
{D{\'{\i}}az-Garc{\'{\i}}a} L.~A.,  et~al., 2015, \mn@doi [A\&A]
  {10.1051/0004-6361/201425582}, 582, A14

\bibitem[\protect\citeauthoryear{{Doi} et~al.,}{{Doi} et~al.}{2010}]{Doi2010}
{Doi} M.,  et~al., 2010, \mn@doi [AJ] {10.1088/0004-6256/139/4/1628}, 139, 1628

\bibitem[\protect\citeauthoryear{{Dupke}, {Benitez}, {Moles}, {Sodre}  \&
  {J-PAS Collaboration}}{{Dupke} et~al.}{2015}]{Dupke2015}
{Dupke} R.~a.,  {Benitez} N.,  {Moles} M.,  {Sodre} L.,   {J-PAS Collaboration}
  2015, IAU General Assembly, 29, \#2257789

\bibitem[\protect\citeauthoryear{{Faber}}{{Faber}}{1972}]{Faber1972}
{Faber} S.~M.,  1972, A\&A, 20, 361

\bibitem[\protect\citeauthoryear{{Fagotto}, {Bressan}, {Bertelli}  \&
  {Chiosi}}{{Fagotto} et~al.}{1994a}]{Fagotto1994a}
{Fagotto} F.,  {Bressan} A.,  {Bertelli} G.,   {Chiosi} C.,  1994a, A\&AS, 104

\bibitem[\protect\citeauthoryear{{Fagotto}, {Bressan}, {Bertelli}  \&
  {Chiosi}}{{Fagotto} et~al.}{1994b}]{Fagotto1994b}
{Fagotto} F.,  {Bressan} A.,  {Bertelli} G.,   {Chiosi} C.,  1994b, A\&AS, 105

\bibitem[\protect\citeauthoryear{{Falc{\'o}n-Barroso},
  {S{\'a}nchez-Bl{\'a}zquez}, {Vazdekis}, {Ricciardelli}, {Cardiel}, {Cenarro},
  {Gorgas}  \& {Peletier}}{{Falc{\'o}n-Barroso} et~al.}{2011}]{Falcon2011}
{Falc{\'o}n-Barroso} J.,  {S{\'a}nchez-Bl{\'a}zquez} P.,  {Vazdekis} A.,
  {Ricciardelli} E.,  {Cardiel} N.,  {Cenarro} A.~J.,  {Gorgas} J.,
  {Peletier} R.~F.,  2011, \mn@doi [A\&A] {10.1051/0004-6361/201116842}, 532,
  A95

\bibitem[\protect\citeauthoryear{{Gallazzi}, {Charlot}, {Brinchmann}, {White}
  \& {Tremonti}}{{Gallazzi} et~al.}{2005}]{Gallazzi2005}
{Gallazzi} A.,  {Charlot} S.,  {Brinchmann} J.,  {White} S.~D.~M.,   {Tremonti}
  C.~A.,  2005, MNRAS, 362, 41

\bibitem[\protect\citeauthoryear{{Gallazzi}, {Brinchmann}, {Charlot}  \&
  {White}}{{Gallazzi} et~al.}{2008}]{Gallazzi2008}
{Gallazzi} A.,  {Brinchmann} J.,  {Charlot} S.,   {White} S.~D.~M.,  2008,
  \mn@doi [MNRAS] {10.1111/j.1365-2966.2007.12632.x}, 383, 1439

\bibitem[\protect\citeauthoryear{{Giavalisco} et~al.,}{{Giavalisco}
  et~al.}{2004}]{Giavalisco2004}
{Giavalisco} M.,  et~al., 2004, \mn@doi [ApJ] {10.1086/379232}, 600, L93

\bibitem[\protect\citeauthoryear{{Girardi}, {Bressan}, {Chiosi}, {Bertelli}  \&
  {Nasi}}{{Girardi} et~al.}{1996}]{Girardi1996}
{Girardi} L.,  {Bressan} A.,  {Chiosi} C.,  {Bertelli} G.,   {Nasi} E.,  1996,
  A\&AS, 117, 113

\bibitem[\protect\citeauthoryear{{Gonz{\'a}lez}~Delgado
  et~al.,}{{Gonz{\'a}lez}~Delgado et~al.}{2014}]{Gonzalez2014}
{Gonz{\'a}lez}~Delgado R.~M.,  et~al., 2014, \mn@doi [ApJ]
  {10.1088/2041-8205/791/1/L16}, 791, L16

\bibitem[\protect\citeauthoryear{{Gonzalez}, {Faber}  \& {Worthey}}{{Gonzalez}
  et~al.}{1993}]{Gonzalez1993}
{Gonzalez} J.~J.,  {Faber} S.~M.,   {Worthey} G.,  1993, in American
  Astronomical Society Meeting Abstracts. p.~1355

\bibitem[\protect\citeauthoryear{{Greggio}}{{Greggio}}{1997}]{Greggio1997}
{Greggio} L.,  1997, \mn@doi [MNRAS] {10.1093/mnras/285.1.151}, 285, 151

\bibitem[\protect\citeauthoryear{{Guo} et~al.,}{{Guo} et~al.}{2016}]{Guo2016}
{Guo} Q.,  et~al., 2016, \mn@doi [MNRAS] {10.1093/mnras/stw1525}, 461, 3457

\bibitem[\protect\citeauthoryear{{Hansson}, {Lisker}  \& {Grebel}}{{Hansson}
  et~al.}{2012}]{Hansson2012}
{Hansson} K.~S.~A.,  {Lisker} T.,   {Grebel} E.~K.,  2012, \mn@doi [MNRAS]
  {10.1111/j.1365-2966.2012.21659.x}, 427, 2376

\bibitem[\protect\citeauthoryear{{Hayward} \& {Smith}}{{Hayward} \&
  {Smith}}{2015}]{Hayward2015}
{Hayward} C.~C.,  {Smith} D.~J.~B.,  2015, \mn@doi [MNRAS]
  {10.1093/mnras/stu2195}, 446, 1512

\bibitem[\protect\citeauthoryear{{Howell}}{{Howell}}{2006}]{Howell2006}
{Howell} S.~B.,  2006, {Handbook of CCD Astronomy}.
Cambridge University Press

\bibitem[\protect\citeauthoryear{{Kauffmann}}{{Kauffmann}}{1996}]{Kauffmann1996}
{Kauffmann} G.,  1996, \mn@doi [MNRAS] {10.1093/mnras/281.2.487}, 281, 487

\bibitem[\protect\citeauthoryear{{Kauffmann} et~al.,}{{Kauffmann}
  et~al.}{2003}]{Kauffmann2003}
{Kauffmann} G.,  et~al., 2003, \mn@doi [MNRAS]
  {10.1046/j.1365-8711.2003.06291.x}, 341, 33

\bibitem[\protect\citeauthoryear{{Kobayashi}, {Inoue}  \& {Inoue}}{{Kobayashi}
  et~al.}{2013}]{Kobayashi2013}
{Kobayashi} M.~A.~R.,  {Inoue} Y.,   {Inoue} A.~K.,  2013, \mn@doi [ApJ]
  {10.1088/0004-637X/763/1/3}, 763, 3

\bibitem[\protect\citeauthoryear{{Koekemoer} et~al.,}{{Koekemoer}
  et~al.}{2011}]{Koekemoer2011}
{Koekemoer} A.~M.,  et~al., 2011, \mn@doi [ApJ] {10.1088/0067-0049/197/2/36},
  197, 36

\bibitem[\protect\citeauthoryear{{Kriek} \& {Conroy}}{{Kriek} \&
  {Conroy}}{2013}]{Kriek2013}
{Kriek} M.,  {Conroy} C.,  2013, \mn@doi [ApJ] {10.1088/2041-8205/775/1/L16},
  775, L16

\bibitem[\protect\citeauthoryear{{Kriek}, {van Dokkum}, {Labb{\'e}}, {Franx},
  {Illingworth}, {Marchesini}  \& {Quadri}}{{Kriek} et~al.}{2009}]{Kriek2009}
{Kriek} M.,  {van Dokkum} P.~G.,  {Labb{\'e}} I.,  {Franx} M.,  {Illingworth}
  G.~D.,  {Marchesini} D.,   {Quadri} R.~F.,  2009, \mn@doi [ApJ]
  {10.1088/0004-637X/700/1/221}, 700, 221

\bibitem[\protect\citeauthoryear{{Kriek} et~al.,}{{Kriek}
  et~al.}{2010}]{Kriek2010}
{Kriek} M.,  et~al., 2010, \mn@doi [ApJ] {10.1088/2041-8205/722/1/L64}, 722,
  L64

\bibitem[\protect\citeauthoryear{{Kroupa}}{{Kroupa}}{2001}]{Kroupa2001}
{Kroupa} P.,  2001, \mn@doi [MNRAS] {10.1046/j.1365-8711.2001.04022.x}, 322,
  231

\bibitem[\protect\citeauthoryear{{Lanz} \& {Hubeny}}{{Lanz} \&
  {Hubeny}}{2003a}]{Lanz2003}
{Lanz} T.,  {Hubeny} I.,  2003a, \mn@doi [ApJS] {10.1086/374373}, 146, 417

\bibitem[\protect\citeauthoryear{{Lanz} \& {Hubeny}}{{Lanz} \&
  {Hubeny}}{2003b}]{Lanzerr2003}
{Lanz} T.,  {Hubeny} I.,  2003b, \mn@doi [ApJS] {10.1086/377130}, 147, 225

\bibitem[\protect\citeauthoryear{{Lanz} \& {Hubeny}}{{Lanz} \&
  {Hubeny}}{2007}]{Lanz2007}
{Lanz} T.,  {Hubeny} I.,  2007, \mn@doi [ApJS] {10.1086/511270}, 169, 83

\bibitem[\protect\citeauthoryear{{Lara-L{\'o}pez} et~al.,}{{Lara-L{\'o}pez}
  et~al.}{2010}]{Lara-Lopez2010}
{Lara-L{\'o}pez} M.~A.,  et~al., 2010, \mn@doi [A\&A]
  {10.1051/0004-6361/201014803}, 521, L53

\bibitem[\protect\citeauthoryear{{Le Borgne} et~al.,}{{Le Borgne}
  et~al.}{2003}]{LeBorgne2003}
{Le Borgne} J.-F.,  et~al., 2003, \mn@doi [A\&A] {10.1051/0004-6361:20030243},
  402, 433

\bibitem[\protect\citeauthoryear{{Lee}, {Idzi}, {Ferguson}, {Somerville},
  {Wiklind}  \& {Giavalisco}}{{Lee} et~al.}{2009}]{Lee2009}
{Lee} S.-K.,  {Idzi} R.,  {Ferguson} H.~C.,  {Somerville} R.~S.,  {Wiklind} T.,
    {Giavalisco} M.,  2009, \mn@doi [ApJS] {10.1088/0067-0049/184/1/100}, 184,
  100

\bibitem[\protect\citeauthoryear{{Lee}, {Ferguson}, {Somerville}, {Wiklind}  \&
  {Giavalisco}}{{Lee} et~al.}{2010}]{Lee2010}
{Lee} S.-K.,  {Ferguson} H.~C.,  {Somerville} R.~S.,  {Wiklind} T.,
  {Giavalisco} M.,  2010, \mn@doi [ApJ] {10.1088/0004-637X/725/2/1644}, 725,
  1644

\bibitem[\protect\citeauthoryear{{Lee} et~al.,}{{Lee} et~al.}{2011}]{Lee2011}
{Lee} K.-S.,  et~al., 2011, \mn@doi [ApJ] {10.1088/0004-637X/733/2/99}, 733, 99

\bibitem[\protect\citeauthoryear{{MacArthur}, {Gonz{\'a}lez}  \&
  {Courteau}}{{MacArthur} et~al.}{2009}]{MacArthur2009}
{MacArthur} L.~A.,  {Gonz{\'a}lez} J.~J.,   {Courteau} S.,  2009, \mn@doi
  [MNRAS] {10.1111/j.1365-2966.2009.14519.x}, 395, 28

\bibitem[\protect\citeauthoryear{{MacArthur}, {McDonald}, {Courteau}  \&
  {Jes{\'u}s Gonz{\'a}lez}}{{MacArthur} et~al.}{2010}]{MacArthur2010}
{MacArthur} L.~A.,  {McDonald} M.,  {Courteau} S.,   {Jes{\'u}s Gonz{\'a}lez}
  J.,  2010, \mn@doi [ApJ] {10.1088/0004-637X/718/2/768}, 718, 768

\bibitem[\protect\citeauthoryear{{Madau} \& {Dickinson}}{{Madau} \&
  {Dickinson}}{2014}]{Madau2014}
{Madau} P.,  {Dickinson} M.,  2014, \mn@doi [ARA\&A]
  {10.1146/annurev-astro-081811-125615}, 52, 415

\bibitem[\protect\citeauthoryear{{Magris}, {Mateu P.}, {Mateu}, {Bruzual A.},
  {Cabrera-Ziri}  \& {Mej{\'{\i}}a-Narv{\'a}ez}}{{Magris}
  et~al.}{2015}]{Magris2015}
{Magris} G.,  {Mateu P.} J.,  {Mateu} C.,  {Bruzual A.} G.,  {Cabrera-Ziri} I.,
    {Mej{\'{\i}}a-Narv{\'a}ez} A.,  2015, \mn@doi [PASP] {10.1086/679742}, 127,
  16

\bibitem[\protect\citeauthoryear{{Maraston}}{{Maraston}}{2005}]{Maraston2005}
{Maraston} C.,  2005, \mn@doi [MNRAS] {10.1111/j.1365-2966.2005.09270.x}, 362,
  799

\bibitem[\protect\citeauthoryear{{Maraston}, {Greggio}, {Renzini}, {Ortolani},
  {Saglia}, {Puzia}  \& {Kissler-Patig}}{{Maraston}
  et~al.}{2003}]{Maraston2003}
{Maraston} C.,  {Greggio} L.,  {Renzini} A.,  {Ortolani} S.,  {Saglia} R.~P.,
  {Puzia} T.~H.,   {Kissler-Patig} M.,  2003, \mn@doi [A\&A]
  {10.1051/0004-6361:20021723}, 400, 823

\bibitem[\protect\citeauthoryear{{Maraston}, {Pforr}, {Renzini}, {Daddi},
  {Dickinson}, {Cimatti}  \& {Tonini}}{{Maraston} et~al.}{2010}]{Maraston2010}
{Maraston} C.,  {Pforr} J.,  {Renzini} A.,  {Daddi} E.,  {Dickinson} M.,
  {Cimatti} A.,   {Tonini} C.,  2010, MNRAS, 407, 830

\bibitem[\protect\citeauthoryear{{Marin-Franch}, {Taylor}, {Cenarro},
  {Cristobal-Hornillos}  \& {Moles}}{{Marin-Franch}
  et~al.}{2015}]{Marin-Franch2015}
{Marin-Franch} A.,  {Taylor} K.,  {Cenarro} J.,  {Cristobal-Hornillos} D.,
  {Moles} M.,  2015, IAU General Assembly, 29, 2257439

\bibitem[\protect\citeauthoryear{{Martins}, {Gonz{\'a}lez Delgado},
  {Leitherer}, {Cervi{\~n}o}  \& {Hauschildt}}{{Martins}
  et~al.}{2005}]{Martins2005}
{Martins} L.~P.,  {Gonz{\'a}lez Delgado} R.~M.,  {Leitherer} C.,  {Cervi{\~n}o}
  M.,   {Hauschildt} P.,  2005, \mn@doi [MNRAS]
  {10.1111/j.1365-2966.2005.08703.x}, 358, 49

\bibitem[\protect\citeauthoryear{{Mitchell}, {Lacey}, {Baugh}  \&
  {Cole}}{{Mitchell} et~al.}{2013}]{Mitchell2013}
{Mitchell} P.~D.,  {Lacey} C.~G.,  {Baugh} C.~M.,   {Cole} S.,  2013, \mn@doi
  [MNRAS] {10.1093/mnras/stt1280}, 435, 87

\bibitem[\protect\citeauthoryear{{Moles} et~al.,}{{Moles}
  et~al.}{2008}]{Moles2008}
{Moles} M.,  et~al., 2008, \mn@doi [AJ] {10.1088/0004-6256/136/3/1325}, 136,
  1325

\bibitem[\protect\citeauthoryear{{Morgan}}{{Morgan}}{1956}]{Morgan1956}
{Morgan} W.~W.,  1956, \mn@doi [PASP] {10.1086/126988}, 68, 509

\bibitem[\protect\citeauthoryear{{Naab} \& {Ostriker}}{{Naab} \&
  {Ostriker}}{2017}]{Naab2017}
{Naab} T.,  {Ostriker} J.~P.,  2017, \mn@doi [ARA\&A]
  {10.1146/annurev-astro-081913-040019}, 55

\bibitem[\protect\citeauthoryear{{Ocvirk}, {Pichon}, {Lan{\c c}on}  \&
  {Thi{\'e}baut}}{{Ocvirk} et~al.}{2006}]{Ocvirk2006a}
{Ocvirk} P.,  {Pichon} C.,  {Lan{\c c}on} A.,   {Thi{\'e}baut} E.,  2006,
  \mn@doi [MNRAS] {10.1111/j.1365-2966.2005.09182.x}, 365, 46

\bibitem[\protect\citeauthoryear{{Oyaizu}, {Lima}, {Cunha}, {Lin}, {Frieman}
  \& {Sheldon}}{{Oyaizu} et~al.}{2008}]{Oyaizu2008}
{Oyaizu} H.,  {Lima} M.,  {Cunha} C.~E.,  {Lin} H.,  {Frieman} J.,   {Sheldon}
  E.~S.,  2008, \mn@doi [ApJ] {10.1086/523666}, 674, 768

\bibitem[\protect\citeauthoryear{{Pacifici}, {Charlot}, {Blaizot}  \&
  {Brinchmann}}{{Pacifici} et~al.}{2012}]{Pacifici2012}
{Pacifici} C.,  {Charlot} S.,  {Blaizot} J.,   {Brinchmann} J.,  2012, \mn@doi
  [MNRAS] {10.1111/j.1365-2966.2012.20431.x}, 421, 2002

\bibitem[\protect\citeauthoryear{{Panter}, {Jimenez}, {Heavens}  \&
  {Charlot}}{{Panter} et~al.}{2008}]{Panter2008}
{Panter} B.,  {Jimenez} R.,  {Heavens} A.~F.,   {Charlot} S.,  2008, \mn@doi
  [MNRAS] {10.1111/j.1365-2966.2008.13981.x}, 391, 1117

\bibitem[\protect\citeauthoryear{{Pforr}, {Maraston}  \& {Tonini}}{{Pforr}
  et~al.}{2012}]{Pforr2012}
{Pforr} J.,  {Maraston} C.,   {Tonini} C.,  2012, \mn@doi [MNRAS]
  {10.1111/j.1365-2966.2012.20848.x}, 422, 3285

\bibitem[\protect\citeauthoryear{{Pforr}, {Maraston}  \& {Tonini}}{{Pforr}
  et~al.}{2013}]{Pforr2013}
{Pforr} J.,  {Maraston} C.,   {Tonini} C.,  2013, \mn@doi [MNRAS]
  {10.1093/mnras/stt1382}, 435, 1389

\bibitem[\protect\citeauthoryear{{Prugniel}, {Vauglin}  \& {Koleva}}{{Prugniel}
  et~al.}{2011}]{Prugniel2011}
{Prugniel} P.,  {Vauglin} I.,   {Koleva} M.,  2011, \mn@doi [A\&A]
  {10.1051/0004-6361/201116769}, 531, A165

\bibitem[\protect\citeauthoryear{{Rauch}}{{Rauch}}{2003}]{Rauch2003}
{Rauch} T.,  2003, \mn@doi [A\&A] {10.1051/0004-6361:20030412}, 403, 709

\bibitem[\protect\citeauthoryear{{Renzini}}{{Renzini}}{2006}]{Renzini2006}
{Renzini} A.,  2006, \mn@doi [ARA\&A] {10.1146/annurev.astro.44.051905.092450},
  44, 141

\bibitem[\protect\citeauthoryear{{Rodr{\'{\i}}guez-Merino}, {Chavez}, {Bertone}
   \& {Buzzoni}}{{Rodr{\'{\i}}guez-Merino} et~al.}{2005}]{Rodmer2005}
{Rodr{\'{\i}}guez-Merino} L.~H.,  {Chavez} M.,  {Bertone} E.,   {Buzzoni} A.,
  2005, \mn@doi [ApJ] {10.1086/429858}, 626, 411

\bibitem[\protect\citeauthoryear{{Rosales-Ortega}, {S{\'a}nchez},
  {Iglesias-P{\'a}ramo}, {D{\'{\i}}az}, {V{\'{\i}}lchez}, {Bland-Hawthorn},
  {Husemann}  \& {Mast}}{{Rosales-Ortega} et~al.}{2012}]{Rosales-Ortega2012}
{Rosales-Ortega} F.~F.,  {S{\'a}nchez} S.~F.,  {Iglesias-P{\'a}ramo} J.,
  {D{\'{\i}}az} A.~I.,  {V{\'{\i}}lchez} J.~M.,  {Bland-Hawthorn} J.,
  {Husemann} B.,   {Mast} D.,  2012, \mn@doi [ApJ]
  {10.1088/2041-8205/756/2/L31}, 756, L31

\bibitem[\protect\citeauthoryear{{Salim}, {Lee}, {Ly}, {Brinchmann},
  {Dav{\'e}}, {Dickinson}, {Salzer}  \& {Charlot}}{{Salim}
  et~al.}{2014}]{Salim2014}
{Salim} S.,  {Lee} J.~C.,  {Ly} C.,  {Brinchmann} J.,  {Dav{\'e}} R.,
  {Dickinson} M.,  {Salzer} J.~J.,   {Charlot} S.,  2014, \mn@doi [ApJ]
  {10.1088/0004-637X/797/2/126}, 797, 126

\bibitem[\protect\citeauthoryear{{Salpeter}}{{Salpeter}}{1955}]{Salpeter1955}
{Salpeter} E.~E.,  1955, \mn@doi [ApJ] {10.1086/145971}, 121, 161

\bibitem[\protect\citeauthoryear{{S{\'a}nchez-Bl{\'a}zquez}
  et~al.,}{{S{\'a}nchez-Bl{\'a}zquez} et~al.}{2006}]{SanchezBlazquez2006}
{S{\'a}nchez-Bl{\'a}zquez} P.,  et~al., 2006, \mn@doi [MNRAS]
  {10.1111/j.1365-2966.2006.10699.x}, 371, 703

\bibitem[\protect\citeauthoryear{{S{\'a}nchez} et~al.,}{{S{\'a}nchez}
  et~al.}{2012}]{Sanchez2012}
{S{\'a}nchez} S.~F.,  et~al., 2012, \mn@doi [A\&A]
  {10.1051/0004-6361/201117353}, 538, A8

\bibitem[\protect\citeauthoryear{{S{\'a}nchez} et~al.,}{{S{\'a}nchez}
  et~al.}{2013}]{Sanchez2013}
{S{\'a}nchez} S.~F.,  et~al., 2013, \mn@doi [A\&A]
  {10.1051/0004-6361/201220669}, 554, A58

\bibitem[\protect\citeauthoryear{{Smith} \& {Hayward}}{{Smith} \&
  {Hayward}}{2015}]{Smith2015}
{Smith} D.~J.~B.,  {Hayward} C.~C.,  2015, \mn@doi [MNRAS]
  {10.1093/mnras/stv1727}, 453, 1597

\bibitem[\protect\citeauthoryear{{Sorba} \& {Sawicki}}{{Sorba} \&
  {Sawicki}}{2015}]{Sorba2015}
{Sorba} R.,  {Sawicki} M.,  2015, \mn@doi [MNRAS] {10.1093/mnras/stv1235}, 452,
  235

\bibitem[\protect\citeauthoryear{{Strateva} et~al.,}{{Strateva}
  et~al.}{2001}]{Strateva2001}
{Strateva} I.,  et~al., 2001, \mn@doi [AJ] {10.1086/323301}, 122, 1861

\bibitem[\protect\citeauthoryear{{Taylor}, {Franx}, {Brinchmann}, {van der Wel}
   \& {van Dokkum}}{{Taylor} et~al.}{2010}]{Taylor2010}
{Taylor} E.~N.,  {Franx} M.,  {Brinchmann} J.,  {van der Wel} A.,   {van
  Dokkum} P.~G.,  2010, \mn@doi [ApJ] {10.1088/0004-637X/722/1/1}, 722, 1

\bibitem[\protect\citeauthoryear{{Tinsley}}{{Tinsley}}{1972}]{Tinsley1972}
{Tinsley} B.~M.,  1972, A\&A, 20, 383

\bibitem[\protect\citeauthoryear{{Tinsley}}{{Tinsley}}{1980}]{Tinsley1980}
{Tinsley} B.~M.,  1980, Fund. Cosmic Phys., 5, 287

\bibitem[\protect\citeauthoryear{{Tojeiro}, {Heavens}, {Jimenez}  \&
  {Panter}}{{Tojeiro} et~al.}{2007}]{Tojeiro2007}
{Tojeiro} R.,  {Heavens} A.~F.,  {Jimenez} R.,   {Panter} B.,  2007, \mn@doi
  [MNRAS] {10.1111/j.1365-2966.2007.12323.x}, 381, 1252

\bibitem[\protect\citeauthoryear{{Tremonti} et~al.,}{{Tremonti}
  et~al.}{2004}]{Tremonti2004}
{Tremonti} C.~A.,  et~al., 2004, \mn@doi [ApJ] {10.1086/423264}, 613, 898

\bibitem[\protect\citeauthoryear{{Vogelsberger} et~al.,}{{Vogelsberger}
  et~al.}{2014}]{Vogelsberger2014}
{Vogelsberger} M.,  et~al., 2014, \mn@doi [MNRAS] {10.1093/mnras/stu1536}, 444,
  1518

\bibitem[\protect\citeauthoryear{{Walcher}, {Groves}, {Budav{\'a}ri}  \&
  {Dale}}{{Walcher} et~al.}{2011}]{Walcher2011}
{Walcher} J.,  {Groves} B.,  {Budav{\'a}ri} T.,   {Dale} D.,  2011, \mn@doi
  [Ap\&SS] {10.1007/s10509-010-0458-z}, 331, 1

\bibitem[\protect\citeauthoryear{{Westera}, {Lejeune}, {Buser}, {Cuisinier}  \&
  {Bruzual}}{{Westera} et~al.}{2002}]{Westera2002}
{Westera} P.,  {Lejeune} T.,  {Buser} R.,  {Cuisinier} F.,   {Bruzual} G.,
  2002, \mn@doi [A\&A] {10.1051/0004-6361:20011493}, 381, 524

\bibitem[\protect\citeauthoryear{{White} \& {Frenk}}{{White} \&
  {Frenk}}{1991}]{White1991}
{White} S.~D.~M.,  {Frenk} C.~S.,  1991, \mn@doi [ApJ] {10.1086/170483}, 379,
  52

\bibitem[\protect\citeauthoryear{{Wolf}, {Meisenheimer}, {Rix}, {Borch}, {Dye}
  \& {Kleinheinrich}}{{Wolf} et~al.}{2003}]{Wolf2003}
{Wolf} C.,  {Meisenheimer} K.,  {Rix} H.-W.,  {Borch} A.,  {Dye} S.,
  {Kleinheinrich} M.,  2003, \mn@doi [A\&A] {10.1051/0004-6361:20021513}, 401,
  73

\bibitem[\protect\citeauthoryear{{Wood}}{{Wood}}{1966}]{Wood1966}
{Wood} D.~B.,  1966, \mn@doi [ApJ] {10.1086/148737}, 145, 36

\bibitem[\protect\citeauthoryear{{Worthey}}{{Worthey}}{1994}]{Worthey1994}
{Worthey} G.,  1994, \mn@doi [ApJS] {10.1086/192096}, 95, 107

\bibitem[\protect\citeauthoryear{{Wuyts}, {Franx}, {Cox}, {Hernquist},
  {Hopkins}, {Robertson}  \& {van Dokkum}}{{Wuyts} et~al.}{2009}]{Wuyts2009}
{Wuyts} S.,  {Franx} M.,  {Cox} T.~J.,  {Hernquist} L.,  {Hopkins} P.~F.,
  {Robertson} B.~E.,   {van Dokkum} P.~G.,  2009, \mn@doi [ApJ]
  {10.1088/0004-637X/696/1/348}, 696, 348

\bibitem[\protect\citeauthoryear{{York} et~al.,}{{York}
  et~al.}{2000}]{York2000}
{York} D.~G.,  et~al., 2000, \mn@doi [AJ] {10.1086/301513}, 120, 1579

\bibitem[\protect\citeauthoryear{{Zibetti}, {Gallazzi}, {Charlot}, {Pierini}
  \& {Pasquali}}{{Zibetti} et~al.}{2013}]{Zibetti2013}
{Zibetti} S.,  {Gallazzi} A.,  {Charlot} S.,  {Pierini} D.,   {Pasquali} A.,
  2013, \mn@doi [MNRAS] {10.1093/mnras/sts126}, 428, 1479

\makeatother
\end{thebibliography}

\appendix
\section{Building the mock sample}\label{sec:mockpar}

\begin{table*}
\caption{Star formation history and stellar population parameters of a fraction of the mock galaxy sample.
}\label{tab:mock-parameters}
\begin{tabular}{rrrrcccrrccrrrcc}
  \hline
\mlc{1}{c}{(1)} & \mlc{1}{c}{(2)} & \mlc{1}{c}{(3)} & \mlc{1}{c}{(4)} & \mlc{1}{c}{(5)} & \mlc{1}{c}{(6)} & \mlc{1}{c}{(7)} & \mlc{1}{c}{(8)} & \mlc{1}{c}{(9)} & \mlc{1}{c}{(10)} & \mlc{1}{c}{(11)} & \mlc{1}{c}{(12)\ } & \mlc{1}{c}{(13)\ } & \mlc{1}{c}{(14)\ } & \mlc{1}{c}{(15)\ } & \mlc{1}{c}{(16)\ } \\
\mlc{1}{c}{$t_\text{form}$} & \mlc{1}{c}{$1/\tau$} & \mlc{1}{c}{$t_\text{burst}$} & \mlc{1}{c}{$t_\text{ext}$} & \mlr{2}{*}{$A$} & \mlc{1}{c}{$t_\text{trunc}$} & \mlc{1}{c}{$\tau_\text{trunc}$} & \mlc{1}{c}{$Z$} & \mlc{1}{c}{$\sigma_v$} & \mlr{2}{*}{$\tau_V$} & \mlr{2}{*}{$\mu$} & \mlc{1}{c}{$\log M_{*}$} & \mlc{1}{c}{$\left<\log{t_*}\right>_M$} & \mlc{1}{c}{$\left<\log{t_*}\right>_L$} & \mlc{1}{c}{$\extv$} & \mlc{1}{c}{$u-r$} \\
\mlc{1}{c}{(Gyr)} & \mlc{1}{c}{($\text{Gyr}^{-1}$)} & \mlc{1}{c}{(Gyr)} & \mlc{1}{c}{(Myr)} & & \mlc{1}{c}{(Gyr)} & \mlc{1}{c}{(Myr)} & \mlc{1}{c}{(Z$_\odot$)} & \mlc{1}{c}{(km s$^{-1}$)} & & & \mlc{1}{c}{(M$_\odot$)} & \mlc{1}{c}{(yr)} & \mlc{1}{c}{(yr)} & \mlc{1}{c}{(mag)} & \mlc{1}{c}{(ABmag)} \\
\hline
    1.90 &   0.033 &    1.04 &  202.56 &    0.32 &     ... &     ... &    1.78 &  141.11 &   0.540 &   0.176 & 	  -1.100 &    8.458 &   8.880 &  0.145 &   0.971 \\
    1.61 &   0.657 &    1.14 &  286.12 &    0.18 &    0.81 &  233.31 &    1.33 &  392.79 &   0.743 &   0.324 & 	  -0.063 &    9.331 &   9.846 &  0.616 &   1.721 \\
    6.00 &   0.651 &    4.81 &  255.00 &    0.04 &     ... &     ... &    1.55 &  371.21 &   0.217 &   0.406 & 	   0.147 &    9.178 &   9.424 &  0.294 &   1.879 \\
    2.50 &   0.090 &    2.18 &  202.56 &    0.66 &    1.28 &  314.41 &    0.27 &  251.19 &   1.247 &   0.432 & 	   0.581 &    8.908 &   9.167 &  0.708 &   1.905 \\
    4.50 &   0.581 &    2.63 &  255.00 &    0.98 &     ... &     ... &    1.40 &  101.57 &   2.443 &   0.115 & 	   0.222 &    9.269 &   9.405 &  0.016 &   1.970 \\
    2.60 &   0.234 &    2.60 &   64.05 &    0.14 &    0.81 &  232.86 &    1.35 &  142.10 &   1.193 &   0.469 & 	   0.008 &    9.412 &   9.633 &  0.098 &   1.979 \\
    6.00 &   0.522 &    2.58 &  202.56 &    1.62 &    2.00 &  287.74 &    2.08 &  141.42 &   0.633 &   0.313 & 	   0.027 &    9.888 &   9.918 &  0.352 &   2.434 \\
    9.50 &   0.616 &    4.06 &  127.80 &    1.37 &     ... &     ... &    1.35 &  287.55 &   0.568 &   0.342 & 	   0.300 &    9.510 &   9.586 &  0.157 &   2.455 \\
    8.25 &   0.378 &    7.73 &  101.52 &    0.14 &    1.02 &  715.37 &    1.37 &   54.11 &   1.989 &   0.492 & 	   0.034 &    9.568 &   9.762 &  1.068 &   2.580 \\
   12.25 &   0.990 &    4.00 &  255.00 &    0.17 &     ... &     ... &    2.32 &   78.53 &   3.090 &   0.666 & 	   0.068 &   10.016 &  10.034 &  0.852 &   3.414 \\
  \hline
\end{tabular}
\end{table*}

\begin{table*}

\caption{Spectral properties of different stellar atlases relevant in the UV, as
published.}\label{tab:spprop}

\begin{tabular}{lcr@{\,---\,}lr@{}rl}
\hline
\mlc{1}{l|}{Stellar} & \mlc{1}{c|}{Stellar} & \mlc{2}{c|}{Wavelength}  & \mlc{2}{c}{\mlr{2}{*}{$R_{\textsc{STEP}}=\frac{\lambda}{\Delta\lambda}$}} & \mlr{2}{*}{Reference} \\
\mlc{1}{l|}{Library} & \mlc{1}{c|}{Type}    & \mlc{2}{c|}{Range}       &                                                                           &                       \\
\hline
Tlusty        & O stars     &  45\,\AA & 300\,$\mu$m  &   26,000\,---\,& 38,000  & \citet{Lanz2003,Lanzerr2003} \\
Tlusty        & B stars     &  54\,\AA & 300\,$\mu$m  &  100,000\,---\,& 200,000 & \citet{Lanz2007}             \\
Martins et al.& A stars     &     3000 & 7000\,\AA    &   10,000\,---\,& 23,000  & \citet{Martins2005}          \\
UVBlue        & F,G,K stars &      850 & 4700\,\AA    & \mlc{2}{r}{50,000} & \citet{Rodmer2005}           \\
Rauch         & T>55MK      &        5 & 2000\,\AA    &   50\,---\,& 20,000      & \citet{Rauch2003}            \\
\hline
\end{tabular}
\end{table*}

\subsection{Mock galaxy parameters}

The Synthetic Spectral Atlas of Galaxies (SSAG) built by \citet{SSAG2014} is a collection of $10$k
Monte Carlo realisations of the recipe for the star formation rate $\Psi(t)$ introduced by
\citet[][\citetalias{Chen2012} hereafter]{Chen2012}. For details on the probability distribution
function (PDF) and the physical motivation supporting the choice of the 11 parameters entering the
\citetalias{Chen2012} definition of $\Psi(t)$, see \citetalias{Chen2012} and \citet{Kauffmann2003}.
In summary, the \citetalias{Chen2012} SFR's have three main components, namely: an underlying
exponentially declining rate, with initial onset at $t=t_\text{form}$, and $e$-folding time $\tau$
(a $\tau$-model); a burst of constant star formation parametrised by a random amplitude
$A\equiv\mburst/M_\text{cont}$, which blends with the $\tau$-model at a random time $t_\text{burst}$
between $t_\text{form}$ and the present, and lasts for $t_\text{ext}$; and a truncated regime, where
$\Psi(t)$ starts declining faster than before at $t=t_\text{trunc}$, with $e$-folding time
$\tau_\text{trunc}<\tau$. The stellar metallicity $Z$ of the mock galaxy is selected from the range
allowed by the \citetalias{Bruzual2003} models, $0.0001 \le Z \le 0.05$. Following
\citetalias{Chen2012}, we use the two-phase starlight dust extinction model by \citet{Charlot2000},
defined by the optical depth $\tau_V$ in the $V$-band when the stellar population is still in the
birth cloud ($\unit[t<10^7]{yr}$), and the fraction $\mu$ of $\tau_V$ that characterises the
interstellar medium after the birth cloud is dissipated ($\unit[t>10^7]{yr}$). To account for
stellar kinematics effects, the resulting SED is broadened using a Gaussian kernel with velocity
dispersion $\sigma_v$. The PDF of the \citetalias{Chen2012} SFH parameters is built assuming they
are uncorrelated. We recognise this as a weakness of the  \citetalias{Chen2012} recipe, since, for
instance, the random selection of the values of $Z,\ \sigma_v,\ \tau_V,\ \mu$, and  $t_\text{form}$
most likely will not reproduce the known correlations among these parameters. The values of the 11
\citetalias{Chen2012} parameters, $t_\text{form}$, $\tau$, $t_\text{burst}$, $t_\text{ext}$, $A$,
$t_\text{trunc}$, $\tau_\text{trunc}$, $Z$, $\sigma_v$, $\tau_V$, and $\mu$ defining the SFH for a
fraction of the galaxies in our mock sample are listed in Table~\ref{tab:mock-parameters}, columns
(1) to (11).

Once $\Psi(t)$ has been specified, the SED $F_\lambda(t)$ of the composite population in the mock
galaxy is computed as in \cite[][\citetalias{Bruzual2003} hereafter]{Bruzual2003}\footnote{The
software to perform this convolution for the \citetalias{Chen2012} $\Psi(t)$ is available at
\url{http://www.bruzual.org/src.tgz}},
\begin{equation}\label{eq:csp}
F_\lambda(t) = \int_0^t \Psi(t-t')\,f_\lambda(t',Z)\,dt',
\end{equation}
where $f_\lambda(t,Z)$ is the SED at age $t$ of a simple stellar population (SSP) for the chosen
initial mass function (IMF) and metallicity $Z$. For $f_\lambda(t,Z)$ we use the \textsc{xmiless}
version (BC03xm hereafter) of the \citetalias{Bruzual2003} models, described below. We note that the
implementation of the SSAG by \citet{SSAG2014} used by \citetalias{Magris2015} is based on the
standard \citetalias{Bruzual2003} models. The following properties of each mock galaxy, $\logm$ (log
stellar mass), $\mwla$ (mass-weighted mean log age), $\lwla$ (luminosity-weighted mean log age), and
$\extv$ (extinction in the $V$-band) are listed in columns (12) to (15) of
Table~\ref{tab:mock-parameters}, in order of increasing present-day $u-r$ colour (column 16). The
values of $\logm$, $\mwla$, $\lwla$, $Z$, and $\extv$ in Table~\ref{tab:mock-parameters} are the
\emph{true values} used to compute the residuals of these properties in this paper
(Eq.~\ref{eq:residual}). We note that $\logm$, $\mwla$, and $Z$ are independent of the SSP models in
use, whereas $\lwla$, $\extv$, and $u-r$ do depend on these models\footnote{The complete version of
Table~\ref{tab:mock-parameters} and the full set of SEDs for the mock galaxy sample can be
downloaded in digital form from \url{http://www.bruzual.org/mn_etal_2017/}}.

\subsection{The BC03 \textsc{xmiless} models}\label{sec:bc03xm}

We have built new evolutionary population synthesis models based on the \citetalias{Bruzual2003}
\emph{Padova 1994} set of stellar evolutionary tracks \citep{Alongi1993, Bressan1993, Fagotto1994a,
Fagotto1994b, Girardi1996}, but using updated libraries of theoretical and empirical stellar
spectra. Table~\ref{tab:spprop} lists the spectral characteristics of the model spectra relevant in
the UV range for stars of different $T_\text{eff}$, as published. In the 4$^\text{th}$ column of
this table, we indicate $R$ as $R_{\textsc{STEP}}$ to emphasise that in this case we use the
\textit{wavelength step} $\Delta\lambda$ to measure $R$, as defined in \citet{Coelho2014}. To
express the UV spectra in a common wavelength scale, we downgrade in resolution the spectra listed
in the first two rows of Table~\ref{tab:xmiless} using a Gaussian smoothing function centred at each
wavelength point $\lambda_i$, sampled according to the step in the 2$^\text{nd}$ column of
Table~\ref{tab:xmiless},
\begin{equation}\label{eqA1}
f(\lambda) = \frac{1}{\sigma_\lambda \sqrt{2\uppi}}\exp{\left[-\frac{(\lambda-\lambda_i)^2}{2\sigma_\lambda^2}\right]},
\end{equation}
where,
\begin{equation}\label{eqA2}
\sigma_\lambda = \frac{\Delta\lambda}{2\sqrt{2\log{2}}} = 0.4247\Delta\lambda.
\end{equation}
In this case $\Delta\lambda$ is the FWHM listed in the 3$^\text{rd}$ column of Table
~\ref{tab:xmiless}. We resample the smoothed spectra with a uniform step $\unit[=0.9]{\AA}$ for
$\lambda$ in the range $5.6\,$---$\,\unit[911]{\AA}$, and $\unit[=0.5]{\AA}$ for $\lambda$ in the
range $911\,$---$\,\unit[3540.5]{\AA}$, as indicated in Table~\ref{tab:xmiless}. The spectral
resolution in the UV range is then $R=\lambda/\Delta\lambda=\lambda/2.5$. The stellar spectra listed
in the last three rows of Table~\ref{tab:xmiless} are used as distributed, with their original
sampling and with no smoothing. The set of spectra described in Table~\ref{tab:xmiless} has been
named \textsc{xmiless}, to indicate the use of an extended version of the Miles+Stelib empirical
libraries. Spectra of the corresponding $[Z/\text{Z}_\odot]$ from all the sources listed in
Table~\ref{tab:xmiless} enter in the construction of the new \citetalias{Bruzual2003} models,
denoted BC03xm hereafter. In this paper we use the BC03xm models computed for the
\citet{Chabrier2003} initial mass function (IMF).\footnote{The \citetalias{Bruzual2003} and BC03xm
models for the \citet{Salpeter1955}, \citet{Kroupa2001}, and \citet{Chabrier2003} IMFs are available
at \url{http://www.bruzual.org/bc03/Updated_version_2016/}}

\begin{table*}

\caption{Wavelength coverage versus spectral atlas adopted in the \citetalias{Bruzual2003}
\textsc{xmiless} models.}\label{tab:xmiless}

\begin{tabular}{r@{\,---\,}lccll}
\hline
\mlc{2}{c}{Wavelength}  & \mlc{1}{c}{Sampling}   & \mlc{1}{c}{FWHM}                  & \mlr{2}{*}{Stellar Library} & \mlr{2}{*}{Reference} \\
\mlc{2}{c}{Range (\AA)} & \mlc{1}{c}{Step (\AA)} & \mlc{1}{c}{$\Delta\lambda$ (\AA)} &                             &                       \\
\hline
5.6    & 911    &   0.9    &   2.0    & Tlusty, Martins et al., UVBlue, Rauch  & Table~\ref{tab:spprop}                \\
911    & 3540.5 &   0.5    &   1.0    & Tlusty, Martins et al., UVBlue, Rauch  & Table~\ref{tab:spprop}                \\
3540.5 & 7351   &   0.9    &   2.5    & Miles                                  & \citet{SanchezBlazquez2006,Falcon2011};\\
\mlc{2}{c}{}    &          &          &                                        & \citet{Prugniel2011}             \\
7351   & 8750   &   1.0    &   3.0    & Stelib                                 & \citet{LeBorgne2003}                   \\
8750   & 36000  & variable & variable & BaSeL 3.1                              & \citet{Westera2002}; Treatment of TP-AGB stars as in \citetalias{Bruzual2003}\\
\hline
\end{tabular}
\end{table*}

\section{A closer view on the TTA}\label{sec:individual-residuals}

\begin{figure*}
  \includegraphics[scale=1.0]{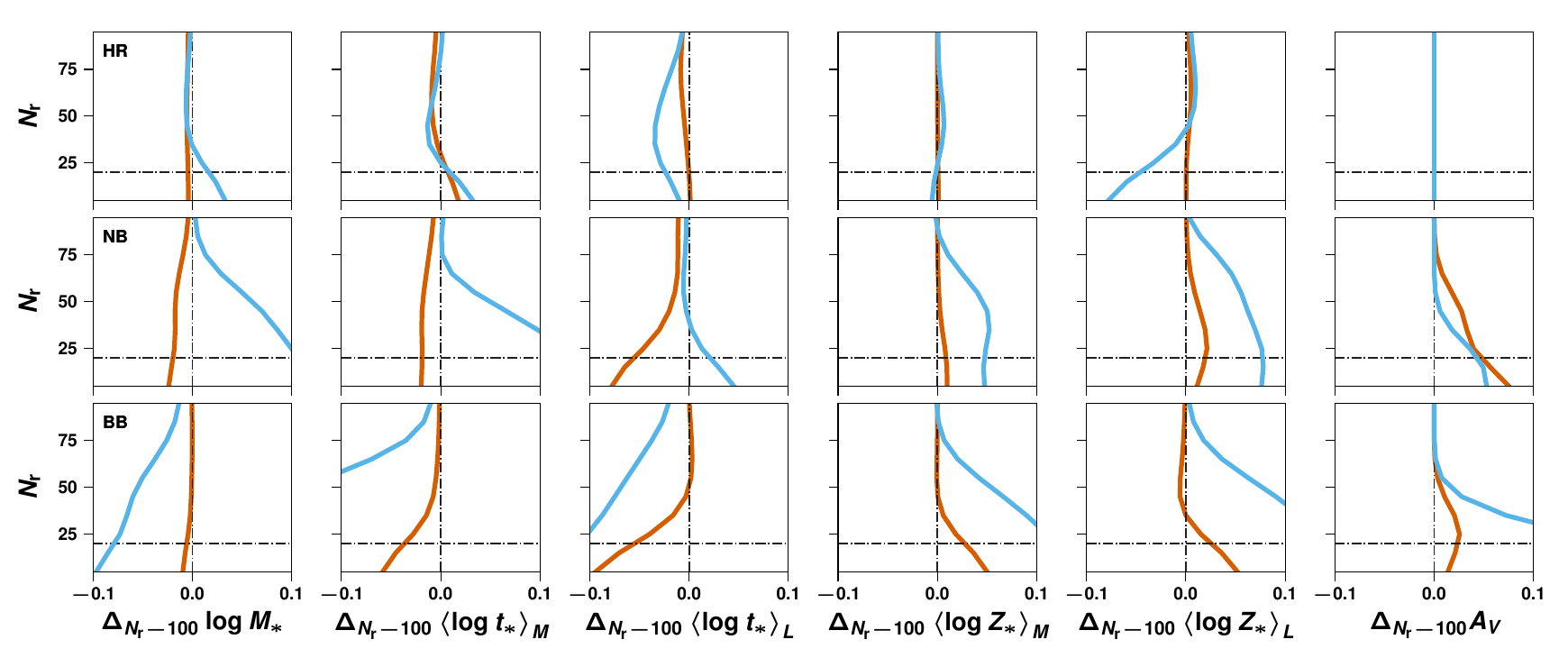}

  \caption{The bias difference between a $100$-realisations and a $N_\text{r}$-realisations
  experiment (labelled $\Delta_{{N_\text{r}}-100}x$) for a SFG (blue) and a PaG (orange) is shown
  for the studied physical properties retrieved from fitting SEDs at the HR, NB and BB spectral
  resolutions. The bias for the PaG converge quickly towards the $100$-realisations bias, however in
  the case of BB results it is clear that $N_\text{r}>20$ noise realisations are needed to reach
  statistical stability. For the SFG the results are even worse in most cases at all spectral
  resolutions. Interestingly, the convergence of dust extinction residuals at the HR resolution is
  very fast regardless of the galaxy type, a result consistent with a robust estimation (cf.
  Fig.~\ref{fig:residual-corrs}). In summary, a conservative figure for the number of realisations
  would be $80$ to reach statistical stability at all spectral resolutions shown. See text for
  details.}

  \label{fig:n-realisations}

\end{figure*}

In \S\ref{sec:main-tta} we explored the residuals behaviour by treating each SED in the mock sample,
which was built upon $134$ SFHs and $N_\text{r}=20$ noise realisations, as a different galaxy. While
several correlations arised resembling what is expected from the presence of the several
degeneracies, it remains to be seen if such correlations are a consequence of the physical
properties of the sample as a whole, i.\,e., different SFHs are located at different regions in the
space of residuals, or if these correlations are intrinsic to the physical properties of the
individual SFHs, i.\,e., the noise realisations of the same SFH scatter across the space of
residuals according to the observed correlations; under the suposition that $20$ noise realisations
yield a representative sample around the maximum likelihood. In Fig.~\ref{fig:degeneracies} we
already showed that the strength of the several degeneracies is independent of the presence of noise
in the data, a result that hints towards an intrinsic origin.

We emphasise that these distributions are not related to the posterior PDF in a Bayesian sense, in
which case those would carry information on the whole parameter space, restricted only by the prior
PDF. In this case, by SED-fitting the $N_\text{r}$ noise realisations we are just sampling the
likelihood around the $\hhi$ of each mock SFH, which statistical dispersion will tell us (at most)
how flat or sharp the \emph{absolute maximum} of the corresponding likelihood is. The best and worst
fitting solutions among the $N_\text{r}$ will have in general very similar $\chi^2\left(\hhi\right)$
values, whereas the properties $\hhi$ may indeed be considerably different (cf.
Fig.~\ref{fig:individual-residuals}). This fact is the main reason why we assess our results in
terms of the recovered physical properties and not in terms of the goodness-of-fit.

\subsection{On the number of noise realisations}

To test the effects of the assumed $N_\text{r}=20$ noise realisations, we select two galaxies
representative of the SFGs and the PaGs. We compute for each spectral resolution (HR, NB and BB)
$N_\text{r}=100$ noise realisations and apply our SED-fitting procedure (as for the mock sample) to
the $600$ spectra. For the resulting physical properties we compute the bias as defined in
\S\ref{sec:derphys} using the $100$ realisations, and then compare this bias to the one obtained
from several values of $N_\text{r}$ in the range $5\,$---$\,95$. We note that $N_\text{r}=100$ is
also an arbitrary choice and as such it may not yield statistical stability to the residual
distributions. Nonetheless, this experiment may allow us to develope some insight into the pace of
convergence of the statistical results. In Fig.~\ref{fig:n-realisations} we show the bias
difference, $\Delta_{{N_\text{r}}-100}x$, for all the physical properties of interest in this paper,
at all spectral resolutions, as a function of the $N_\text{r}$. It is clear that $N_\text{r}=20$ is
not enough in most cases, particularly at the NB and BB resolutions for the PaG. For the SFG, the
results are even worse, specially at the BB resolution. We conclude that, in order to study the mock
sample \emph{on a SFH per SFH basis}, at least $N_\text{r}=80$ to reach statistical stability at the
HR, NB and BB spectral resolutions. Since in \S\ref{sec:main-tta} we explored the residual
distribution statistics on a SED by SED basis, the assumed $N_\text{r}=20$ probably has a negligible
impact.

\subsection{On the residuals of invidual SFHs}

\begin{figure*}
\includegraphics[scale=1.0]{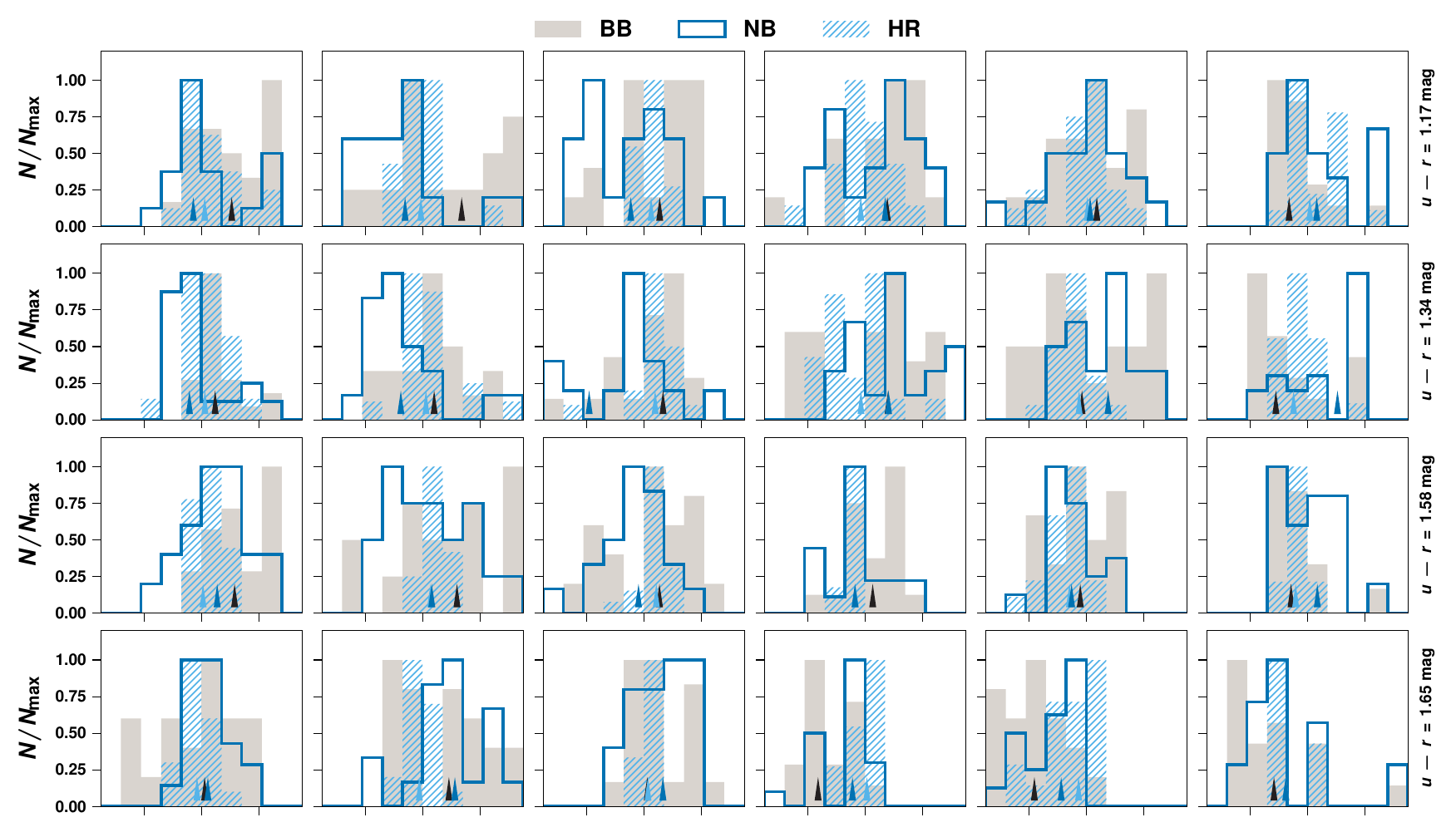}
\includegraphics[scale=1.0]{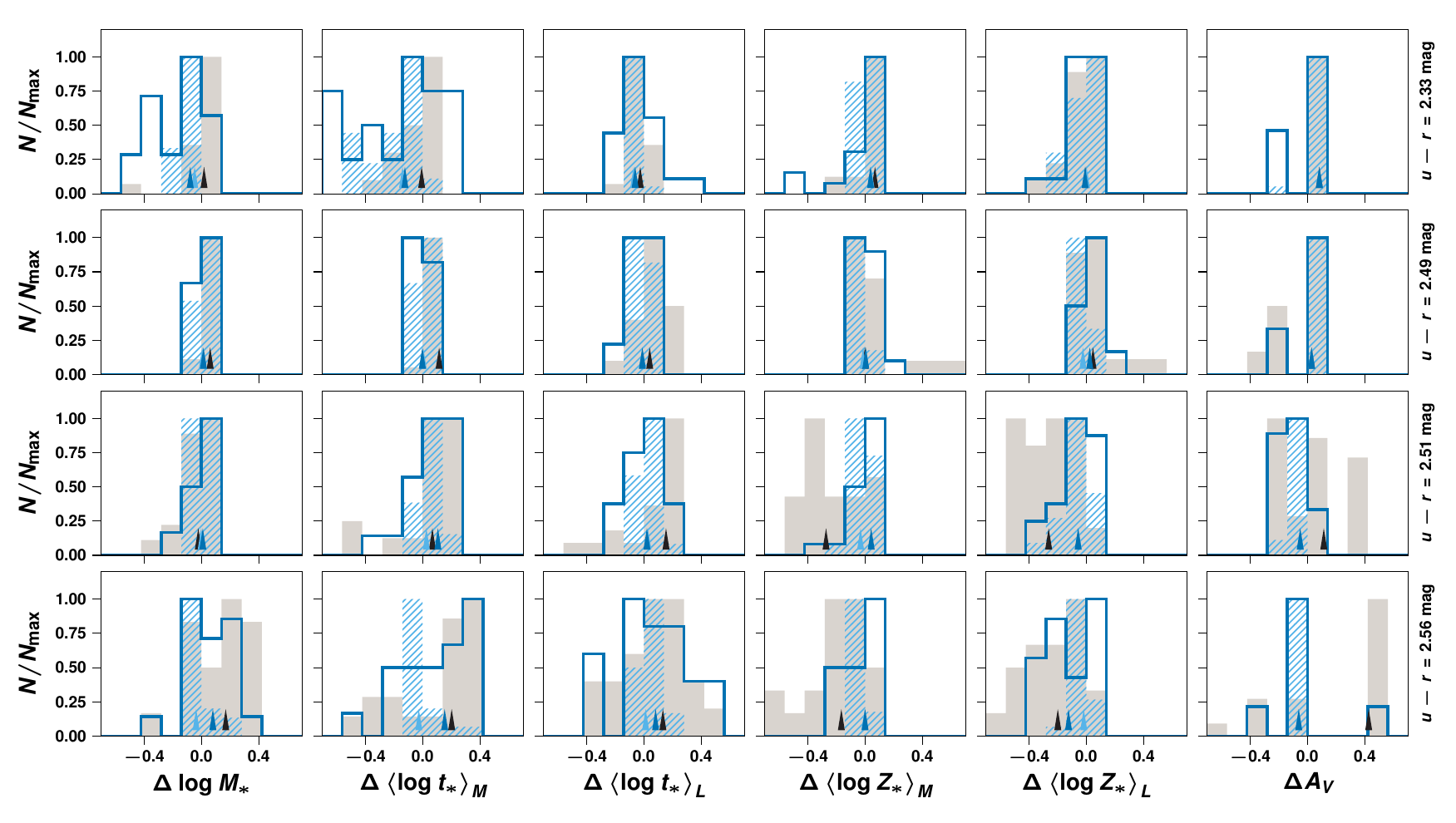}

\caption{\textit{Top four rows:} The residual distributions over the $20$ noise realisations for
four SFGs in the several physical properties we study in this paper. \textit{Bottom four rows:} Same
as top frame, but for PaGs. HR spectroscopy shows high residuals with low statistical dispersion in
all physical properties, with a mild increment of the precision towards PaGs, whereas lower
resolution NB and BB residuals are usually disperse in the observed range, with a remarkable
increment in the precision towards PaGs. See text for details. The median residuals at the three
spectral resolutions represented by the arrows in each plane are usually low in HR determinations,
whereas NB and BB exhibit larger biases for SFGs, and biases comparable to HR for
PaGs.}\label{fig:individual-residuals}

\end{figure*}

\begin{figure*}
\includegraphics[scale=1.0]{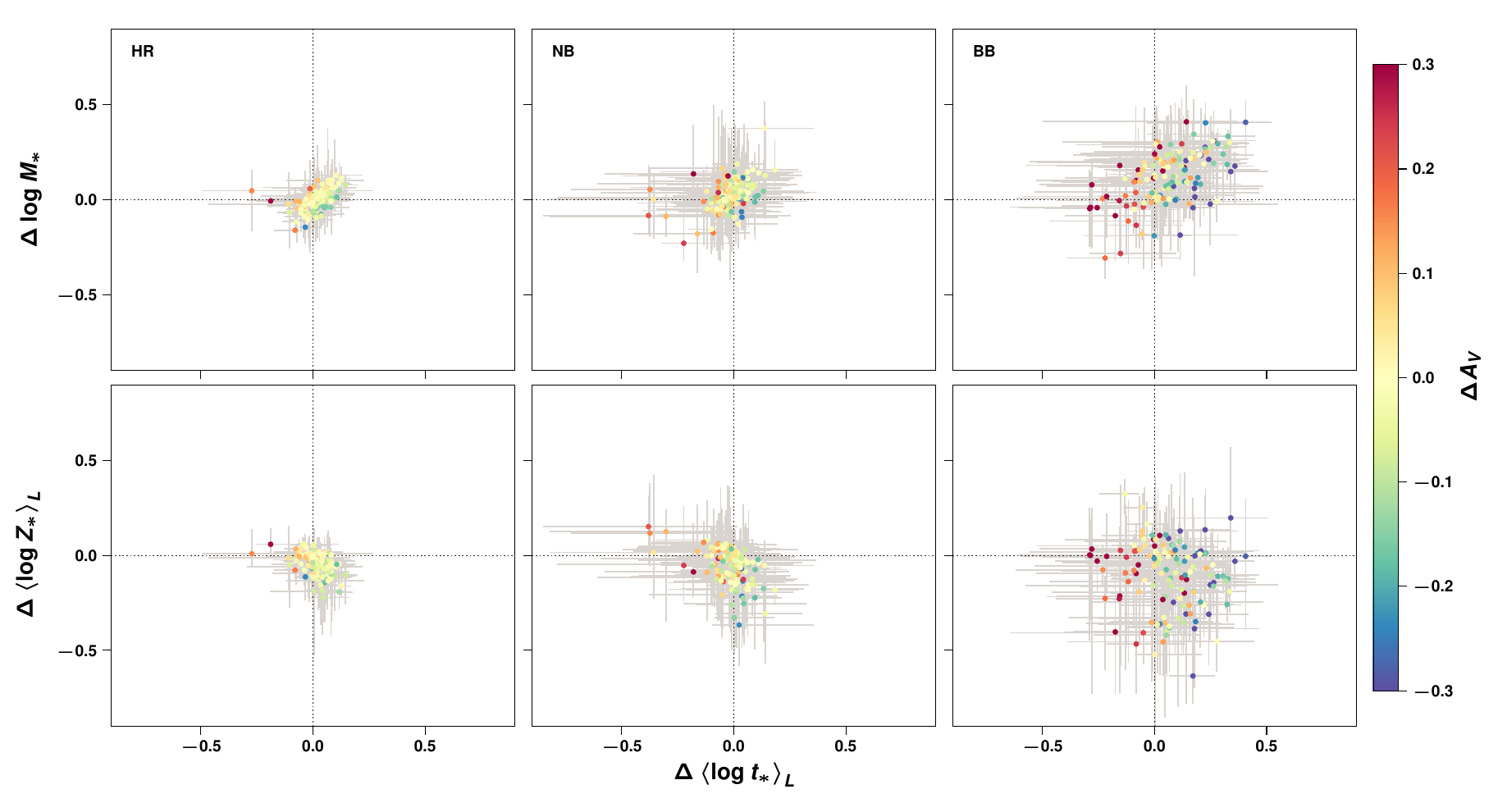}
\includegraphics[scale=1.0]{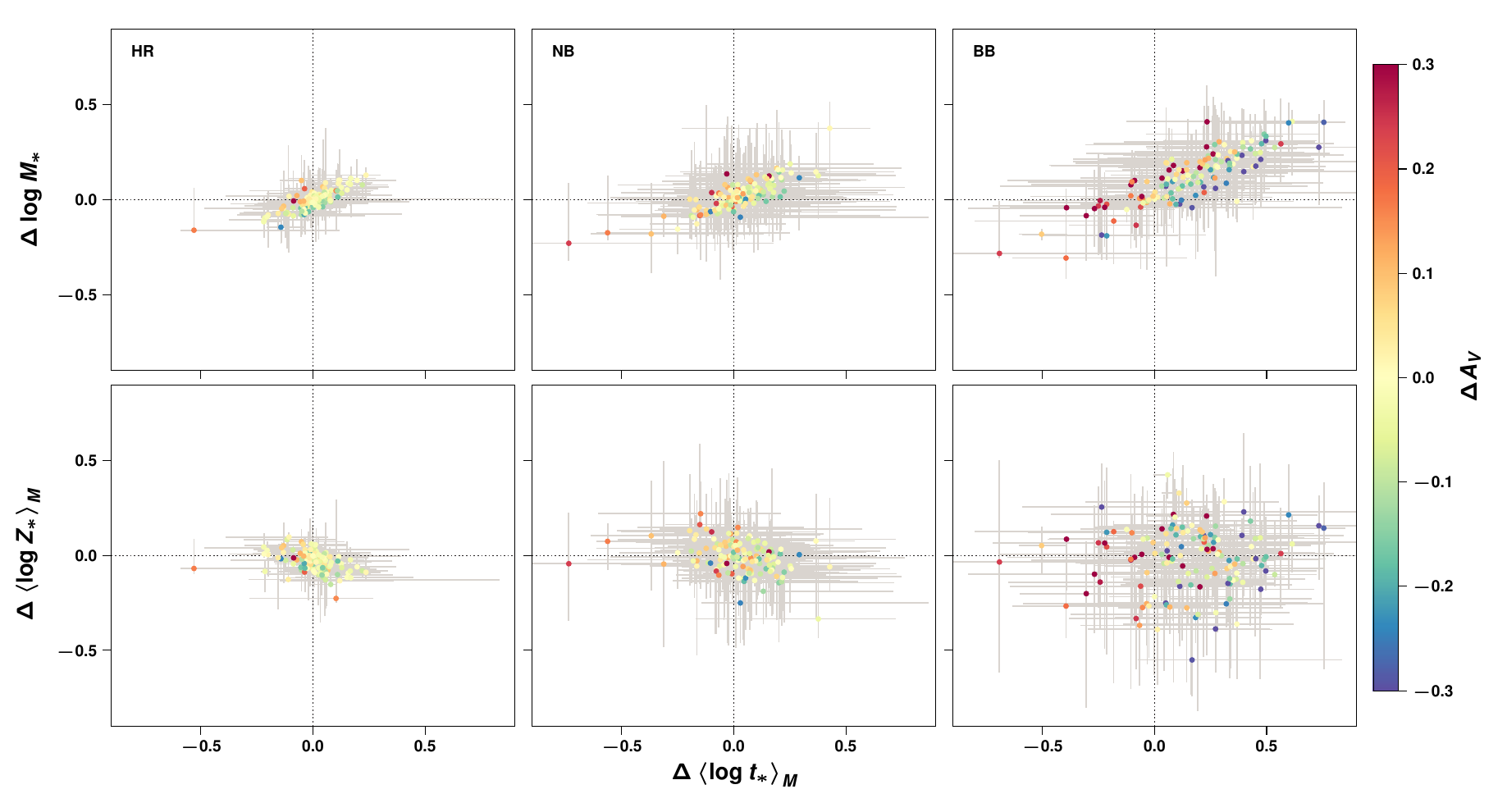}

\caption{\textit{Top:} The residuals for the several physical properties we discuss in this paper.
Each point and cross represents the median and the $16\,$---$\,84$th percentile range of the
residuals for each galaxy in the mock sample over the $20$ noise realisations. It is clear that the
trends (whenever present) remain the same as in Fig.~\ref{fig:residual-corrs}. \textit{Bottom:} Same
as top frame, but for the mass-weighted mean age and metallicity. See text for details.}

\label{fig:individual-residual-corrs}

\end{figure*}

Fig.~\ref{fig:individual-residuals} shows the residual distributions of four SFGs (top rows) and
four PaGs (bottom rows) over their corresponding $20$ noise realisations. The median residual is
represented by the arrows at the HR, NB and BB spectral resolutions (light blue, dark blue and grey,
respectively). The HR determinations are, in general, less biased and more precise, as signaled by
the dispersion of the distributions, with a tendency to slightly improve towards PaGs. The NB and
BB, on the other hand, show larger biases and imprecisions in SFGs, with a remarkable improvement
towards PaGs. From the histograms in Fig.~\ref{fig:individual-residuals} we compute the median and
the $16$ and $84$th percentiles. In Fig.~\ref{fig:individual-residual-corrs} we represent these
median values in $134$ dots, along with the $16\,$---$\,84$th percentile range (cross), at the three
spectral resolutions HR, NB and BB. The observed behaviour resembles the one in
Fig.~\ref{fig:residual-corrs} to great extent, which then again demonstrates the intrinsic origin of
these correlations.

\end{document}